\documentstyle[epsfig,hhline,amsmath,amssymb]{elsart}

\newcommand{\be}{\begin{eqnarray}}
\newcommand{\ee}{\end{eqnarray}}
\newcommand{\ba}{\begin{array}{rcl}}
\newcommand{\ea}{\end{array}}
\newcommand{\bma}{\begin{displaymath}}
\newcommand{\ema}{\end{displaymath}}
\newcommand{\bfi}{\begin{figure}}
\newcommand{\efi}{\end{figure}}
\newcommand{\bc}{\begin{center}}
\newcommand{\ec}{\end{center}}

\newcommand{\clearemptydoublepage}{\newpage{\pagestyle{empty}\cleardoublepage}}

\newcommand{\boldpi}{\mbox{\boldmath $\pi$}}
\newcommand{\boldrho}{\mbox{\boldmath $\rho$}}
\newcommand{\boldaz}{\mbox{\boldmath $a_0$}}
\newcommand{\boldomega}{\mbox{$\omega$}}
\newcommand{\boldsigma}{\mbox{$\sigma$}}
\newcommand{\boldtau}{\mbox{\boldmath $\tau$}}
\newcommand{\boldT}{\mbox{\boldmath $T$}}
\newcommand{\barr}[1]{\not\mathrel #1}
\newcommand{\dis}{\displaystyle}

\begin{document}
\hfill FZJ-IKP(TH)-2003-16
\begin{frontmatter}
\title{Meson production in nucleon--nucleon collisions close to the threshold}

\author{C. Hanhart}

{\small Institut f\"{u}r Kernphysik, Forschungszentrum J\"{u}lich GmbH,}\\ 
{\small D--52425 J\"{u}lich, Germany} \\

\begin{abstract}
  The last decade has witnessed great experimental progress that has led to
  measurements of near threshold cross sections---polarized as well as
  unpolarized---of high accuracy for various inelastic nucleon--nucleon
  collision channels. These data, naturally, pose challenges to theorists to
  develop methods by which they can be understood and explained in
  commensurate detail.

In this work we review the status of the present theoretical
understanding of this class of reactions  with special emphasis on
model--independent methods.  We discuss in detail not only the many
observables involved in the reactions, but also the physical questions that
can be addressed by studying them in the various reaction channels.  The
special advantages of nucleon--nucleon induced reactions are stressed.
Foremost among these is the use of the initial and final states as a
spin/isospin filter.  This opens, for example, a window into the spin
dependence of the hyperon--nucleon interaction and the dynamics of the light
scalar mesons.

\end{abstract}


\end{frontmatter}


\pagenumbering{roman}

\include{titel-n}
\tableofcontents

\clearemptydoublepage
\pagenumbering{arabic}

\setcounter{page}{1}

\section{Introduction}
  
\subsection{Strong Interactions at low and medium Energies}

Quantum Chromo Dynamics (QCD), as the well--accepted theory of the
Strong Interaction, unfolds its impressive predictive power in
particular at high energies.  However, with decreasing momentum
transfer the mathematical structure of QCD becomes increasingly
complicated because the perturbative expansion in the coupling
constant no longer converges.  The fact that particles carrying strong
charge (color) have never been observed in an isolated state is just
one prominent example of the non--perturbative character of QCD at
low energy.  Al\-though a large amount of new data is available from
measurements with {\it electromagnetic} probes (e.g., from MAMI at Mainz,
ELSA at Bonn, and JLAB at Newport News), there is still much to be
learned about the physics with {\it hadronic} probes at intermediate
energies, comprising the investigation of production, decay, and
interaction of hadrons. An important class of experiments in this
context is meson production in nucleon--nucleon, nucleon--nucleus, and
nucleus--nucleus collisions close to the production thresholds.
Recently, two reviews on this subject \cite{machnerrep,oelertrep} were
published, however both had their main emphasis on the experimental
aspects. In this report we will focus on recent theoretical
developments and insights. In addition in neither of these reviews
was the potential of using polarization---which will be the main
emphasis of this work---discussed in detail.

It is interesting to ask {\it a priori} what physics questions can be
addressed with the production of various mesons near threshold in $pp$,
$pd$, and $dd$ collisions with stored and extracted polarized beams.
We therefore present here a brief list, naturally influenced by
the personal taste of the author, that is not aimed at completeness,
but more at giving the reader a flavor of the potential 
of meson production reaction in nucleonic collisions for gaining insight into
the strong interaction physics at intermediate energies.

\begin{itemize}
\item {\it Final state interactions}: In practice it is difficult
to prepare secondary beams of unstable particles with sufficient
accuracy and intensity that allow to study experimentally
the scattering of those particles off nucleons. Here
production reactions are an attractive alternative. One prominent example in
this context is the hyperon--nucleon interaction that can
be studied in the reaction $pp\to pYK$, where $Y$ denotes
a hyperon ($\Lambda$, $\Sigma$ etc.) and $K$ denotes
a kaon. From the invariant mass distributions of the $YN$ system, 
information about the on--shell $YN$ interaction can be extracted.
We shall return to this point in sec. \ref{secdisp}.
\item {\it Baryon resonances in a nuclear environment}: By design, in
  nucleon--nucleon and nucleon--nucleus collisions we study systems with
  baryon number larger than 1. This allows to study particular resonances in
  the presence of other baryons and excited through various exchanged particles.
  One example is the $N^*$(1535), which is clearly visible as a bump in
  any $\eta$ production cross section on a single nucleon. A lot is known
  about this resonance already, however, many new questions can be
  answered by a detailed study of the reaction $NN\to NN\eta$. In the
  close--to--threshold regime, kinematics and the conservation of total angular
  momentum constrains the initial state to only a few partial waves. This, in
  combination with the dominance of the $N^*(1535)$ in the $\eta$ production
  mechanism, allows for a detailed study of the $NN\to N^*N$ transition
  potential. This is relevant to understanding the behavior of the $N^*$ in any
  nuclear environment as well as might reveal information on the structure of
  the resonance. For a detailed discussion we refer to sec. \ref{eta}.

 In hadronic reactions there is additionally a large number of
excitation mechanisms for intermediate resonances possible.
Besides pseudoscalars and vector mesons, which are
more cleanly studied in pion and photon induced reactions,
scalar excitations are possible. It was shown recently that,
for instance, the Roper resonance is rather easily excited by
a scalar source \cite{osetroper}, and thus hadronic reactions might be the
ideal place to study this controversial resonance \cite{morsch}.

\item{\it Charge symmetry breaking} (CSB): Having available several possible
  initial isospin states ($pp$, $pn$, $dd$, etc.), which all have several
  possible spin states, allows experiments that make the study of symmetry
  breaking easily accessible. One example is the impact of charge symmetry
  breaking on the reaction $pn\to d\pi^0$: it leads to a forward--backward
  asymmetry in the differential cross section (c.f. section \ref{syms} and
  Refs. \cite{csbrep,allenapi0}).  It is important to note that in case of the
  pion production the leading
  charge symmetry breaking operators are linked to the up--down mass
  difference.  

A forward--backward asymmetry in the reaction $pn\to
  d(\pi^0\eta)_{(s-wave)}$ should allow extraction of the CSB $f_0-a_0$ mixing
  matrix element---a quantity believed to give insights into the structure of
  the light scalar mesons.

 Once the leading symmetry breaking
mechanism is identified, one can use the specific signals
caused by CSB to extract information on symmetry conserving matrix
elements. For details on the idea as well as an application---extracting information on
the existence of mesonic bound states---we refer the reader to Ref. \cite{mix}.

\item{\it Effective field theory in large momentum transfer reactions}:
Pion production in nucleon--nucleon collisions is still amenable to treatment within
 chiral perturbation theory---if the expansion is adapted to
the large momentum transfer typical of those reactions. This
research is still in progress, however once completed it will
allow not only to study systematically the CSB pion production
mentioned above, but also to 
pin down the size of three body forces relevant for 
a quantitative understanding of $pd$ scattering, and to
include the dispersive corrections
to $\pi d$ scattering in a chiral perturbation theory analysis.
The latter is necessary for an accurate extraction
of the isoscalar $\pi N$ scattering length from deuteron
reactions, which are otherwise difficult  to access. This issue will
be discussed in chapter \ref{cpt}.
\end{itemize}
The wealth of information comes with the drawback that, apart from rare cases,
it is difficult to extract a particular piece of information from the data.
For example, resonances and final state interactions modify in a coherent
superposition the invariant mass plot.  Fortunately,  polarization
can act as a spin filter and different contributions to the
interaction can be singled out because they show up in the angular
distributions of different spin combinations.  The various polarization
observables will be discussed in detail in sec. \ref{pol}.

\begin{figure}[t]
\begin{center}
\epsfig{file=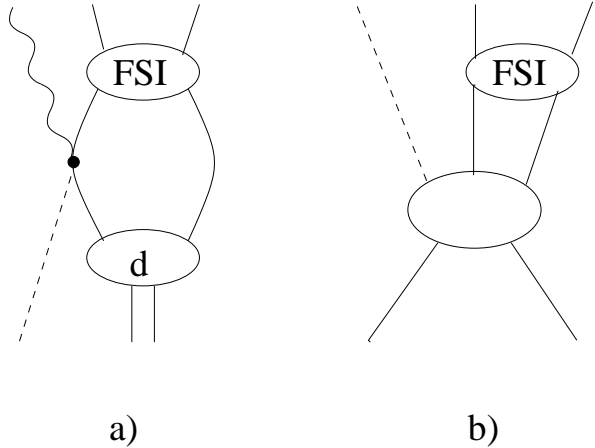, height=6cm}
\caption{\it Comarison of two different reactions that can be used to extract
  parameters of a strong final state interaction. Shown is a typical low
  momentum transfer reaction (diagram a), e.g.,  $K^- d\to \gamma \Lambda n$)
  and a large momentum transfer reaction (diagram b), e.g.,  $pp\to
  pK^+\Lambda$). }
\label{lmstvslmst}
\end{center}
\end{figure}

One characteristic feature of meson production in $NN$ collisions at the first
glance looks like a disadvantage with respect to a possible theoretical
analysis: the large momentum transfer makes it difficult to reliably construct
the production operator. However, as we will discuss in detail, due to this
the production operator is largely independent of the relative energy of a
particular particle pair in the final state\footnote{If there are resonances
  near by this statement no longer true. See the discussion in sec.
  \ref{pol}.} if we stay in the regime of small invariant masses.  Because of
this, dispersion relations can be used to extract low energy scattering
parameters of the final state interaction and at the same time give a
model--independent error estimate.  In contrast, many reactions dominated by
one--body currents that are characterized by small momentum transfers (c.f.
Fig.  \ref{lmstvslmst}). This makes a quantitatively controlled,
model--independent analysis difficult. For example, it was shown in Refs.
\cite{gibbsnn,gibbsln} that value for the neutron--neutron ($\Lambda N$)
scattering length extracted from $\pi^-d\to \gamma nn$ ($K^-d\to \gamma \Lambda n$),
where the initial state is in an atomic bound state, is
sensitive to the short--range behavior of the baryon--baryon interaction
used\footnote{This implies that meson exchange currents
  should contribute at the same order of magnitude (c.f. discussion in sec. \ref{offshell}).}.


\subsection{Theoretical approaches to the reaction $NN\to B_1B_2x$}
\label{thapp}

In this chapter we will briefly describe the theoretical developments 
for the reactions $NN\to B_1B_2x$ close to the threshold, where the $B_i$ denote
the two outgoing baryons (either $NN$ or, e.g., in case of the production of a
strange meson $YN$, where $Y$ denotes a hyperon).
 
The development of theoretical models especially for the reaction $NN\to NNx$ has a long
history. For a review of earlier works we refer the reader to the book by H.
Garcilazo and T. Mizutani \cite{GM}.  Most of the models can be put into one
of two classes: namely those that use the distorted wave Born approximation,
where a production operator that is constructed within some perturbative
scheme is convoluted with nucleonic wavefunctions, and truly non--perturbative
approaches, where integral equations are solved for the full ($NN$,$NNx$)
coupled--channel problem.

We start this section with a brief review of the second class.  The obvious
advantage of this kind of approach is that no truncation scheme is required
with respect to multiple rescattering.  This is a precondition to truly
preserve three--body unitarity \cite{3dnnpi1,3dnnpi2,3dnnpi3}.  However, as we
will see, this approach is technically rather involved and there is
significant progress still to be made that will allow for quantitative
predictions, especially for reactions with unbound few--nucleon systems in the
initial and final state.

One obvious complication is to preserve the requirements of chiral symmetry,
which are very important---at least for the reaction $NN\to NN\pi$ close to
the threshold, as will be discussed in the following sections, since in its
non linear realization it is necessary to consistently take into account
one--, two-- and more pion vertices as soon as loops are considered.
Naturally, this can only be done within a truncation scheme that is consistent
with chiral symmetry.  As a possible solution, in Ref. \cite{blank2} it was
suggested to change to a linear representation, as given, e.g., by the linear
sigma model. To our knowledge this area is not yet very well developed.

All the calculations so far carried out for pion\footnote{To be concrete in
  this section we will talk about pions only. They are, in fact, the only
  mesons that have so far been studied with integral equation approaches.}
production within the full ($NN$,$NNx$) coupled--channel system, where
performed within time ordered perturbation theory. Within this scheme, one way
to impose unitarity is it to demand, that the number of pions in any given
intermediate state should not to exceed a given number.  This, in combination
with that there is no particle number conservation, implies the fact, that the
$\pi NN$ coupling constant generated by the equations at some intermediate
time depends on the number of pions in flight at that time.  The same is true
for the renormalization of the nucleon pole. Put to the extreme: in the
nuclear matter limit the dressed pion nucleon coupling vanishes \cite{ssf}
(see also the discussion in Ref. \cite{blankleider2}).  Another problem
inherently present in time--ordered formalisms was pointed out by Jennings
\cite{jennings}: namely that it is impossible simultaneously to fulfill the
Pauli--Principle and the unitarity condition within these integral equations.
A consequence of this is a wrong sign for $T_{20}$ for elastic $\pi d$
scattering at backward angles \cite{lamot}.  Both of these problems can be
avoided within a covariant formulation of the integral equations as is given
in Refs. \cite{blank,afnan}.  Unfortunately, not only are those equations
rather involved, they also need covariant $T$ matrices for both subsystems.
Although these exist (e.g., Ref.  \cite{andrew} for the $\pi N$ system and
Ref. \cite{tjon} for the $NN$ system), they were not constructed consistently.
Also the three dimensional convolution approach of Refs. \cite{blankleider1,blankleider2}
does not suffer from the above mentioned problems, but also within this scheme
  so far no results for the ($NN$,$NNx$) were published.
\begin{figure}
\begin{center}
\vskip 5cm          
\includegraphics{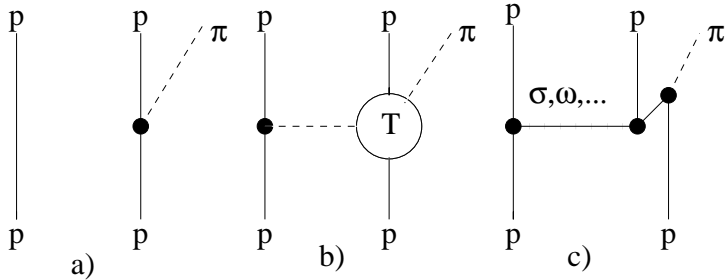} 
\caption{\it Some possible contributions to
the production operator for neutral pion production in $pp$ collisions.
Solid lines are nucleons, dashed lines are pions.}
\label{kur}
\end{center}
\end{figure}

On the other hand there are the approaches that treat the production
operator perturbatively (typically diagrams
of the type shown in Fig. \ref{kur} are considered),
 while including the nucleon--nucleon or,
more generally, the baryon--baryon interaction in the final and
initial state non--perturbatively.
Therefore in this formalism three--body unitarity can be achieved only
perturbatively,
however, 
this formalism at least fulfills the minimal requirement:
to take fully into account the non--perturbative nature
of the $NN$ interaction. This is indeed  necessary,
as was confirmed experimentally, since
a proper inclusion of the $NN$ final state interaction
is required to describe the energy dependence of the
reaction $pp\to ppX$ \cite{machnerrep,oelertrep}.
A production operator that is constructed perturbatively
allows in addition to treat the interactions between the
subsystems without approximation and consistent with chiral symmetry
(c.f. discussion in sec. \ref{cpt}). 
A priori, it is unclear
according to what rules the production operator should be constructed
and only a comparison with experiment can tell if the
approach is appropriate or not. 
There is, however, one observation in favor of a perturbative
treatment of the production operator: in the case of pion
production, effective field theory methods can be applied to show
 that as long as there is a meson exchange current
in leading order it dominates the low energy amplitude. Loops
undergo strong cancellations and thus might well be neglected.
If, however, there is no meson exchange current at leading order
(as is the case for the reaction $pp\to pp\pi^0$) the situation
is significantly more involved and loops might well be
significant\footnote{Note, however, even then a perturbative treatment of the
  production operator is still justified.}.
This issue will be discussed in detail in sec. \ref{pionprod}.
In this introductory chapter we only wish to stress, that to 
our present knowledge the case of $\pi^0$ seems to be an exception
and, in addition to the reaction $pp\to pn\pi^+$
 all heavy meson production cross sections close to the threshold
are indeed dominated by meson exchange currents\footnote{Here
this phrase is to be understood to include resonance excitations as well.}.
We must add,
however, that the of heavier 
mesons can also give a significant contribution.


An approach that lies somewhere between those that solve the integral
equations for the whole $NN\to NN\pi$ system and those that treat the
production operator perturbatively is that of Refs. \cite{tjon,EHSM,OBEPQB}.
These works document the attempt to extend straight--forwardly what is known
about the phenomenology of nucleon--nucleon scattering (see for example
\cite{MHE}) to energies above the pion threshold. The three--body
singularities stemming from both the energy dependent pion exchange as well as
the nucleon and Delta self energies were taken into account. Via the optical
theorem the pion production cross sections can then be calculated in a
straight forward way. It turned out, however, that the models could not
describe the close to threshold data. The reasons for this will become clear in chapter
\ref{pionprod}. In sec. \ref{pionlessons} we will discuss how the recent
progress in our understanding of pion production shows what is needed in order
to improve, for example, the work of Ref. \cite{EHSM}.

%

The construction of theoretical models for the production of mesons heavier
than the pion started only recently, after the accelerator COSY came into
operation. Most of these models are one meson exchange models. Those
approaches will be described to some extend in section \ref{remheavy}.

\subsection{Specific aspects of hadronic meson production close to the threshold}


In the near--threshold regime the available phase space changes
very quickly. Thus, especially when comparing different reactions,
an appropriate measure of the energy 
with respect to that particular threshold 
is required. In pion production traditionally the
variable $\eta$, defined as the maximum pion momentum in units of the
pion mass, is used to measure the excess above threshold. For all heavier 
mesons the so--called excess energy $Q$, defined as
$$
Q = \sqrt{s}-\sqrt{s^{(thres)}} \ ,
$$
is normally used. In Appendix \ref{kin} we give the explicit formulas that
relate $Q$ to $\eta$.  It is useful to understand the physical meaning of the
two different quantities: $Q$ directly gives the energy available for the
final state. The interpretation of $\eta$ is somewhat more involved.  In a
non--relativistic, semiclassical picture the maximum angular momentum allowed
can be estimated via the relation $l_{max} \simeq Rq'$, where $q'$ denotes the
typical momentum of the corresponding particle and $R$ is a measure of the
range of forces.  If we identify $R$ with the Compton wavelength of the meson
of mass $m_x$, we find $l_{max} \simeq q'/m_x \simeq \eta$.  This
interpretation was given for example in Refs. \cite{rosenfeld,gmw}.

When trying to compare the cross sections of reactions with
different final states one has to choose carefully the variable
that is used for the energy. In Ref. \cite{pawel_etap} it
is shown that the relative strength of the total cross sections
for $pp\to pp\pi^0$, $pp\to pp\eta$ and $pp\to pp\eta '$ is strikingly different 
when comparing them at equal $\eta$ or at equal $Q$. Since the dominant final state
interaction in all of these reactions is the $pp$ interaction,
it appears most appropriate to compare the cross sections at equal $Q$,
for then at any given excess energy the impact of the final state
interaction is equal for all reactions, which is not the case for
equal values of $\eta$. In Ref. \cite{pawel_fsicomp}
a different energy variable was suggested, namely the phase space volume.

In order to produce a meson in nucleon--nucleon collisions the 
kinetic energy of the initial particles needs to be sufficiently large
to put the outgoing meson on its mass shell. To produce a meson of
mass $m_x$ an initial momentum larger than the threshold value 
\begin{equation}
|\vec p_i \, ^{(thres)}|^2=m_xM_N+\frac{m_x^2}{4} 
\end{equation}
is necessary.
As long as we stay in a regime close to the production threshold the
momenta of all particles in the final state are small and  therefore
$p_i^{(thres)}$ also sets the scale for the typical momentum transfer.
In a non--relativistic picture a
large momentum transfer translates to a small reaction volume, defined
by a size parameter $R \sim 1/p_i$. Thus, already for pion production $R$ is
as small as
 $R \sim 0.5$ fm and is getting even smaller for heavier mesons.

\begin{figure}[t]
\begin{center}
\epsfig{file=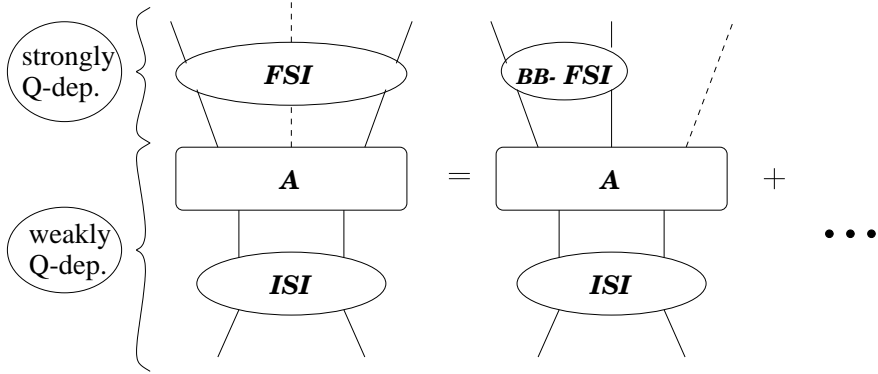, height=5cm}
\caption{\it Sketch of the production reaction showing the
initial state interaction amongst the two nucleons (ISI), the
final state interaction of all outgoing particles (FSI) as
well as the production operator $A$ as defined in the
text. The left diagram indicates the complexity possible, whereas
the right diagram shows the first and potentially leading term
for the FSI only.}
\label{isifsi}
\end{center}
\end{figure}

Therefore the two nucleons in the initial state have to approach 
each other very closely before the production of the meson
can happen. It then should come as no surprise that
it is important for quantitative predictions to understand the
elastic as well as inelastic $NN$ interaction in the initial state.
However, since for all reactions we will be looking at the
initial energy is significantly larger than the excess energy
$Q$, the initial state interaction should at most mildly influence
the energy dependence.
The energy dependence of the production operator should also 
be weak, for it should be controlled by the typical momentum transfer,
which is significantly larger than the typical outgoing momenta.

On the other hand, in the near--threshold regime all particles
in the final state have low relative momenta and thus 
undergo potentially strong final state interactions that can induce strong
energy dependences. The different parts of the matrix element are
illustrated in Fig. \ref{isifsi}. 

In the next subsections we will briefly sketch some properties of
meson production in nucleon--nucleon collisions in rather
general terms with the emphasis on the gross features.
Details, as well as selected results for various reactions,
will be presented in the subsequent sections.

\subsection{Remarks on the production operator}
\label{remprodop}

As was mentioned above one of the characteristics of meson production
in nucleon--nucleon collisions is the large momentum transfer. This
leads to a large momentum mismatch for any one--body operator that might
contribute to the production reaction. 

\begin{figure}[t!]
\begin{center}
\epsfig{file=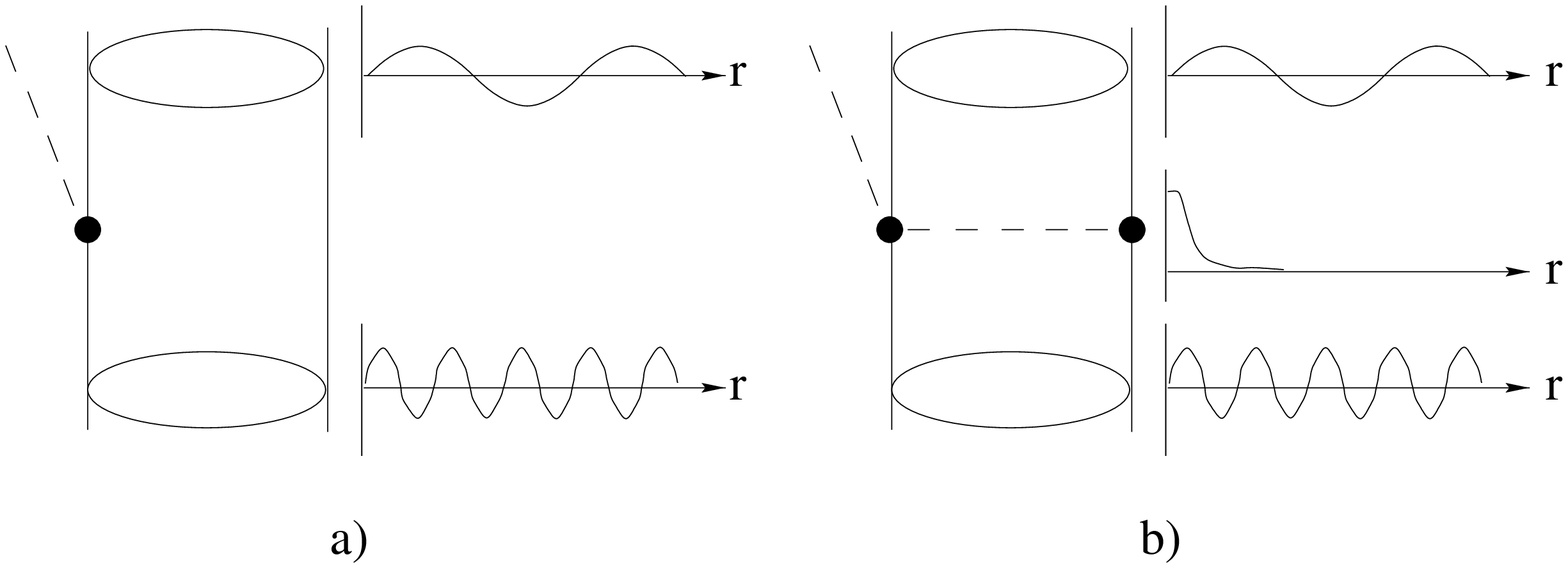, height=5cm}
\caption{\it Illustration of the momentum mismatch. The picture
shows why one should a priori expect meson exchange currents
to be very important close to the threshold.}
\label{missmatch}
\end{center}
\end{figure}

This reasoning becomes most transparent when looking at it in real space,
as shown in Fig. \ref{missmatch}. In case of the one--body operator
(panel a) the
evaluation of the matrix element involves the convolution of a rapidly
oscillating function with a mildly oscillating function, which  
generally will lead to significant cancellations. 
If there is a meson exchange present, as sketched in panel b, the cancellations
will not be as efficient. Thus
 one should expect that if there is a meson
exchange current possible at leading order, it should dominate the production
process. This picture was confirmed in explicit model calculations (see e.g.
Ref. \cite{KuR}).

As mentioned above one should in any case expect the production operator to be
controlled by the large momentum transfer. Thus variations of the individual
amplitudes with respect to the total energy should be suppressed by
$(p'/p)^2$, where $p'$ ($p$) is the relative momentum of an outgoing to
particle pair (of the initial nucleons).  Therefore it should be negligible close to the
production threshold. This will be used for model--independent analyses of the
production of heavy mesons and was checked within the meson exchange picture
(see discussion in Ref. \cite{ynfsi} and sec. \ref{secdisp}).

\subsection{The role of different isospin channels}
\label{isospinrole}

One big advantage of meson production in nucleon--nucleon collisions as
well as nucleon--nucleus and nucleus--nucleus collisions is, that
besides the spin of the particles there is another internal degree
of freedom to be manipulated: the isospin.

Nucleons are isospin--(1/2) particles in an $SU(2)$ doublet (isospin $t=1/2$)
with $t_3=+1/2$ and $t_3=-1/2$ for the proton ($p$) and neutron ($n$), respectively.
Thus a two nucleon pair can be either
in an isotriplet state (total isospin $T=1$ with the three possible projections of the
total isospin $T_3=+1$ for a $pp$ state, $T_3=0$ for a $pn$ state and
$T_3=-1$ for a $nn$ state) or in an isosinglet state ($T=0$ with $T_3=0$ for a $pn$ 
state).

Let us first concentrate on the isospin conserving situation.
Even then various transitions are possible, depending on whether
an isovector or an isoscalar particle is produced. 
Let us denote the allowed
 transition amplitudes by $A_{T_iT_f}$, where $T_i$($T_f$) denote the
total isospin of the initial (final) $NN$ system \cite{rosenfeld}.
Then, in the case of the production of an isovector, the amplitudes 
$A_{11}$, $A_{10}$, and $A_{01}$ are possible. They could
be extracted individually from $pp\to ppx^0 \propto |A_{11}|^2$,
 $pp\to pnx^+ \propto |A_{11}+A_{10}|^2$, and
 $pn\to ppx^- \propto 1/2|A_{11}+A_{01}|^2$, where
the factor of $1/2$ in the latter case stems from the
isospin factor of the initial state (we have $|pn\rangle = 1/\sqrt{2}(|10\rangle
+|00\rangle)$ in which the isospin states are labeled by both
the total isospin as well as their projection). On the
other hand, for the production of an isoscalar there
are two transitions possible, namely $A_{00}$ and $A_{11}$, where
 $pp\to ppx \propto |A_{11}|^2$ and
 $pn\to pnx \propto 1/2|A_{11}+A_{00}|^2$.
In this case the measurement of two different reaction
channels allows extraction of all the production amplitudes.

The deuteron is an isoscalar and thus acts as an isospin filter. 
Accordingly,
one finds for the production of any isovector particle
$x$, $A(pp\to dx^+)=2A(pn\to dx^0)$, as long as isospin is conserved.
The consequences of isospin breaking will be discussed in sec. \ref{syms}.

The relative strength of the different transition amplitudes
proved to provide significant information about the production
operator. As an example let us look at the production of an
isoscalar particle. If the production operator is dominated by 
an isovector exchange the corresponding isospin structure
of the exchange current is $\hat O_{iv}=(\vec \tau_1\cdot \vec \tau_2)$,
where $\tau_i$ denotes the isospin operator of nucleon
 $i$, which is clearly distinguished
from the structure corresponding to an isoscalar exchange $\hat O_{is}=1$.
Thus one finds\footnote{The easiest way to see this is to
express $\hat O_{iv}$ in terms of the Casimir operators of the underlying
group:
$$\vec \tau_1\cdot \vec \tau_2 = 4\hat t_1 \cdot \hat t_2 = 2(\hat T ^2-\hat
t_1^2-\hat t_2^2) \ ,$$
with $< T ^2>=T(T+1)$ and $<t_i^2>=3/4$.}

$$
\langle TT_3|\hat O_{iv}|TT_3 \rangle = 2T(T+1)-3 =
\begin{cases}
\phantom{-}1 & \text{for $T=1$} \ , \\ -3 & \text{for $T=0$}  \ .
\end{cases}
$$
On the other hand
$$
\langle TT_3|\hat O_{is}|TT_3 \rangle = 1 .
$$
For two given $NN$ states a different total isospin implies that also either
the total spin or the angular momentum are different (c.f. next section). 
Therefore, different isospin states can not interfere in the total cross
section and
one may expect  
$$
\frac{\sigma_{tot}(pn\to pn x)}{\sigma_{tot}(pp \to pp x)}\sim
\begin{cases}
5 & \text{if isovector exchanges dominate} \ , \\ 1 & \text{if isoscalar
exchanges dominate}  \ .
\end{cases}
$$
For $x=\eta$ a recent measurement at Uppsala \cite{calen2} gave
for this ratio a value of 6.5, clearly indicating a dominance
of isovector exchanges. It should be stressed that the estimates
given here only hold if effects from initial and final state interactions,
the spin dependence of the production operator,
as well as all other dynamical effects are completely
neglected. However, even in the most detailed
studies, the bulk of the ratio stems from the isospin factors \cite{vadim,kanzoeta,colineta}.

\subsection{Selection Rules for $NN\to NNx$}
\label{sr}

In this section we will present the selection rules relevant
for nucleon--nucleon induced meson production. Those rules
are based on the symmetries of the strong interaction that imply
 the conservation of  parity, total angular momentum
 and isospin. 
 
 A two--nucleon system has to obey the Pauli Principle, which implies $
 (-)^{L+S+T} = (-1)$ where $L$, $S$ and $T$ denote the angular momentum,
 total spin and total isospin of the two nucleon system respectively. In the case of
 a two proton system, for example, where $T$=1, $L+S$ needs to be even. Consequently, for
 $T=1$ all even angular momenta, and therefore all even parity states, are
 spin singlet ($S=0$) states and consequently have $J=L$. On the other hand,
 for $T=0$ states it is the odd parity states that are spin singlets.

In addition, for a reaction of the type $NN\to NNx$ we find from
parity conservation 
\begin{equation}
(-)^L=\pi_x(-)^{(L'+l')} \ ,
\label{parity}
\end{equation}
where $\pi_x$ denotes the intrinsic parity of particle $x$ and $L$, $L'$ and
$l'$ denote, respectively, the angular momentum of the incoming two nucleon system, of the
outgoing two--nucleon system and of particle $x$ with respect to 
the outgoing two--nucleon system. 

We may now combine the two criteria to write
\begin{equation}
(-)^{(\Delta S+ \Delta T)}=\pi_x(-)^{l'} \ ,
\label{parpauli}
\end{equation}
where $\Delta S$ ($\Delta T$) denotes the change in total (iso)spin
when going from the initial to the final $NN$ system.

\begin{table}
\begin{center}
\begin{tabular}{|c|c|c|ccc|cccc|}
\hline
 $L_{NN}l_x$ &  
\, $(NN)_i$ \,& 
\, $(NN)_fl'$ \, & \, S \, & \, L \, & \, J \, & \, S' \, & \, L' \, & \, j' \, & \, l' \,\\
\hline
\hline
\multicolumn{10}{|l|}
{$NN|_{(T=1)}\to NN|_{(T=1)}+\mbox{pseudo--scalar}$ (e.g.,  $pp\to pp\pi^o$)} \\
\hline
 \, Ss \, & $^3P_0$  & $^1S_0 \, s$ & 1 & 1 & 0 & 0 & 0 & 0 & 0    \\
\hline
 \, Sp \, & \multicolumn{9}{c|}{\it not allowed}    \\
\hline
 \, Ps \, & $^1S_0$  & $^3P_0 \, s$ & 0 & 0 & 0 & 1 & 1 & 0 & 0    \\
 & $^1D_2$    & $^3P_2 \, s$ & 0 & 2 & 2 & 1 & 1 & 2 & 0    \\
\hline
\hline
\multicolumn{10}{|l|}
{$NN|_{(T=1)}\to NN|_{(T=0)}+\mbox{pseudo--scalar}$ (e.g.,  part of $pp\to pn\pi^+$)} \\
\hline
 \, Ss \, & $^3P_1$  & $^3S_1 \, s$ & 1 & 1 & 1 & 1 & 0 & 1 & 0    \\
\hline
 \, Sp \, & $^1S_0$  & $^3S_1 \, p$ & 0 & 0 & 0 & 1 & 0 & 1 & 1    \\
 & $^1D_2$    & $^3S_1 \, p$ & 0 & 2 & 2 & 1 & 0 & 1 & 1    \\
\hline
 \, Ps \, & \multicolumn{9}{c|}{\it not allowed}    \\
\hline
\hline
\multicolumn{10}{|l|}
{$NN|_{(T=0)}\to NN|_{(T=1)}+\mbox{pseudo--scalar}$ (e.g.,  part of $pn\to pp\pi^-$)} \\
\hline
 \, Ss \, & \multicolumn{9}{c|}{\it not allowed}    \\
\hline
 \, Sp \,
 & $^3S_1$  & $^1S_0 \, p$ & 1 & 0 & 1 & 0 & 0 & 0 & 1    \\
 & $^3D_1$  & $^1S_0 \, p$ & 1 & 2 & 1 & 0 & 0 & 0 & 1    \\
\hline
 \, Ps \, 
& $^3S_1$  & $^3P_1 \, s$ & 1 & 0 & 1 & 1 & 1 & 1 & 0    \\
& $^3D_J$  & $^3P_J \, s$ & 1 & 2 & J & 1 & 1 & J & 0    \\
\hline
\hline
\multicolumn{10}{|l|}
{$NN|_{(T=0)}\to NN|_{(T=0)}+\mbox{pseudo--scalar}$ (e.g.,  part of $pn\to pn\eta$)} \\
\hline
 \, Ss \, & $^1P_1$  & $^3S_1 \, s$ & 0 & 0 & 0 & 1 & 0 & 1 & 1    \\
\hline
 \, Sp \, 
& $^3S_1$  & $^3S_1 \, p$ & 1 & 0 & 1 & 1 & 0 & 1 & 1    \\
& $^3D_J$  & $^3S_1 \, p$ & 1 & 2 & J & 1 & 0 & 1 & 1    \\
\hline
 \, Ps \, 
& $^3S_1$  & $^1P_1 \, s$ & 1 & 0 & 1 & 0 & 1 & 1 & 0    \\
& $^3D_1$  & $^1P_1 \, s$ & 1 & 2 & 1 & 0 & 1 & 1 & 0    \\
\hline
\end{tabular} 
\end{center}
\caption{
{\it The lowest partial waves for the production of a pseudo--scalar
particle in $NN\to NNx$.}}
\label{tab_secr}
\end{table}

\begin{table}
\begin{center}
\begin{tabular}{|c|c|c|ccc|cccc|}
\hline
 $L_{NN}l_x$ &  
\, $(NN)_i$ \,& 
\, $(NN)_fl'$ \, & \, S \, & \, L \, & \, J \, & \, S' \, & \, L' \, & \, j' \, & \, l' \,\\
\hline
\hline
\multicolumn{10}{|l|}
{$NN|_{(T=1)}\to NN|_{(T=1)}+\mbox{scalar}$ (e.g.,  $pp\to ppa_0^0$)} \\
\hline
 \, Ss \, & $^1S_0$  & $^1S_0 \, s$ & 1 & 0 & 0 & 0 & 0 & 0 & 0    \\
\hline
 \, Sp \, & $^3P_1$  & $^1S_0 \, p$ & 1 & 1 & 1 & 0 & 0 & 0 & 1    \\
\hline
 \, Ps \, 
& $^3P_J$  & $^3P_J \, s$ & 1 & 1 & J & 1 & 1 & J & 0    \\
& $^3F_2$  & $^3P_2 \, s$ & 1 & 3 & 2 & 1 & 1 & 2 & 0    \\
\hline
\hline
\multicolumn{10}{|l|}
{$NN|_{(T=1)}\to NN|_{(T=0)}+\mbox{scalar}$ (e.g.,  part of $pp\to pna_0^+$)} \\
\hline
 \, Ss \, & \multicolumn{9}{c|}{\it not allowed}    \\
\hline
 \, Sp \, & $^3P_J$  & $^3S_1 \, p$ & 1 & 1 & J & 1 & 0 & 1 & 1    \\
 & $^3F_2$    & $^3S_1 \, p$ & 1 & 3 & 2 & 1 & 0 & 1 & 1    \\
\hline
 \, Ps \, 
& $^3P_J$  & $^3P_J \, s$ & 1 & 1 & J & 1 & 1 & J & 0    \\
& $^3F_2$  & $^3P_2 \, s$ & 1 & 3 & 2 & 1 & 1 & 2 & 0    \\
\hline
\hline
\multicolumn{10}{|l|}
{$NN|_{(T=0)}\to NN|_{(T=1)}+\mbox{scalar}$ (e.g.,  part of $pn\to ppa_0^-$)} \\
\hline
 \, Ss \, & \multicolumn{9}{c|}{\it not allowed}    \\
\hline
 \, Sp \,
 & $^1P_1$  & $^1S_0 \, p$ & 0 & 1 & 1 & 0 & 0 & 0 & 1    \\
\hline
 \, Ps \, 
& $^1P_1$  & $^3P_1 \, s$ & 0 & 1 & 1 & 1 & 1 & 1 & 0    \\
\hline
\hline
\multicolumn{10}{|l|}
{$NN|_{(T=0)}\to NN|_{(T=0)}+\mbox{scalar}$ (e.g.,  part of $pn\to pnf_0$)} \\
\hline
 \, Ss \, 
& $^3S_1$  & $^3S_1 \, s$ & 1 & 0 & 1 & 1 & 0 & 1 & 0    \\
& $^3D_1$  & $^3S_1 \, s$ & 1 & 2 & 1 & 1 & 0 & 1 & 0    \\
\hline
 \, Sp \, 
& $^1P_1$  & $^3S_1 \, p$ & 0 & 1 & 1 & 1 & 0 & 1 & 1    \\
\hline
 \, Ps \, 
& $^1P_1$  & $^1P_1 \, s$ & 0 & 1 & 1 & 0 & 1 & 1 & 0    \\
\hline
\end{tabular} 
\end{center}
\caption{
{\it The lowest partial waves for the production of a scalar
particle in $NN\to NNx$.}}
\label{tab_secr2}
\end{table}

As an example let us look at the reaction $pn\to d(\pi \eta)_{s-wave}$ --- a
reaction that should provide valuable information on the light scalar
resonances (note: the $\pi \eta$ channel is the dominant decay channel of the
scalar--isovector meson $a_0(980)$, c.f.  chapter \ref{a0f0}).  Based on Eq.
(\ref{edep}), the near--threshold regime is dominated by the lowest partial
waves which in this case would be $S'=1$ and $T'=0$ (deuteron), together with
a relative $s$ wave between the two meson system and the deuteron ($l'=0$).
In addition, the $(\pi \eta)_{s-wave}$ system is a scalar--isovector system.
Therefore we have to use $\pi_x=+$, $l'=0$ and $\Delta T=1$ in Eq.
(\ref{parpauli}), yielding $\Delta S=0$. On the other hand the initial state
needs to be of even parity, but even parity $T=1$ states have $S=0$. Thus, for
the production of a scalar--isovector state the $s$--wave deuteron with
respect to the two meson system is prohibited and, as long as isospin is
conserved, this reaction has to have at least one $p$--wave in the final
state.  On the other hand, for the production of an isoscalar scalar particle,
the $s$--wave final state is allowed.

In a straight--forward way the corresponding selection rules for systems with
nuclei in the initial and/or final state can be easily derived along the same
lines. Bose symmetry, for example, demands a $dd$ system to be symmetric under
exchange of the two deuterons, forcing $L+S$ to be even. Therefore the lowest
partial waves contributing to $dd\to \alpha \eta$ are $^3P_0\to s$ and
$^5D_1\to p$.

Another example of selection rules at work in a system other than $NN\to NNx$
will be given in sec. \ref{hnpol}.

%

%

\section{The final state interaction}
\label{fsisec}

The role of final state interactions in production reactions has been
discussed in the literature for a long time and at various levels of
sophistication.  Extensive discussions can be found in Refs.
\cite{gw,fsibook}. For more recent discussions focusing on the production of
mesons in nucleon--nucleon collisions we refer to Refs.
\cite{saschafsi,withkanzo,ynfsi}. In this section we will approach the
question of the role of the final state interactions from two approaches:
first we will present a heuristic approach that should make the physics clear,
but has only limited quantitative applicability; then, using dispersion
integrals, quantitative expressions will be derived, but these are less
transparent. This investigation will show under what circumstances scattering
parameters can be extracted from high momentum transfer production reactions,
and will address questions about the accuracy  of this extraction.

\subsection{Heuristic approach}
\label{heur}

\begin{figure}[t]
\begin{center}
\epsfig{file=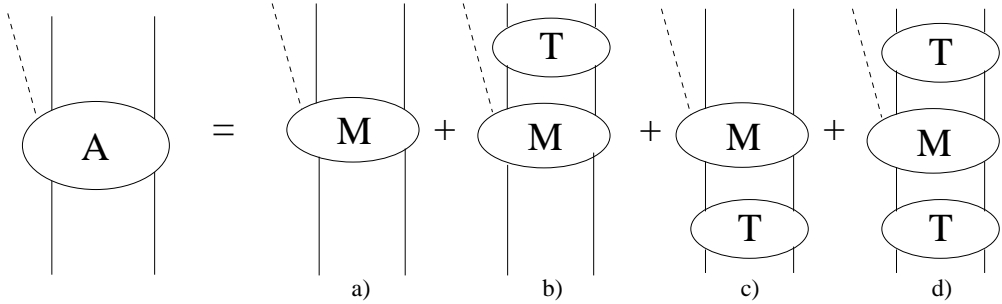, height=4cm}
\caption{\it Diagrammatic presentation of the distorted wave Born
  approximation.}
\label{pwpfsi}
\end{center}
\end{figure}

As early as  1952 Watson pointed out  the circumstances under which one would expect the
final state interaction to strongly modify the energy dependence of the total
production cross--section $NN \to NNx$ \cite{watson}. His rather convincing
argument went as follows: assume in contrast to the production reaction the
pion absorption on a free two nucleon pair. Due to time reversal invariance
the matrix elements for the two reactions are equal. Based on the observation
that the absorption takes place in a small volume, it is obvious that the
reaction is a lot more probable if, in a first step, the two nucleons
come together rather closely due to their attractive interaction and that
this significantly more compact object then absorbs the pion. In this picture the
influence of the two--nucleon system on the energy dependence of the
cross section is very natural. Based on this argument, one should
expect\footnote{Here all final state particles are treated
  non--relativistically for simplicity.}
\begin{equation}
\label{watsonstart}
\sigma_{NN \to NNx}(\eta) \propto \int_0^{m_x \eta} d^3q'p'd\Omega_{p'}
|\Phi_x ^\dagger (q')\Psi_{NN}(p')|^2 \ ,
\end{equation}
where $q'$ ($p'$) denotes the momentum of the outgoing meson (the relative momentum of
the outgoing two nucleon pair) and $d^3q'=dq'd\Omega_{q'}$; energy conservation implies $p'\,
^2=M_N(E-\omega_{q'})$, where $\omega_{q'}$ denotes the energy of the meson.
In Eq. \eqref{watsonstart} $\Psi(p',p')$ is the $NN$ wavefunction
 at the energy, $2E(p')$, of the final $NN$ subsystem,
 where  $E(p') \equiv {{p'}^2\over 2 M_N}$ and $\Phi_x$ denotes
the wavefunction for the outgoing meson.
If we now neglect the impact of the meson--nucleon interaction
on the energy dependence, the meson
wave function is given by a plane wave. For its component in the
$l$--th partial wave we get \cite{joachin}
\begin{equation}
\label{tphas}
\Phi_x ^l(q') \propto j_l(q'R) \simeq q' \, ^l \ ,
\end{equation}
where $j_l(q'R)$ denotes the $l$th Bessel function.  The 
term ${q'}^l$ is the first term in an expansion of $q'R$, where $R$
represents the range of forces. Therefore, Eq. (\ref{tphas}) should
appropriately describe the $q'$ dependence of the corresponding partial wave
amplitude as long as $(q'R)\ll 1$.
In addition, when the
final state interaction is very strong and shows a strong
energy dependence (as, for example, the nucleon--nucleon interaction
in the $^1S_0$ for small relative energies), we may replace
the $NN$ wave function by the on--shell $T$ matrix, which can
be written in the $L$--th partial wave
\begin{equation}
T^L(E_{p'}) = T_0^L e^{i\delta_L(E_{p'})}\frac{\sin{\delta_L(E_{p'})}}
{p' \, ^{L+1}} \ , \label{phsft}
\end{equation}
where $\delta_L(E_{p'})$ denotes the corresponding phase shift as 
deduced from elastic $NN$ scattering data.  In practice this means
that we neglect the term with the non--interacting outgoing two nucleon pair
(Fig. \ref{pwpfsi}a) und c)) compared to that with the pair interacting (Fig.
\ref{pwpfsi}b) and d)).  In case of the $NN$ final state interaction this is indeed 
justified. In general, however, the plane wave piece is not negligible.
Fortunately in this case dispersion integrals may be used to fix the relative
strength of diagrams $a$ and $b$ of Fig.  \ref{pwpfsi}. This will be discussed
in detail in the next section.

At low energies the $T$ matrix may be replaced by the
first term in the effective range expansion\footnote{Unfortunately, in
the literature there are two different conventions for the sign
of the scattering length. The convention used here is that
of Goldberger and Watson \cite{gw};
the usual convention for baryon--baryon interactions has a minus sign on the right hand side.}.
\begin{equation}
p' \, ^{2L+1} \cot{\delta_L(E_{p'})} = \frac{1}{a_L} \ ,
\label{effrexp}
\end{equation}
where $a_L$ denotes the scattering length of the $NN$ interaction
in the partial wave characterized by $L$. Thus we can write
\begin{equation}
\sigma_{NN \to NN x}^{(L,l)}(\eta) \propto \int_0^{m_x\eta}dq' \, 
q' \, ^{2l+2} 
\frac{p' \, ^{2L+1}}{1+ a_L^2 p' \, ^ {4L+2}} \ .
\label{sigendep}
\end{equation}
Since Eq. (\ref{tphas}) only holds for $(q'R)\ll 1$, in Eq. (\ref{sigendep})
the
$a_L$--term has to be kept in the denominator only if $a_L\gg R ^ {2L+1}$
In case of the $NN$ interaction, where $R\sim m_\pi$ this condition holds for
the $S$--waves only.
 When data for the reaction $pp \to pp\pi^0$ close to threshold became
available \cite{meyer1,meyer2} Eq. \ref{sigendep} indeed turned out to give
the correct energy dependence of the total cross--section \cite{MuS}.
Eq. (\ref{sigendep}) contains two very important messages.  First of
all that a strong final state interaction will significantly change
the invariant mass distribution of the corresponding two particle
subsystem---the relevant scale parameter is $(a_L{p'})$.  This
observation was used in Ref. \cite{jan3} to extract information on the
$\Lambda N$ interaction. We come back to this in section \ref{fsisec}.
Secondly that due to the centrifugal barrier higher partial waves are
suppressed in the close to threshold regime.  As a rough estimate we can take
non relativistic kinematics to derive
\begin{equation}
\sigma_{NN \to NN x}^{(L,l)}(\eta) \propto  \left(\frac{\mu_{NN}}{\mu_{(NN)x}}
\right)^{L}(m_x\eta)^{2l+2L+4} \propto \mu_{NN}^L\mu_{(NN)x}^l
Q^{l+L+2} \ ,
\label{edep}
\end{equation}
for all those partial waves where there is no strong FSI.
Here we used the relation between $Q$ and $\eta$ as given in Eq.
(\ref{etarelat}) and $\mu_{NN}=M_N/2$ and $\mu_{(NN)x}=m_x/(1+m_x/(2M_N))$
 denote
the reduced masses of the $NN$ or more generally the baryon--baryon system
and of the particle $x$ with respect to the $NN$ system respectively.

\begin{figure}[t]
\begin{center}
\epsfig{file=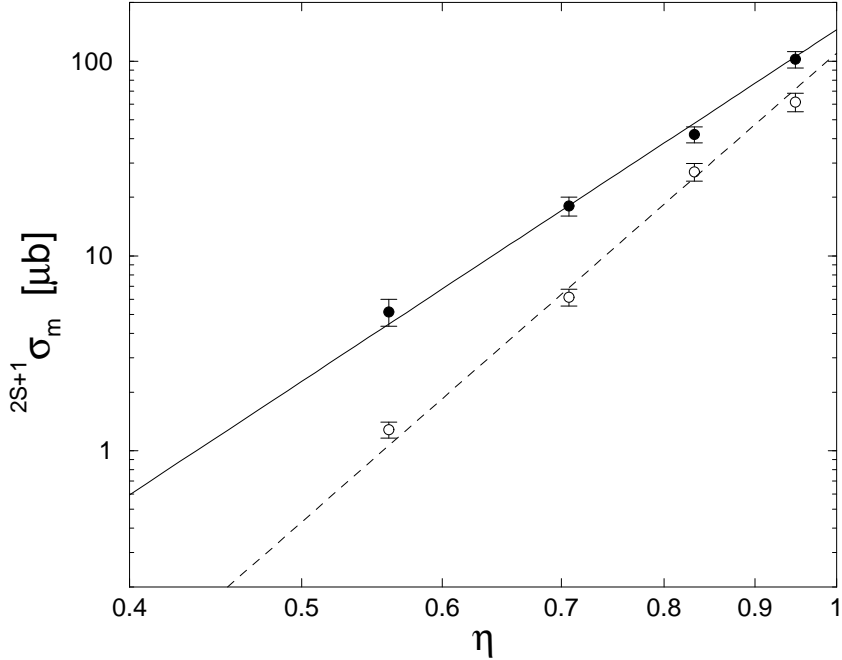, height=9cm}
\caption{\it Energy  dependence of two different spin cross sections
for the reaction $pp\to pp\pi^0$.
The data for
 $^1\sigma_0$ ($^3\sigma_1$), shown as solid (opaque) dots, are
compared to a fit of the form $y=A\eta^\tau$ with $\tau=6$ ($\tau=8$).
The data are from table~V of Ref. \protect{\cite{meyerpol}} with an
additional error of 10 \% on the total cross section.}
\label{partxs}
\end{center}
\end{figure}

Eq. (\ref{edep}) contains two pieces of information: first of all it
shows that, for a given value of $Q$,
 heavier systems are more likely to be in a higher partial.
The reason is that their maximum momentum allowed by energy conservation is
larger than that of light particles.
Secondly it should give a reasonable estimate of
the energy dependence of the partial cross sections for all those
partial waves where the $NN$ system is in a partial wave higher than
the $S$--wave.  In case of the pion production this finding was
confirmed recently in Ref. \cite{meyerpol} for the reaction $pp\to
pp\pi^0$ as is demonstrated in Fig. \ref{partxs}, where the spin cross
sections $^1\sigma_0$ and $^3\sigma_1$ are shown as a function of
$\eta$. We use the notation of Ref. \cite{meyerpol} as
$^{2S+1}\sigma_m$, where $S$ ($m$) denotes the total spin (spin
projection) of the initial state. For the definition we
refer to chapter \ref{polobs}.  It is easy to show that the lowest
partial waves in the final state that contribute to $^1\sigma_0$ 
($^3\sigma_1$) are $Ps$ ($Pp$ and $Ds$), where capital letters
denote the relative angular momentum of the two nucleons and the small
letters that of the pion with respect to the two--nucleon system. Here
we used that pion $d$--waves should be suppressed 
compared to $NN$ $D$--waves (c.f. Eq. (\ref{edep})).
Accordingly one should expect the $Ps$ states to have an energy dependence
given by $\eta^6$ whereas
the $Pp$ states should follow an  $\eta^8$ dependence (c.f. Eq. (\ref{edep})). The corresponding fits
are shown in the Figure as a solid and a dashed line, respectively.
The good agreement of these simple forms for the energy
dependence of the different partial waves supports the conjecture
that neither the initial state interaction nor the 
production operator and not even the FSI, for partial waves higher than
$S$--waves,
induce a significant energy dependence beyond the centrifugal barrier.

As we have seen, the energy dependence factorizes from the total amplitude. The
situation is somewhat more complicated, however, when it comes to quantitative
predictions for the cross section: in the presence of very strong final state
interactions it is not possible to separate model independently production
operator and final state interaction.  The reason for this is that what sets
the scale for the size of the matrix element is the convolution integral of
the production operator with the final state half off--shell $NN$ T--matrix.
This convolution consists of a unitarity cut contribution in which the
intermediate two nucleons are on their mass shell, and a principal value
piece that is sensitive to the off--shell behavior of both the $NN$ T--matrix
as well as the production operator. This already shows that it is impossible
to separate the production operator from the final state interaction in a
model independent way. Since off--shell effects are not observable, a
consistent construction of production operator and final state interaction is
required to get meaningful results for the observables.

In order to keep the formulas simple
so far we have not discussed at all the effect of the Coulomb
interaction, although its effect especially on near threshold meson production 
with two protons in the final state is known to be quite strong \cite{MuS}.
To account for the Coulomb interaction in the formulas given above one should
use the Coulomb modified effective range expansion instead of
Eq. (\ref{effrexp}). The corresponding formulas are e.g. given in Ref. \cite{couleffrange}.

\subsection{Dispersion theoretical approach}
\label{secdisp}

The fact that meson production in nucleon--nucleon collisions is a high
momentum transfer reaction allows for a rigorous treatment of strong final
state interactions within a dispersion theoretical approach\footnote{Actually,
  the formalism applies to all large momentum transfer reactions.  In addition
  to meson production in $NN$ collisions, one may study e.g. $\gamma d\to
  B_1B_2+meson$. This reaction is also discussed in Refs.
  \cite{gammad1,gammad2,gammad3,gammad4,gammad5,gammad6}.}. As we will see, in
those cases where the effective range of a particular final state interaction
is of the order of the scattering length, the modification of the invariant
mass spectrum induced by the corresponding final state interaction is no
longer proportional to the elastic scattering situation. However, the
corrections that arise can be accounted for systematically.  In addition to
things that are are well known and presented in various text books
\cite{gw,fsibook}, we derive an integral representation for the scattering
length in terms of an observable \cite{ynfsi}.  In addition one gets at
the same time an integral representation of the uncertainty of the method. The
latter can be estimated using scale arguments or by doing a model calculation.
For example, in the case of the hyperon--nucleon interaction the scattering
lengths as extracted from the available data on elastic scattering are of the
order of a few fermi, but with an uncertainty of a few fermi---especially for
the spin singlet scattering length (c.f. discussion in section \ref{yn}).
Using the kinematics for associated strangeness production to estimate of the
theoretical uncertainty, we find a value of 0.3 fm. Thus it is feasible from
the theoretical point of view to improve significantly our knowledge of the
hyperon--nucleon scattering parameters through an analysis of the reaction
$pp\to pK\Lambda$.

To be definite, we will discuss the reaction $pp\to pK\Lambda$, however---as
should be obvious---the complete discussion also applies to any subsystem with
a sufficiently strong interaction.  Let us assume the $\Lambda p$ system is in
a single partial wave (this can be easily extracted from
polarized experiments as will be discussed in sec. \ref{hnpol}) and the third
particle---here the kaon---is produced with a definite momentum transfer
$t=(p_1-p_K)^2$, where $p_1$ denotes the beam momentum.  The full
production amplitude may then be written as \be
A(s,t,m^2)=\frac1\pi\int_{-\infty}^{\tilde m\, ^2} \frac{D(s,t,m' \, ^2)}{m'
  \, ^2-m^2}dm' \, ^2 +\frac1\pi\int_{m_0^2}^\infty \frac{D(s,t,m' \, ^2)}{m'
  \, ^2-m^2}dm' \, ^2,
\label{dispers}
\ee where $m^2=\tilde m\, ^2$ is the lefthand singularity closest to the
physical region, $m_0=M_N+M_\Lambda$ corresponds to the first right hand cut,
and 
\be D(s,t,m^2) = \frac{1}{2i}(A(s,t,m^2+i0)-A(s,t,m^2-i0)).  \ee
 Note that, in
contrast to the representation of the amplitude in case of one open channel
only as in $\bar NN \to \pi\pi$, $D$ is not simply given by the imaginary part
of $A$, since the initial state interaction also induces a phase in $A$
(c.f. discussion in the next section).  Thus $D$ has to be written explicitly as
the discontinuity of $A$ along the appropriate two--particle cut.

\begin{figure}[t]
\begin{center}
\epsfig{file=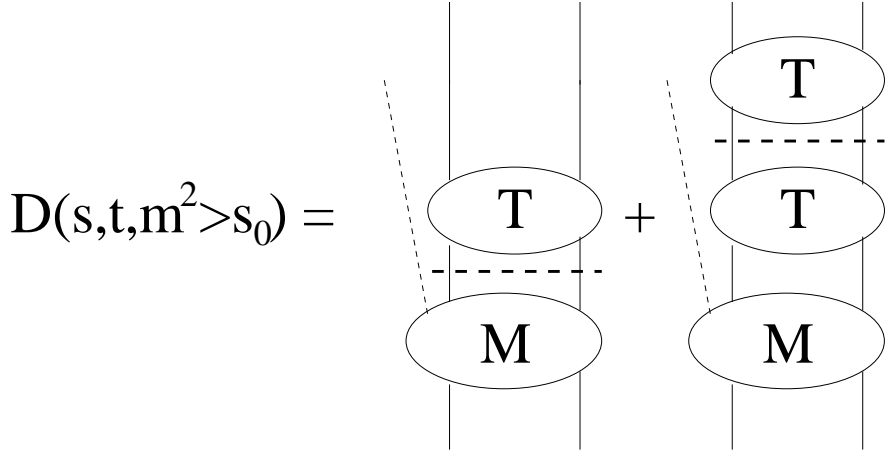, height=6cm}
\caption{\it Illustration of the sources of a discontinuity
in the amplitude $A$ from two baryon intermediate states. The
dashed horizontal lines indicate the presence of a two baryon unitarity cut
(Diagrams with initial state interaction are not shown explicitly (c.f.
Fig. \protect\ref{pwpfsi})).}
\label{disca}
\end{center}
\end{figure}

The second integral in Eq. (\ref{dispers}) gets contributions
from the various possible final state interactions, namely $\Lambda K$,
$NK$ and $\Lambda N$, where the latter is the strongest.
We may thus neglect the former two for a moment to get,
for $m^2>m_0^2$,
\be
D(s,t,m^2) = A(s,t,m^2)e^{-i\delta}\sin{\delta},
\label{discaf}
\ee where $\delta$ is the elastic $\Lambda N$ phase shift.  To see where this
relation comes from we refer to Fig. \ref{disca}: the appearance of the
various unitarity cuts due to the baryon--baryon final state interaction leads
to a discontinuity in the amplitude along the corresponding branch cut. We
thus may read almost directly off the figure that, indeed, the discontinuity in
the amplitude is given by the production amplitude times the two--baryon phase
space density times the (complex conjugate of the) on--shell $\Lambda N$
$T$--matrix\footnote{The appearance of the complex conjugation here follows
  from the direct evaluation of the discontinuity of the $T$--matrix by
employing the unitarity of the $S$ matrix: from $S^\dagger S=1$ it follows,
that $$\mbox{disc}(T):=\frac{1}{2i}(T(m^2+i0)-T(m^2-i0))=-\kappa |T(m^2)|^2\
.$$}.
 Eq. (\ref{discaf})
follows by simply using the definition of the elastic phase shift:
$\kappa T(m^2) = e^{i\delta}\sin{\delta}$, where $\kappa=p'\pi\mu$ denotes the
phase space density here expressed in terms of the reduced mass of the
$\Lambda N$ system $\mu=(M_NM_\Lambda)/(M_N+M_\Lambda)$.

We will discuss in sec. \ref{yn} how to control the possible influence
of the $K$ interactions. 
The solution of \eqref{dispers} in the 
physical region can then be written as (see  \cite{fsibook} and references therein)
\be
A(s,t,m^2)=e^{\dis{u(m^2+i0)}}
\Phi (s,t,m^2) \ ,
\label{Frazer}
\ee
where
\be
\Phi (s,t,m^2) = \frac1\pi\int_{-\infty}^{\tilde m\, ^2} 
\frac{dm' \, ^2D(s,t,m' \, ^2)}{m' \, ^2-m^2}e^{\dis{-u(m' \, ^2)}}
\ee
and, in the absence of bound states,
\be
u(z)=\frac1\pi\int_{m_0^2}^\infty\frac{\delta(m' \, ^2)}{m' \, ^2-z}dm' \, ^2.
\label{udef}
\ee The $m^2$ dependence of $\Phi$ is dominated by the $m^2$ dependence of the
production operator. The momentum transfer in the production operator,
however, is controlled by the initial momentum and therefore one should expect
the $m^2$ dependence of $\Phi$ to be weak as long as the corresponding
relative momentum of the $\Lambda$--nucleon pair is small. Thus, in a large
momentum transfer reaction, the $m^2$ dependence of the production amplitude
$A$ is governed by the elastic scattering phase shifts of the dominant two
particle reaction in the final state! The relation between the phase
  shifts and the $m^2$ dependence of the amplitude as can be extracted from
  an invariant mass distribution {\it is fixed by analyticity and unitarity!}

 So far
we are in line with the reasoning of the previous section, however
 the exponential factor in Eq. \eqref{Frazer}
is in general not simply the elastic scattering amplitude.
For illustration we investigate the form of $A(m^2)$ for a final
state interaction
that is fully described by the first two terms in the effective range
expansion (c.f. Eq. \eqref{effrexp}), $p'$ctg$(\delta(m^2)) = 1/a+(1/2)rp' \, ^2$, where $p'$ is the
relative momentum of the final state particles under consideration
in their center of mass system. Then $A$ can be given in closed
form as \cite{gw}
\be
A(s,t,m^2) = \frac{(p' \, ^2+\alpha^2)r/2}{1/a+(r/2)p' \, ^2-ip'}\Phi(s,t,m^2) \ ,
\label{arform}
\ee where $\alpha = (1/r)(1+\sqrt{1+2r/a})$. Note:
The case of an attractive interaction without a bound state
in this convention
 corresponds to a positive value of $a$
and a positive value of $r$.  In the limit $a\to \infty$, as is
almost realized in $NN$ scattering, the energy dependence of $A(m^2)$ is given
by $1/(1-iap')$ as long as $p' \ll 1/r$. This, however, exactly agrees with the
energy dependence for $NN$ on--shell scattering.  This prediction was
experimentally confirmed by the near--threshold measurements of the reaction
$pp\to pp\pi^0$ \cite{meyer1,meyer2}.  However, for interactions where the
effective range is of the order of the scattering length, the numerator of Eq.
(\ref{arform}) plays a role and thus the full production amplitude is no
longer given by the on--shell elastic scattering times terms whose energy
dependence is independent of the scattering parameters.

\begin{figure}[t]
\vspace{5cm}
\includegraphics{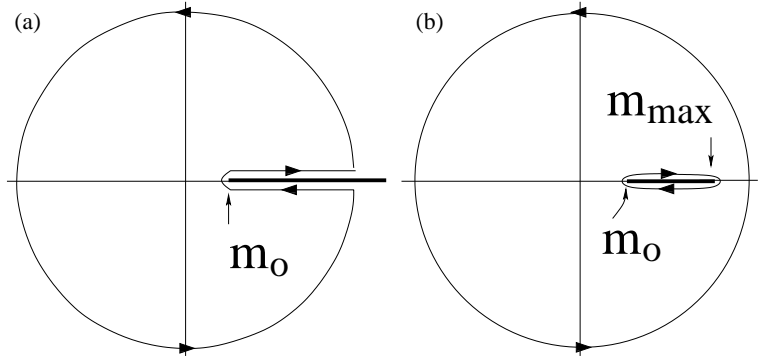}
\caption{\it The integration contours in the complex $m' \, ^2$ plane to be
  used in the original treatment by Geshkenbein
  \protect\cite{Geshkenbein1969,Geshkenbein1998} (a) and in the derivation of
  Eq. (\protect{\ref{dis3}}) (b).  The thick lines indicate the branch cut
  singularities.}
\label{intcont} 
\end{figure}

 It is often argued that since
the full production amplitude is given by a term with a plane wave final
state as well as one with the strong interactions, that an appropriate
parameterization is given by a Watson term (proportional to the elastic
scattering of the outgoing particles) plus a constant term, and their relative
strength should be taken as a free parameter. The considerations in
the previous paragraphs show, however, that this is not the case:
Eq. \eqref{Frazer} (and thus
also the special form given in Eq. \eqref{arform}) describes the $m^2$
dependence of the full invariant mass spectrum.

One more comment is in order: In the previous section we stressed that,
although the shape of the invariant mass spectra as well as the energy
dependence of the total cross section are governed by the on--shell
interactions in the subsystems, the overall normalization is not.  This
statement, based on effective field theory arguments, is illustrated in
Appendix \ref{fsistuff}. On the other hand, Eq. \eqref{Frazer} gives a closed
expression for the invariant mass spectrum for arbitrary values of $m^2$.
On the first glance this look like a contradiction.
However, it should be stressed that Eq. \eqref{Frazer} does in
general not
allow one to relate the asymptotic form of the full production amplitude with
information on the scattering parameters of the strongly interacting subsystems
in the final state to the amplitude in the close to threshold regime. Besides
the trivial observation that Eq. \eqref{Frazer} holds for each amplitude
individually and that far away from the threshold there should be many partial
waves contributing, there is no reason to believe that the function $\Phi$ is
constant over a wide energy range. Therefore, we want to emphasize that the $m^2$
dependence of Eq. \eqref{Frazer} is controlled by the scattering phase shifts
of the most strongly interacting subsystem only in a very limited range of
invariant masses. In fact, the leading singularity that contributes to an
$m^2$ dependence of $\Phi$ is that of the $t$--channel meson exchange in the
production operator. The $m^2$ dependence that originates from this singularity
can be estimated through a Taylor expansion of the momentum transfer and
thus turns out to be governed by $(p'/p)^2$, where $p'$ denotes the relative
momentum of the final state particle pair of interest. For $p$ we may use that of the
initial state particles at the production threshold,
all to be taken in the over all center of mass system.
As a consequence, for values of $m^2$ that do not significantly
deviate from $m_0^2$ (or equivalently $p'\ll p$) the assumption of $\Phi$
being constant is justified. However $\Phi$ can show
a significant $m^2$ dependence over a large range of invariant masses.

Let us now return to our main goal: namely to derive from \eqref{Frazer} a formula
that allows extraction of the elastic scattering phase shifts from
an invariant mass spectrum. In this course we will derive
 a  dispersion relation in terms of 
the function $|A(m^2)|^2$. Here we will use a method similar to 
the one used in \cite{Geshkenbein1969,Geshkenbein1998}. However, in contrast
to the formulas derived in these references, we will present a integral
representation for the elastic scattering phase shifts from production data
that involves a finite integration range only.

For this 
we first observe that
Eq. \eqref{udef} holds for purely elastic scattering only and thus is of very
limited practical use.  On the other hand a
significant contribution stems from large values of $m' \, ^2$ which depends
only weakly on $m^2$ in the near--threshold region, and therefore can be
absorbed into the function $\Phi$. Let
\begin{eqnarray}
A(m^2)=\exp\left[{\frac1\pi\int_{m_0^2}^{m_{max}^2}\frac{\delta(m' \, ^2)}{m' \, ^2-m^2-i0}dm' \, ^2}\right]
\tilde \Phi(m_{max}^2,m^2) \label{C_int} \ ,
\end{eqnarray}
where $\tilde \Phi(m_{max}^2,m^2)=\Phi(m^2)\Phi_{m_{max}^2}(m^2)$, with
\begin{eqnarray}
 \Phi_{m_{max}^2}(m^2)=\exp\left[{\frac1\pi\int_{m_{max}^2}^{\infty}\frac{\delta(m' \, ^2)}{m' \, ^2-m^2-i0}dm' \, ^2}\right].
\label{dis3}
\end{eqnarray}
The quantity $m_{max}^2$ is to be chosen by physical arguments in such a way
that both $\Phi(m^2)$, and $\Phi_{m_{max}^2}(m^2)$ vary slowly on the interval
$(m_0^2,m_{max}^2)$.  Obviously, it needs to be sufficiently large that the
structure in the amplitude we are interested in can be resolved\footnote{One
  immediately observes that the integral given in Eq. \eqref{final} goes to
  zero for $m_{max}\to m_0$. At the same time the theoretical uncertainty
  $\delta a^{m_{max}}$, defined in Eq. \eqref{dam}, tends to the value of the
  scattering length.}. On the other hand, it should be as small as possible,
since this will keep the influence of the inelastic channels small.  In order
to estimate the minimal value of $m_{max}$ we see from Eq. \eqref{arform} that
the values of $p'$ that enter in the integral given in Eq. \eqref{final}
should be at least large enough that the scattering length term plays a
significant role. Thus we require $p'_{max}\sim 1/a_{typ}$.  In case of the
$pn$ interaction (the spin triplet $pn$ scattering length is 5.4 fm) this
would correspond to a value of $p'_{max}$ of only 10 MeV corresponding to a
value for $\epsilon_{max}=m_{max}-m_0$--- the maximum excess energy that
should occur within the integral in Eq. \eqref{C_int} (as well as Eq.
\eqref{final} below)---as small as 2 MeV.  On the other hand, if we want to
study the hyperon--nucleon interaction, we want to 'measure' scattering
lengths of the order of a few fermi we choose $a_{typ} = 1$ fm, leading to a
value of 40 MeV for $\epsilon_{max}$.  Note that 40 MeV is still significantly
smaller than the thresholds for the closest inelastic channels (that for the
$K \Sigma N$ final state is at 75 MeV, that for the $\pi \Lambda K N$ channel
at 140 MeV).

The integral in \eqref{C_int} contains 
an unphysical  singularity of the type $\log{(m_{max}^2-m^2)}$, which is 
canceled by the one in $\tilde \Phi(m_{max}^2,m^2)$, but this
does not affect the region near threshold.

Notice next that the 
function 
\begin{eqnarray}
\nonumber
\frac{\log{\{A(m^2)/\tilde \Phi(m_{max}^2,m^2)\}}}{\sqrt{(m^2-m_0^2)(m_{max}^2-m^2)}}
&=&\frac1{\sqrt{(m^2-m_0^2)(m_{max}^2-m^2)}} \\
& & \qquad  \times \
\frac1\pi\int_{m_0^2}^{m_{max}^2}\frac{\delta(m' \, ^2)}{m' \, ^2-m^2-i0}dm' \, ^2
\end{eqnarray}
has no singularities in the complex plane except
the cut from $m_0^2$ to $m_{max}^2$ (c.f. Fig. \ref{intcont}b) and its value
below the cut equals the negative of the complex conjugate 
from above the cut.
Hence, 
\begin{eqnarray}
\nonumber
\frac{\delta(m^2)}{\sqrt{m^2-m_0^2}}&=& \\
& & \!\!\!\!\!\!\!\!\!\!\!\!\!\!\!\!\!\!\!\!\!\!\!\!\!-\frac1{2\pi}{\bf P}
\int_{m_0^2}^{m_{max}^2}\frac{\log{|A(m' \, ^2)/
\tilde \Phi(m_{max}^2,m' \, ^2)|^2}}
{\sqrt{m' \, ^2-m_0^2}(m' \, ^2-m^2)}
\sqrt{\frac{m_{max}^2-m^2}{m_{max}^2-m' \, ^2}}dm' \, ^2.
\label{almostfinal}
\end{eqnarray}
It is an important point to stress that 
\be
{\bf P}
\int_{m_0^2}^{m_{max}^2}\frac{1}
{\sqrt{m' \, ^2-m_0^2}(m' \, ^2-m^2)}\sqrt{\frac{m_{max}^2-m^2}{m_{max}^2-m'
    \, ^2}}dm' \, ^2 = 0 \ 
\label{pvc}
\ee as long as $m^2$ is in the interval between $m_0^2$ and $m_{max}^2$.
Therefore, if the function $\Phi$ only weakly depends on $m^2$, as it should
in large momentum transfer reactions, it can well be dropped in the above
equation. In addition, up to kinematical factors, $|A|^2$ agrees with the cross
section for the production of a $\Lambda p$ pair of invariant mass $m$. We
can therefore replace it in Eq.  (\ref{almostfinal}) by the cross section
where, because of Eq. \eqref{pvc}, all constant prefactors can be dropped.
Thus we get
\begin{eqnarray}
\nonumber
a_S&=&\lim_{{m}^2\to m_0^2}\frac1{2\pi}\left(\frac{M_\Lambda+M_p}
{\sqrt{M_\Lambda M_p}}\right){\bf P}
\int_{m_0^2}^{m_{max}^2}dm'\, ^2\sqrt{\frac{m_{max}^2-{m}^2}{m_{max}^2-m'\, ^2}}\\
& & \qquad \qquad \qquad \times \ \frac1{\sqrt{m'\, ^2-m_0^2}(m'\, ^2-{m}^2)}
\log{\left\{\frac{1}{p'}\left(\frac{d^2\sigma_S}{dm'\, ^2dt}\right)\right\}}
\ .
\label{final}
\end{eqnarray}
This is the desired formula: namely, the scattering length is expressed in
terms of an observable. Note that Eq. (\ref{final}) is applicable only if it is
just a single $YN$ partial wave that contributes to the cross section
$\sigma_S$. In sec. \ref{hnpol} we will discuss how the use of
polarization in the initial state can be used to project out a particular spin
state in the final state.

Up to the neglect of the kaon--baryon interactions,
Eq. (\ref{almostfinal}) is exact. Thus, $\delta a^{(th)}$---the
theoretical uncertainty of the scattering length extracted using
Eq. (\ref{final})---is given by the integral
\begin{eqnarray}
\nonumber \delta a^{(th)}&=& -\lim_{{m}^2\to
m_0^2}\frac1{2\pi}\left(\frac{M_\Lambda+M_p} {\sqrt{M_\Lambda
M_p}}\right) \\ & & \qquad \times
{\bf P}\int_{m_0^2}^{m_{max}^2}\frac{\log{|\tilde \Phi(m_{max}^2,m' \, ^2)|^2}}
{\sqrt{m' \, ^2-m_0^2}(m' \, ^2-m^2)}\sqrt{\frac{m_{max}^2-m^2}{m_{max}^2-m'
\, ^2}}dm' \, ^2.
\label{deltaath}
\end{eqnarray}
Since $\log{|\tilde
  \Phi(m_{max}^2,m^2)|^2}=\log{|\Phi(m^2)|^2}+\log{|\Phi_{m_{max}^2}(m^2)|^2}$
we may write $\delta a^{(th)}=\delta a^{(lhc)}+\delta a^{m_{max}}$, where the
former, determined by $\Phi(m^2)$, is controlled by the left hand cuts, and
the latter, determined by $\Phi_{m_{max}^2}(m^2)$, by the large--energy
behavior of the $\Lambda N$ scattering phase shifts.  The closest left hand
singularity is that introduced by the production operator, which is governed
by the momentum transfer. Up to an irrelevant overall constant, we may
therefore estimate the variation of $\Phi \sim 1+\delta(p'/p)^2$, where we
assume $\delta$ to be of the order of 1. Evaluation of the integral
(\ref{deltaath}) then gives $$\delta a^{(lhc)} \sim \delta(p'_{max}/p^2)\sim
0.05 \ \mbox{fm} \ , $$
where we used $p'_{max}\, ^2=2\mu \epsilon_{max}$ with
$\epsilon_{max} \sim 30$ MeV and the threshold value for $p \sim 900$ MeV.  On
the other hand, using the definition of $\Phi_{m_{max}^2}(m^2)$ given in Eq.
(\ref{dis3}) one easily derives \be |\delta a^{m_{max}}|=\frac2{\pi
  p'_{max}}\left|\int_0^\infty \frac{\delta(y)dy}{(1+y^2)^{(3/2)}}\right| \leq
\frac2{\pi p'_{max}} |\delta_{max}| \label{dam} \ , \ee where
$y^2=(m-m_{max})/\epsilon_{max}$. Thus, in order to estimate $\delta
a^{m_{max}}$ we need to make an assumption about the maximum value of the
elastic $\Lambda N$ phase for $m^2 \ge m_{max}^2$. Recall that we
implicitly assume the inelastic channel not to play a role. This assumption
was confirmed in Ref. \cite{ynfsi} within a model calculation.
The denominator in the integral
appearing in Eq. (\ref{dam}) strongly suppresses large values of $y$. Since for
none of the existing $\Lambda N$ models does $\delta_{max}$ exceed 0.4 rad, we
estimate $$\delta a^{m_{max}}\sim 0.2 \ \mbox{fm} \ .$$
When
using the phase shifts as given by the models directly in the integral, the
value for $\delta a^{m_{max}}$ is significantly smaller, since for all models
the phase changes sign at energies above $m_{max}^2$.  Combining the two
error estimates, we conclude
$$
\delta a \lesssim 0.3 \ \mbox{fm} \ .
$$

In the considerations of the theoretical uncertainty of the parameters
extracted so far we did not talk about the possible effect of the kaon--baryon
interaction. Unfortunately it can not be quantified a priori; however, it can
be controlled experimentally.  As will be discussed in sec. \ref{pol}, a
significant meson--baryon interaction, due, for example, to the presence of a
resonance, will show up as a band in the Dalitz plot. If this band overlaps
with the region of final state interactions we are interested in, interference
effects might heavily distort the signal \cite{saschaeyrich}.  By choosing a
different beam energy, the FSI region and the resonance band move away from
each other. Thus there should be an energy regime where the FSI can be
studied undistorted. It is therefore necessary for a controlled extraction of
FSI parameters to do a Dalitz plot analysis at the same time. An additional
possible cross--check of the influence of a meson--baryon interaction would be to
do the full analysis at two different beam energies separated in energy by at
least the typical hadronic width (about 100--200 MeV). If the parameters
extracted agree with each other, there was no substantial influence from other
subsystems.

 The possible gain in experimental accuracy for the extraction
of the $\Lambda N$ scattering lengths is illustrated and discussed in detail
in sec. \ref{yn}.

To our knowledge so far the corresponding formalism including the Coulomb interaction 
is not yet derived. 

\section{The initial state interaction}
\label{secisi}

As mentioned earlier, the collision energy of the two nucleons in the initial
state is rather large, especially when we study the production of a heavy
meson. However, as long as the excess energy is small, only very few partial
waves contribute in the final state and thus the conservation of total angular
momentum as well as parity and the Pauli principle allow for only a small
number of partial waves in the initial state. In addition, only small total
angular momenta are relevant.  In such a situation the standard methods
developed by Glauber, which include the initial state interaction via an
exponential suppression factor \cite{glauberdist}, cannot be applied.  The role
of the initial state interaction relevant to meson production reactions close
to the threshold was discussed for the first time  in Ref.
\cite{withkanzo}.

It should be clear a priori that for quantitative predictions the ISI has to
play an important role: to allow a production reaction to proceed the nucleons
in the initial state have to approach each other very closely---actually
significantly closer that the range of the $NN$ interaction. Thus, especially
for the production off heavy mesons, a large number of elastic and inelastic
$NN$ reactions can happen before the two--nucleon pair comes close enough
together to allow for the meson production.  Thus the effective initial
current gets reduced significantly.

Technically the inclusion of the initial state interaction in a calculation
for a meson production process requires a convolution of the production
operator with the half off--shell $T$--matrix for the scattering of the
incoming particles (c.f. Fig. \ref{pwpfsi}c) and d)). Contrary to the final
state interaction, for the high initial energies the energy dependence of the
$NN$ interaction is rather weak. One might therefore expect that the real part
of the convolution integral is small and that the effect of the initial state
interaction is dominated by the two nucleon unitarity cut \cite{withkanzo}.
This contribution, however, can be expressed completely in terms of on--shell
$NN$ scattering parameters:
\begin{eqnarray}
\nonumber
|\lambda_L|^2 & = & \left| \frac{1}{2} e^{i\delta_L (p)} 
              \left( \eta_L (p) e^{i\delta_L(p)}+e^{-i\delta_L(p)}\right) \right|^2
 \\     & = & -\eta_{L}(p) \sin ^2(\delta_{L}(p)) + 
              \frac{1}{4}[1+\eta_{L}(p)]^2 \leq \frac{1}{4}[1+\eta_{L}(p)]^2 \ ,
\label{isieff}
\end{eqnarray}
where $p$ denotes the relative momentum of the two nucleons in the initial
state with the total energy $E$, and $\delta_L$ ($\eta_L$) denote the phase
shift (inelasticity) in the relevant partial wave $L$.  Each partial wave
amplitude should be multiplied by $\lambda_L$, defined in Eq. \eqref{isieff},
in order to account for the dominant piece of the ISI. This method was used
e.g. in Ref. \cite{kanzonew}.  Using typical values for phase shifts and
inelasticities at $NN$ energies that correspond to the thresholds of the
production of heavier mesons, Eq. (\ref{isieff}) leads to a reduction factor
of the order of 3, thus clearly indicating that a consideration of the ISI is
required for quantitative predictions.  One should keep in mind, however, that
Eq.  (\ref{isieff}) can only give a rough estimate of the effect of the
initial state interaction and should whenever possible be replaced by a full
calculation. This issue is discussed in detail in Ref. \cite{vadim}.

Unfortunately, the applicability of Eq. (\ref{isieff}) is limited to energies
where scattering parameters for the low $NN$ partial waves are available. 
Due to the intensive program of the EDDA collaboration for elastic
proton--proton scattering \cite{edda} and the 
subsequent partial wave analysis documented in the SAID database \cite{said}, 
scattering parameters in the isospin--one channel are available up to energies
that correspond to the $\phi$ production threshold. The situation is a lot 
worse in the isoscalar channel. Here $pn$ scattering data would be
required. 
At present,
those are available only up to the $\eta$ production threshold.
It is very fortunate that there is a proposal in preparation to measure spin
observables in the $pn$ system in this unexplored energy range at COSY \cite{frankpriv}.


\section{Observables}
\label{pol}

In this section the various observables experimentally accessible
are discussed. After some general remarks about the three body kinematics 
in the final state, in the next subsection we will discuss unpolarized
observables. In the following subsection we will then focus on  polarized
observables.

\begin{figure}
\begin{center}
\vskip 6cm          
\includegraphics{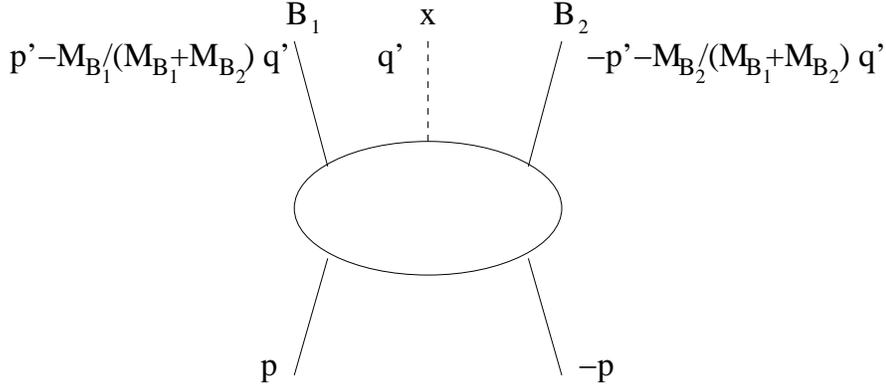} 
\caption{\it Illustration of the choice of variables in the over all center--of--mass
system.
}
\label{coorsys}
\end{center}
\end{figure}

Even if $x$ denotes just a single meson, the reaction $NN\to
B_1B_2x$ is subject to a five dimensional phase: three particles in the final
state introduce $3\times 3=9$ degrees of freedom, but the four--momentum conservation
reduces this number to 5.  As we will restrict ourselves to the 
near--threshold regime, the final state can be treated non--relativistically. The
natural coordinate system is therefore given by the
Jacobi--coordinates in the overall center of mass system (c.f. Ref.
\cite{joachin}), where first the relative momentum of one pair of particles is
constructed and then the momentum of the third particle is calculated as the
relative momentum of the third particle with respect to the two body system.
Obviously there are three equivalent sets of variables possible.
In the center of mass this choice reads
\be
\nonumber
\vec{p}_{ij}\,' &=& \frac{M_i\vec p_j - M_j \vec p_i}{M_i + M_j}  \ , \\
\vec{q}_k\,'  &=& \frac{(M_i+M_j)\vec p_k - M_k( \vec p_i+\vec p_j)}{M_i + M_j+M_k} = \vec p_k
 \ ,
\label{coors}
\ee
where we labeled the three final state  particles as $ijk$; $q_k\,
'=p_k$ holds in the over all center of mass system only.
For simplicity in the following we will drop the subscripts when confusion is excluded.
 For reactions
of the type $NN\to B_1B_2x$ it is common to work with the relative momentum of the
two--nucleon system and to treat the particle $x$ separately. 
From the theoretical point of
view this choice is most convenient, for one is working already with the
relative momentum of the dominant final state interaction and the assignment
of partial waves as used in sec. \ref{sr} is straightforward.
Note that for any given relative energy of the outgoing two nucleon system
$\epsilon = p' \, ^2/M_N$ the modulus of the meson momentum $|\vec q \, '|$ is
fixed by energy conservation; we thus may characterize the phase space by the
5 tuple
\begin{equation}
\xi = \{\epsilon,\Omega_{p'},\Omega_{q'}\} \ ,
\label{xidef}
\end{equation}
where $\Omega_k=(\cos(\theta_k),\phi_k)$ denotes the angular part of vector
 $\vec k$.
 The coordinate system
is illustrated in Fig. \ref{coorsys}.
 The center of mass nucleon momentum in the initial state will be
denoted by $\vec p$.
Throughout this report we choose the beam axis along the $z$ axis.
Explicit expressions for the vectors appearing are given in Appendix
\ref{vecdef}.

Due to the high dimensionality of the phase space it is not possible
to present the full complexity of the data in a single plot. At the end of
sec. \ref{polobs} we will discuss a possible way of presenting the data through
integrations
subject to particular constraints. A different choice would be simply to
present highly differential observables as is done for
 bremsstrahlung\footnote{It should be noted that in the
early  bremsstrahlung experiments the method of integration was not possible,
 because the detectors used had very limited angular acceptance.}, or
at least to use the two dimensional representation of the Dalitz plot to show
some correlations.

\subsection{Unpolarized observables}
\label{unpolobs}

\begin{figure}[t]
\begin{center}
\epsfig{file=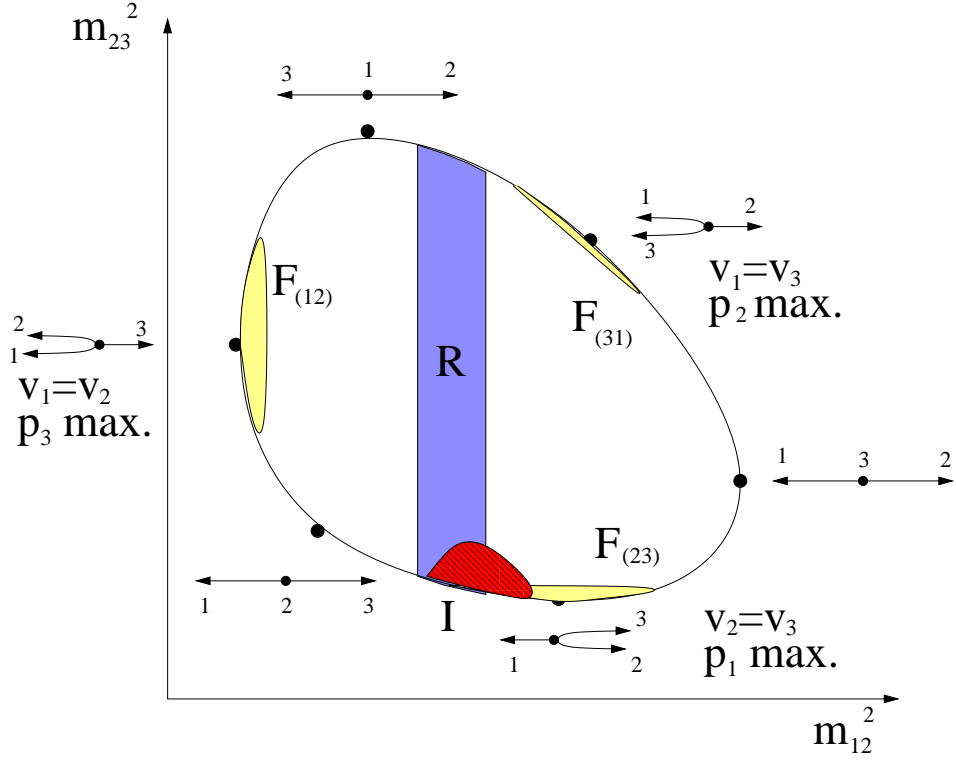, height=10cm}
\caption{\it Sketch of a Dalitz plot for a reaction with a three--body final
  state (particles are labeled as $1,2$ and $3$). Regions of possibly strong
final state interactions in the $(ij)$ system are labeled by $F_{(ij)}$; a
resonance
in the $(12)$--system would show up like the band labeled as $R$. 
The region of a possible interference of the two is labeled as $I$.}
\label{dalitz}
\end{center}
\end{figure}

As long as the initial state is unpolarized, the system has azimuthal
rotation symmetry, reducing the number of degrees of freedom from 5 to 4.

In the case of a three--particle decay (a $1\to 3$ reaction) the physics does not depend on the initial
direction. The same is true for reactions where the initial state
does not define a direction, like in the experiment series for $\bar p p$
annihilation at rest carried out by the Crystal Barrel collaboration at LEAR
(see Ref. \cite{klempt} and references therein). Therefore the number of
degrees of freedom is further reduced from four to two\footnote{This is quite
  obvious, since one can look at a three particle decay as the crossed channel
  reaction of two body scattering which is well known to be characterized by
  two variables in the unpolarized case.}.  Those are best displayed in the
so--called Dalitz plot (see \cite{kinbook} and references therein) that
shows a two dimensional representation in the plane of the various
invariant masses $m_{ik}^2=(p_i+p_k)^2$. In reactions with two particles in
the initial state and three particles in the final state ($2\to 3$ reactions),
however, the initial momentum defines an axis and therefore a single fully
differential plot is no longer possible. Especially in the example given below
it will become clear how the appearance of the additional momentum changes
drastically the situation.

To account for the higher complexity of the $2\to 3$ reactions, if enough
statistics were collected in a particular experiment, one can present
differential Dalitz plots---a new plot for each different orientation of the
reaction plane \cite{kilian}.  However, in what follows we will call Dalitz
plot the representation of a full data set/calculation in the plane of
invariant masses, ignoring the initial direction. Obviously this means throwing
out some correlations, for in general any integration reduces the amount of
information in an observable. 
We will briefly review
the properties of a Dalitz plot, closely following Ref. \cite{kinbook}.  In Fig. \ref{dalitz} a
schematic picture of it is shown. In this plot also different
regimes are specified: those of potentially strong final state interaction for
the subsystem $(ij)$ are labeled $F_{(ij)}$. They should occur when particle
$i$ and $j$ move along very closely. On the other hand, resonances will show
up as bands in the Dalitz plot. Labeled as $R$ in the figure, the effect of
a resonance in the $(12)$ system is shown.  The total area of the Dalitz plot
scales with the phase space volume. Therefore, as the excess energy decreases,
resonance and final state interaction signals or the regions of different
final state interactions might start to overlap, leading to interference
phenomena (c.f. the hatched area $I$ in Fig. \ref{dalitz}).  It was
demonstrated recently \cite{saschaeyrich} that those can rather strongly
distort resonance and final state interaction signals. At least for
known final state properties, these patterns might help to better pin down
resonance parameters \cite{saschaeyrich}.  On the other hand, as the excess
energy increases the different structures move away from each other and 
one should then be able to study them individually. This observation is of great
relevance if one wants to extract parameters of a particular final state
interaction from a production reaction (c.f. discussion in section
\ref{fsisec}).

In case of particle decays of spinless or unpolarized particles, the Dalitz
plot not only contains the information about the occurrence of a resonance, but
also its quantum numbers can be extracted by projecting the
events in the resonance band (labeled as $R$ in Fig. \ref{dalitz}) on the
appropriate axis (for the example of the figure this is the 23 axis).
In the case of $2\to 3$ reactions, however, this projection is not necessarily
conclusive. To explain this statement we have to have a closer look at the
angles of the system. First of all there is a set of angles, the so called
helicity angles, that can be constructed from the final momenta only. One
example is
\be \cos (\theta_{p'q'}) = \frac{\vec p \, '\cdot \vec q \,
  '}{|\vec p \, '||\vec q \, '|} \ , \ee
 were $p'$ and $q'$ where defined in
Eq. \eqref{coors}. These angles that can be extracted from the Dalitz
plot directly.  In addition there are those angles that are related to
the initial momentum---the Jackson angles. One example is
\be \cos (\theta_{p'p}) = \frac{\vec p \, '\cdot \vec p}{|\vec p \, '||\vec p|} \ . \ee
It should be stressed that it is not in  the distribution of the helicity
angles but that of the Jackson
angles that the subsystems reflect their quantum numbers
\cite{kinbook}\footnote{As is stressed in Ref. \cite{kinbook}, only under
  special conditions, namely for peripheral production as it occurs in high
  energy experiments, the information in the helicity
  angles and the Jackson angles agrees. Close to threshold, however, meson
  production is not at all peripheral. }. Therefore a pure Dalitz plot analysis
is insufficient for production reactions and the distributions for the
Jackson angles have to be studied as well. This will be illustrated in an
example in the following subsection.

Note that in the presence of spin there are even additional axes in the
problem. This will be discussed in detail in sec. \ref{polobs}.

\subsubsection{Example: analysis of $pp\to d\bar K^0 K^+$ close to threshold}
\label{a0dat}

\begin{figure}[t]
\begin{center}
\epsfig{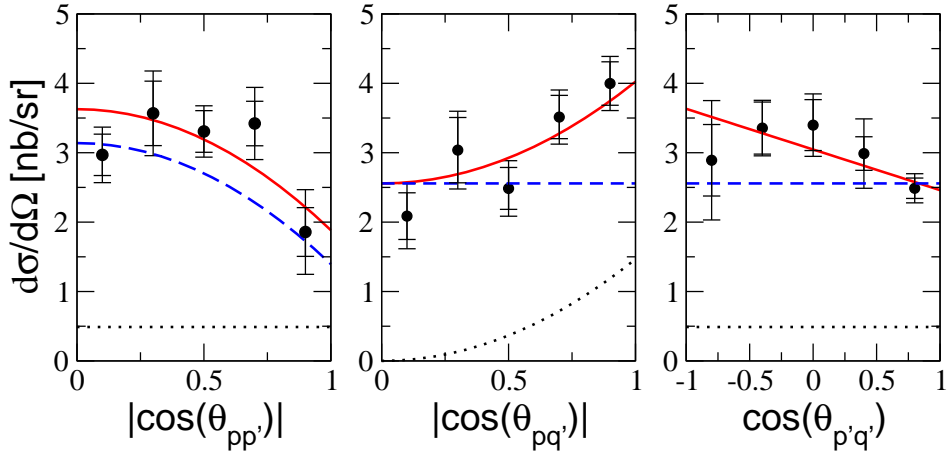}
\caption{\it Angular distributions for the reaction
  $pp\to d\bar K^0 K^+$ measured at $Q=46$ MeV \protect\cite{a0exp}.  The
  solid line shows the result of the overall fit including both $\bar K K$
  $s$--waves as well as $p$--waves. To obtain the dashed (dotted) line, the
  parameters for the $p$--wave ($s$--wave) were set to zero (see text).  The
  small error shows the statistical uncertainty only, whereas the large
  ones contain both the systematic as well as the statistical uncertainty
  added linearly. }
\label{a0angsp}
\end{center}
\end{figure}

Recently, a first measurement of the reaction $pp\to d\bar K^0 K^+$ close to
the production threshold was reported \cite{a0exp} at an excess energy $Q=46$ MeV.
The data, as well as the corresponding theoretical analysis, based on
the assumption that only the lowest partial waves contribute, will now be used
to illustrate the statements of the previous section. It will
become clear, especially, that the information encoded in the distributions of the Jackson
angles and the helicity angles is rather different.

As can be seen in table \ref{tab_secr2}, a final state that contains
$s$--waves only is not allowed in this reaction. For later
convenience\footnote{In sec.  \ref{a0f0} it will be argued, that the reaction
  $pp\to d\bar K^0 K^+$ can be used to study scalar resonance $a_0^+(980)$.
  Thus we are especially interested in the partial waves of the kaon system,
  that should show a strong final state interaction.} we work with the
relative momentum of the kaon system ($\vec p_{\bar K K}\, '\equiv \vec p \,
'$; c.f. Eqs. \eqref{coors}) and the deuteron momentum with respect to this
system ($\vec q_d \, '\equiv \vec q\, ' $).  Given our assumptions, that only
the lowest partial waves contribute, the amplitudes that contribute to the
production reaction are either linear in $q'$ or linear in $p'$.  In Ref.
\cite{unsera0f0} the full amplitude was constructed (c.f. also discussion in
sec. \ref{a0f0}), however it should be clear that the spin averaged square of
the matrix element can be written as \be \nonumber \bar{|{\cal M}|^2} &=&
C_0^{q'}{q'}^2+C_0^{p'}{p'}^2
+C_1^{q'}(\vec {q'}\cdot \hat p)^2+C_1^{p'}(\vec {p'}\cdot \hat p)^2 \\
& & \qquad \qquad \qquad \qquad \qquad +C_2(\vec {p'}\cdot \vec {q'})+C_3(\vec
{p'}\cdot \hat p)(\vec {q'}\cdot \hat p) \ ,
\label{mform}
\ee
since all terms in the amplitude are either linear in $\vec {p'}$
or linear in $\vec {q'}$.
Here $\hat p = \vec p /|\vec p |$ denotes the beam direction.
Since the two protons in the initial state are identical, any observable has
to be symmetric under the transformation $\vec p = -\vec p$. This is why
$\hat p$ appears in even powers only.

\begin{table}[!t]
\begin{center}
\caption{\it Results for the $C$ parameters from a fit to the experimental data.
The parameters are given in units of $C_0^{q'}$.}
\begin{tabular}{ccccc} 
\hline
{$C_0^{p'}$} &
{$C_0^{q'}$} & 
{$C_1^{p'}$} &
{$C_1^{q'}$} &
{$C_2+\frac{1}{3}C_3$}\\
\hline
$0\pm 0.1$ &$1 \pm 0.03$  &$1.26 \pm 0.08$&$-0.6\pm 0.1$ &$-0.36\pm 0.17$\\
\hline
\end{tabular}
\label{cpara}
\end{center}
\end{table}

\begin{figure}[t]
\begin{center}
\epsfig{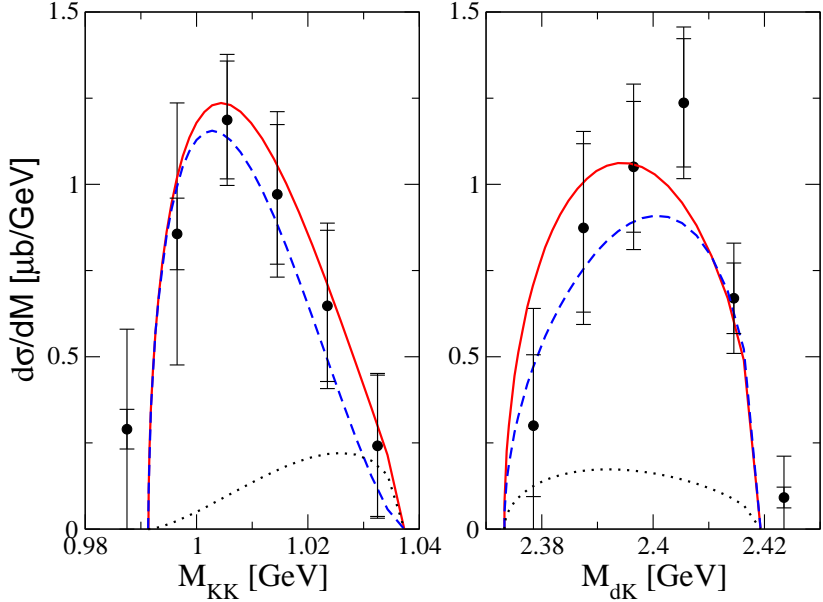}
\caption{\it Various mass distributions for the reaction $pp\to d\bar K^0 K^+$
    (Line code as in Fig. \protect\ref{a0angsp}).  The small error bars
show the statistical uncertainty only, whereas the large ones contain both
    the systematic as well as the statistical uncertainty (c.f. Ref.
    \protect\cite{a0exp}). }
\label{a0masssp}
\end{center}
\end{figure}

Figs. \ref{a0angsp} and \ref{a0masssp} show the data as well as a fit
based on Eq.  \eqref{mform}. The parameters extracted are given in table
\ref{cpara}. The first two panels of Fig. \ref{a0angsp} contain the
distributions of the angles $(\vec {p'}\cdot \hat p)/p'$ and $(\vec {q'}\cdot
\hat p)/q'$.  The last panel contains the distribution of the helicity angle
$(\vec {p'}\cdot \vec {q'})/(q'p')$.  The solid line corresponds to a complete
fit to the data including both $p$--waves in the $\bar K^0 K^+$ system as well as
those in the $d(\bar K^0 K^+)$ system. For the long dashed line the $\bar K^0 K^+$
$p$-waves were set to zero, whereas for the dotted line the $d(\bar K^0 K^+)$
$p$--waves (corresponding to $\bar KK$ $s$--waves)
were set to zero.  Thus, the first two panels truly reflect
the partial wave content of the particular subsystems individually, whereas
the helicity angle (which can also be extracted from the Dalitz plot) shows a
flat distribution only if $both$ subsystems are in a $p$--wave
simultaneously. Therefore, the helicity angle can well be isotropic although
one of the subsystems is in a high partial wave.  Note that this
statement is true even if all particles were spinless.  The only thing that
would change is that $C_0^{q'}$ and $C_0^{p'}$ would vanish (c.f. subsec.
\ref{a0amp}).  In Fig. \ref{a0masssp} two Dalitz plot projections (invariant
mass spectra) are shown. The first one ($d\sigma/dm_{\bar K K}$) is needed to
disentangle $C_0^{q'}$ and $C_0^{p'}$. The second one does not give any
additional information.

What is now the information contained in the two--dimensional  Dalitz plot? Since it does not
contain any information about the initial state, the parts of the squared amplitude 
that can be extracted from the Dalitz plot are easily derived from
Eq. \eqref{mform} by integrating over the beam direction \cite{kinbook}, giving
\be
\nonumber
\int d\Omega_p \bar{|{\cal M}|^2} &=& \left(C_0^{q'}+\frac{1}{3}C_1^{q'}\right){q'}^2+\left(C_0^{p'}
+\frac{1}{3}C_1^{p'}\right){p'}^2 \\
& & \qquad \qquad \qquad \qquad \qquad +\left(C_3+\frac{1}{3}C_4\right)(\vec {p'}\cdot \vec
{q'}) \ .
\label{intmform}
\ee Therefore from the Dalitz plot one can extract the total $\bar K K$
$s$--wave strength $(C_0^{q'}+(1/3)C_1^{q'})$, the total $\bar K K$ $p$--wave
strength $(C_0^{p'}+(1/3)C_1^{p'})$, as well as the strength of the
interference of the two $(C_3+(1/3)C_4)$.  Note that in this
particular example, all the coefficients given in Eq.
\eqref{intmform} (and even the $C_0$ and $C_1$ individually) can as well be
extracted from the angular distributions given in Fig. \ref{a0angsp}
and the $\bar K K$ invariant mass distribution (left panel of
Fig. \ref{a0masssp}) directly; the Dalitz plot here does not provide any
additional information.

Obviously, as we move further away from the threshold, the complexity of the
amplitude increases and the Dalitz plot contains information not revealed in the
projections. To summarize, in order to allow for a complete analysis of
the production data, in addition to the Dalitz plot the angular distributions
of the final momenta on the beam momentum need to be
analyzed as well. The latter distributions are the ones that give most direct
access to the partial wave content of the subsystems.

%
\subsection{Spin dependent observables}
\label{polobs}

Polarization observables for $2\to 2$ reactions are discussed in great detail
in Ref. \cite{ohlsen}. In our case, however, we have one more particle in the
final state and therefore there are more  degrees of freedom
available. Here we will not only derive the expressions for the
observables in terms of spherical tensors but also relate these to the partial
wave amplitudes of the production matrix elements. In this section we closely
follow Refs. \cite{meyerpol,lynn1}.

In terms of the so called Cartesian polarization observables,
the spin--dependent cross section can be written as
\begin{eqnarray}
\nonumber
\sigma (\xi, \vec P_b, \vec P_t, \vec P_f)
&=& \sigma_0(\xi)\left[1+\sum_i ((P_b)_iA_{i0}(\xi)+(P_f)_iD_{0i}(\xi))\right. \\
\nonumber
& &\phantom{\sigma_0(\xi)[1} \ +\sum_{ij}((P_b)_i(P_t)_jA_{ij}(\xi)+(P_b)_i(P_f)_jD_{ij}(\xi)) \\
& &\phantom{\sigma_0(\xi)[1}  \
+\left.\sum_{ijk}(P_b)_i(P_t)_j(P_f)_kA_{ij,k0}(\xi) ... \right] \ .
\label{obsdef}
\end{eqnarray}
where $\sigma_0(\xi)$ is the unpolarized differential cross section, the
labels $i,j$ and $k$ can be either $x,y$ or $z$, and
$P_b$, $P_t$ and $P_f$ denote the polarization vector of beam, target and 
the first one
of the final state particles, respectively. All kinematic variables are
collected in $\xi$, defined in Eq. \eqref{xidef}.
The observables shown explicitly in Eq. \eqref{obsdef} include the beam
analysing powers $A_{i0}$, the corresponding quantities for the final state
polarization $D_{0i}$, the spin correlation coefficients $A_{ij}$, and the spin
transfer coefficients $D_{ij}$.
In this context it is important to note that baryons that decay
weakly, as the hyperons do, have a self analyzing decay. In other
words, the angular pattern of the decay particles depends on the
polarization of the hyperon,  
therefore, the  hyperon polarization in the final state
can be measured without an additional polarimeter (see e.g. Ref.  \cite{kilianslamantilam}).
All those observables that can be defined by just
exchanging $\vec P_b$ and $\vec P_t$, such as the target analyzing power $A_{0i}$,
are not shown explicitly.
From Eq. (\ref{obsdef}) it follows that for example
\begin{equation}
\sigma_0A_{ij,k0}=\frac{1}{(2s_b+1)(2s_t+1)}
{\mbox Tr}( \sigma_k^{(f)}{\cal M} \sigma_i^{(b)}\sigma_j^{(t)}{\cal M}^\dagger) \ ,
\label{obscalc}
\end{equation} 
where the $\sigma_i^{(b)}$ ($\sigma_j^{(t)}$) are the
Pauli matrices acting in the spin space of beam and target, respectively.
The production matrix element is denoted by $\cal M$. In addition, $s_b$
($s_t$) denote the total spin of the beam (target) particles.

\begin{table}
\begin{center}
\begin{tabular}{|c c c c |}
\hline
Cartesian Observable &  ${\cal T}_{k_1q_1k_2q_2}$ & $Q_i=q_1+q_2$ & \\
\hline     
Differential cross section & & & \\
 $\sigma_0$ & $ {\cal T}_{0000}(s,\epsilon )$ & 0 & * \\
Beam analyzing powers & & & \\
 $\sigma_0A_{x0}$ & $-\sqrt{2}\mbox{Re}( {\cal T}_{1100}(s,\epsilon ))$ & 1  & *\\
 $\sigma_0A_{y0}$ & $-\sqrt{2}\mbox{Im}( {\cal T}_{1100}(s,\epsilon ))$ & 1 &  \\
 $\sigma_0A_{z0}$ & $ {\cal T}_{1000}(s,\epsilon )$ & 0 & * \\
Target analyzing powers & & & \\
 $\sigma_0A_{0x}$ & $-\sqrt{2}\mbox{Re}( {\cal T}_{0011}(s,\epsilon ))$ & 1  & (*)\\
 $\sigma_0A_{0y}$ & $-\sqrt{2}\mbox{Im}( {\cal T}_{0011}(s,\epsilon ))$ & 1 &  \\
 $\sigma_0A_{0z}$ & $ {\cal T}_{0010}(s,\epsilon )$ & 0 & (*) \\
Spin correlation parameters & & & \\
 $\sigma_0A_{zz}$ & $ {\cal T}_{1010}(s,\epsilon )$ & 0 & * \\
 $\sigma_0A_\Sigma$ & $-2\mbox{Re}( {\cal T}_{111-1}(s,\epsilon ))$ & 0 & * \\
 $\sigma_0A_\Delta$ & $2\mbox{Re}( {\cal T}_{1111}(s,\epsilon ))$ & 2 & * \\
 $\sigma_0A_{xz}$ & $-\sqrt{2}\mbox{Re}( {\cal T}_{1110}(s,\epsilon ))$ & 1 & (*) \\
 $\sigma_0A_{zx}$ & $-\sqrt{2}\mbox{Re}( {\cal T}_{1011}(s,\epsilon ))$ & 1 & * \\
 $\sigma_0A_{yz}$ & $-\sqrt{2}\mbox{Im}( {\cal T}_{1110}(s,\epsilon ))$ & 1 &  \\
 $\sigma_0A_{zy}$ & $-\sqrt{2}\mbox{Re}( {\cal T}_{1011}(s,\epsilon ))$ & 1 &  \\
 $\sigma_0[A_{xy}+A_{yx}]$ & $2\mbox{Im}( {\cal T}_{1111}(s,\epsilon ))$ & 2 & \\
 $\sigma_0A_\Xi$ & $2\mbox{Im}( {\cal T}_{111-1}(s,\epsilon ))$ & 0 & *\\
\hline
\end{tabular} 
\end{center}
\caption{\it Relations between spherical tensors and some
observables that do not contain the final state polarization following
Ref. \protect{\cite{lynn1}}.
To simplify notation, the indices specifying the final state polarization are dropped.
The symbol * indicates a
 possible set of independent observables.
Note: For $pp$ induced reactions more
observables become equivalent, as described in the text. Those are marked by
{\rm (*)}.
The linear combinations of spin correlation observables
appearing in the table are defined in Eqs. (\protect\ref{lincomdef}).}
\label{tensobs}
\end{table}

\begin{table}
\begin{center}
\begin{tabular}{|c c c c |}
\hline
Cartesian Observable &  ${\cal T}_{k_1q_1k_2q_2}^{k_3q_3k_4q_4}$ & $Q=q_1+q_2-q_3-q_4$ & \\
\hline     
Induced polarization & & & \\
 $\sigma_0D_{0x}$ & $\sqrt{2}\mbox{Re}( {\cal T}^{1-100}_{0000}(s,\epsilon ))$ & 1  & *\\
 $\sigma_0D_{0y}$ & $\sqrt{2}\mbox{Im}( {\cal T}^{1-100}_{0000}(s,\epsilon ))$ & 1 &  \\
 $\sigma_0D_{0z}$ & $ {\cal T}^{1000}_{0000}(s,\epsilon )$ & 0 & * \\
Spin transfer coefficients & & & \\
 $\sigma_0D_{zz}$ & $ {\cal T}^{1000}_{1000}(s,\epsilon )$ & 0 & * \\
 $\sigma_0D_\Sigma$ & $2\mbox{Re}( {\cal T}^{1100}_{1100}(s,\epsilon ))$ & 0 & * \\
 $\sigma_0D_\Delta$ & $-2\mbox{Re}( {\cal T}^{1-100}_{1100}(s,\epsilon ))$ & 2 & * \\
 $\sigma_0D_{xz}$ & $-\sqrt{2}\mbox{Re}( {\cal T}^{1000}_{1100}(s,\epsilon ))$ & 1 & * \\
 $\sigma_0D_{zx}$ & $\sqrt{2}\mbox{Re}( {\cal T}^{1-100}_{1000}(s,\epsilon ))$ & 1 & * \\
 $\sigma_0D_{yz}$ & $-\sqrt{2}\mbox{Im}( {\cal T}^{1000}_{1100}(s,\epsilon ))$ & 1 &  \\
 $\sigma_0D_{zy}$ & $\sqrt{2}\mbox{Re}( {\cal T}^{1-100}_{1000}(s,\epsilon ))$ & 1 &  \\
 $\sigma_0[D_{xy}+D_{yx}]$ & $-2\mbox{Im}( {\cal T}^{1-100}_{1100}(s,\epsilon ))$ & 2 & \\
 $\sigma_0D_\Xi$ & $2\mbox{Im}( {\cal T}^{1100}_{1100}(s,\epsilon ))$ & 0 & *\\
\hline
\end{tabular} 
\end{center}
\caption{\it Relations between spherical tensors and some
observables that do contain the final state polarization. 
The symbol * indicates a
 possible set of independent observables.
The linear combinations of spin correlation observables
appearing in the table are defined in Eqs. (\protect\ref{lincomdef2}).}
\label{tensobs2}
\end{table}

It is straightforward to relate the polarization observables to
the partial wave amplitudes that can be easily extracted from
any model. For this purpose it is convenient to use spherical tensors
defined through
\begin{equation}
{\cal T}_{k_1q_1,k_2q_2}^{k_3q_3,k_4q_4}=\frac{1}{(2s_b+1)(2s_t+1)}
Tr\left[\tau_{k_3q_3}^{(f_1)\, \dagger}\tau_{k_4q_4}^{(f_2)\, \dagger} 
{\cal M} \tau_{k_1q_1}^{(b)}\tau_{k_2q_2}^{(t)}{\cal M}^\dagger \right] \ ,
\label{deft}
\end{equation}
where the $\tau_{kq}$ denote the spherical representation of the spin
matrices 
\be
\tau_{10}=\sigma_z \, , \ \tau_{1 \pm 1}=
\mp\frac{1}{\sqrt{2}}(\sigma_x\pm i\sigma_y) \, , \ \tau_{00}=1 .
\label{taudef}
\ee
To relate the observables to the spherical tensors, the easiest method is
to use the definitions of Eqs. \eqref{taudef} inside the various Eqs. \eqref{deft}. 
For the observables for which the final polarization remains
undetected, the relations between the various ${\cal T}$ and the corresponding observables
are shown in Table \ref{tensobs}.
In Table  \ref{tensobs2} a few of the observables that contain the final state
polarization are listed. Triple polarization observables are not listed
explicitly, but it is straightforward to derive also the relevant expressions
for these, such as
$$
A_{xx,x}+A_{yy,x}-(A_{xy,y}-A_{yx,y}) = 
\sqrt{2}^3\mbox{Re}({\cal T}_{111-1}^{1100}) \ .
$$
After some algebra given explicitly in
Appendix \ref{pw}, one finds
\begin{equation}
{\cal T}_\rho (\hat p, \hat q) = \frac{1}{4}\sum_{\tilde L \tilde l \lambda}
B^Q_{\tilde L \tilde l, \lambda}(\hat q, \hat p) {\cal A}
^\rho
_{\tilde L \tilde l,\lambda} \ ,
\label{tdecomp}
\end{equation}
where $\rho = \{k_1q_1,k_2q_2,k_3q_3,k_4q_4\}$ and $Q=q_1+q_2-q_3-q_4$. All the angular dependence
is contained in
\begin{equation}
B^Q_{\tilde L \tilde l, \lambda}(\hat q, \hat p)
=\sum _{\mu_L,\mu_l}
 \frac{1}{4\pi}\langle \tilde L \mu_L, \tilde l \mu_l|
\lambda Q\rangle 
Y_{\tilde L \mu_L}(\hat p)Y_{\tilde l \mu_l}(\hat q) \ 
\label{bdef}
\end{equation}
and
\begin{equation}
{\cal A}^\rho
_{\tilde L \tilde l,\lambda} = \sum_{\alpha,\bar \alpha}
C^{\alpha,\bar \alpha,\rho}_{\tilde L \tilde l,\lambda}M^\alpha 
(M^{\bar \alpha})\, ^\dagger \ .
\end{equation}
Here $\alpha$ and $\bar \alpha$ are multi--indices for
all the quantum numbers necessary to characterize a 
particular partial wave matrix element and $C$ 
denotes a coupling coefficient that can be expressed
in terms of Clebsch--Gordan coefficients. Its explicit form
is given in Eq. (\ref{cdef}).
From Eq. (\ref{tdecomp}) one can derive the angular dependences
of all observables. 

In what follows it is convenient to define the following quantities:
\begin{eqnarray}
A_\Sigma = A_{xx}+A_{yy} \ , 
A_\Delta = A_{xx}-A_{yy} \ , \ \mbox{and} \
A_\Xi = A_{xy}-A_{yx} \ 
\label{lincomdef}
\end{eqnarray}
and, analogously,
\begin{eqnarray}
D_\Sigma = D_{xx}+D_{yy} \ , 
D_\Delta = D_{xx}-D_{yy} \ , \ \mbox{and} \
D_\Xi = D_{xy}-D_{yx} \ .
\label{lincomdef2}
\end{eqnarray}

Using the conservation of parity and the explicit expression for the $C$
coefficient given in Eq. \eqref{cdef}
\begin{equation}
C^{\alpha,\bar \alpha, \rho}_{\tilde L \tilde l,\lambda} =
(-)^{(k_1+k_2+k_3+k_4)}
C^{\bar \alpha, \alpha, \rho}_{\tilde L \tilde l,\lambda} .
\label{csym}
\end{equation}
Since the parameter $C$ is real,
 the analyzing powers
 are proportional to the imaginary part of 
$M_\alpha M_{\bar \alpha}^*$, whereas the differential cross section
as well as the spin correlation parameters depend on the real part 
of ${\cal M}_\alpha {\cal M}_{\bar \alpha}^*$. Thus, it is either the
real part or the imaginary part of $B$ that contributes to the
angular structure. 
As we will see in the following subsection, this
observation allows for a straight--forward identification
of the possible azimuthal dependences of each observable.
Another obvious consequence of Eq. \eqref{csym} is, that those
 observables
for which $\sum k_i$ is odd have to be small when only a single partial wave
 dominates. Thus, at the threshold, analyzing powers will vanish.
 
 The structure of Eq. (\ref{tdecomp}) is general---no assumption regarding the
 number of contributing partial waves was necessary.  However, if we want to
 make statements about the expected angular dependence of observables, the
 number of partial waves needs to be restricted. For example, if we allow for
 at most $p$--waves for the $NN$ system as well as the particle $x$ with
 respect to the $NN$ system, then the largest value of $\tilde L$ and $\tilde
 l$ that can occur is 2, which strongly limits the possible
 $\theta$--dependences that can occur in the angular function $B$ defined in
 Eq. (\ref{bdef}).

If the spin of the particles is not detected there is no interference between
different spin states in the final state. This leads to a severe selection
rule in case of the reaction $pp \to ppX$: since the final state is
necessarily in a $T=1$ state the Pauli principle demands that different spin
be accompanied by different angular momentum. Therefore, for the a reaction
with a $pp$ final state the partial waves can be grouped into two sets,
namely $\{Ss,Sd,Ds,Dp\}$ and $\{Pp,Ps\}$, where only the members of the individual
sets interfere with each other.
%

\subsubsection{Equivalent observables}

All the angular dependence of the observables
 is contained in the function $B$ defined in Eq. (\ref{bdef}).
 In this subsection we will discuss some
properties of $B$ and relate these to properties of particular observables.

%

The functional form of $B$ enables one immediately to read off the allowed
azimuthal dependences for each observable as well as to identify
equivalent observables.  To see this we rewrite Eq. (\ref{bdef}) as
\begin{equation}
B^Q_{\tilde L \tilde l, \lambda}(\hat q, \hat p)
=\sum _{n}f_{\tilde L \tilde l, \lambda, Q,n}(\theta_{p'},\theta_{q'})
\exp\left\{i[(Q-n)\phi_{p'}+n\phi_{q'}]\right\} \ ,
\label{bsym}
\end{equation}
which directly translates into the following $\phi$--dependences for
the spherical tensors (c.f. Eq. (\ref{tdecomp})):
\begin{equation}
{\cal T}_\rho (\hat p, \hat q) = \sum_{n=-N}^{N}g_{\rho,n}(\theta_{p'},\theta_{q'})
\exp\left\{i[(Q-n)\phi_{p'}+n\phi_{q'}]\right\} \ .
\label{tdecompphi}
\end{equation}
Note that $N$ is given by the highest
partial waves that contribute to the reaction considered:
\begin{equation}
-\tilde L_{max} \le (Q-N) \le \tilde L_{max} \ 
\mbox{and} \  N \le \tilde l_{max} \ ,
\label{mrange}
\end{equation}
where $\tilde L_{max}$ ($\tilde l_{max}$) is given by twice the maximum baryon--baryon
(meson) angular momentum. These limits are inferred by the Clebsch--Gordan
coefficient appearing in the definition of $B$ in Eq. \eqref{bdef}.
 
Eq. (\ref{tdecompphi}) directly relates the real and the imaginary parts of
the spherical tensors:
\begin{equation}
\mbox{Im}({\cal T}_\rho(\theta_{p'},\phi_{p'}
+\pi/(2Q),\theta_{q'},\phi_{q'}+\pi/(2Q))) = 
\mbox{Re}({\cal T}_\rho(\theta_{p'},\phi_{p'},\theta_{q'},\phi_{q'})) \ .
\label{tsym}
\end{equation}
Thus, two observables are equivalent if they are given 
by the real and imaginary parts of the same spherical tensor
with $Q \ne 0$ \ .
In table \ref{tensobs} the relations of the various observables to the 
spherical tensors are given. Thus, using Eq. (\ref{tsym}) we can identify
the following set of pairwise equivalent observables:
\begin{equation}
A_{y0}\equiv A_{x0} \, , \ A_{0y}\equiv A_{0x} \, , \ 
 A_{xx}-A_{yy}\equiv A_{xy}+A_{yx} \ ,
\end{equation}
and analogously for observables for which the final state polarization is measured
as well.
Notice that there is no connection between $ A_{xx}+A_{yy}$
and $A_{xy}-A_{yx}$, for these have $Q=0$ and therefore there
is no transformation, such as the one given in Eq. (\ref{tsym}),
that relates real and imaginary parts of the spherical tensors.

For identical particles in the initial state,
as in $pp$ induced reactions,
all observables should be equivalent under the exchange
of beam and target. This further reduces the number
of independent observables, for now the beam analyzing
powers are equivalent to the target analyzing powers and
$A_{xz}$ is equivalent to $A_{zx}$. 
In tables \ref{tensobs} and \ref{tensobs2}
a possible set of independent observables is marked by a $*$.
Those of these that are not independent for identical particles in the initial state
are labeled as $(*)$.

From the discussion in the previous section it follows (c.f. Eq.
(\ref{csym})), that all those observables with an even (odd) value of
$k_1+k_2+k_3+k_4$ lead to a real (imaginary) value for $g_{\rho,n}$, defined
in Eq. (\ref{tdecompphi}).
As a consequence, for all coefficients appearing in the expansion of the
observables, the $\phi$--dependence is fixed (c.f. table (\ref{tensobs})); for
example, the terms that
contribute to $\sigma_0$, $\sigma_0A_{zz}$, and $\sigma_0A_{\Sigma}$ behave as
$\cos (n(\phi_{q'}-\phi_{p'}))$.


\begin{figure}
\begin{center}
\vskip 10.5cm          
\includegraphics{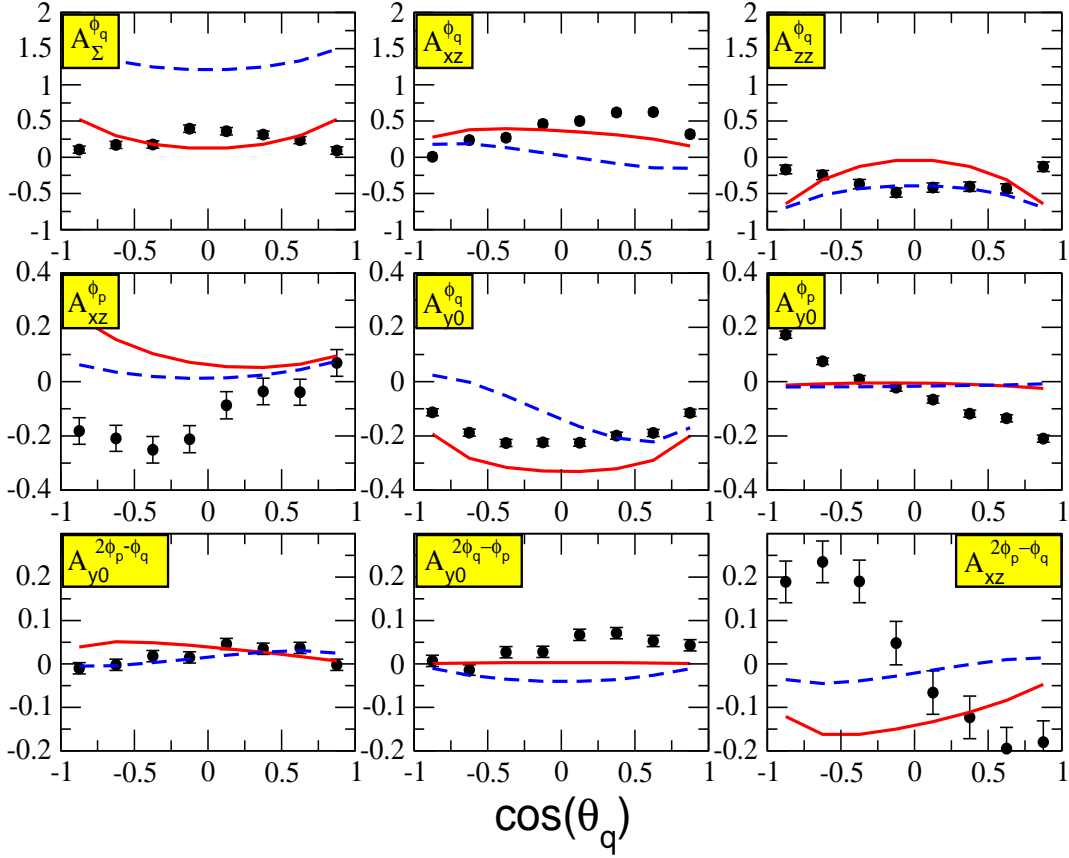} 
\caption{\it Some polarization observables reported
in  Ref. \protect{ \cite{meyerpol}} for the reaction $\vec p\vec p\to pp\pi^0$
at $\eta=0.83$
 as a function of
the pion angle compared to 
predictions of the model of Ref. \protect\cite{unserpol}. The solid lines
show the results for the full model whereas
contributions from the Delta where omitted for 
the dashed lines.}
\label{thpipol}
\end{center}
\end{figure}

\begin{figure}
\begin{center}
\vskip 10.5cm          
\includegraphics{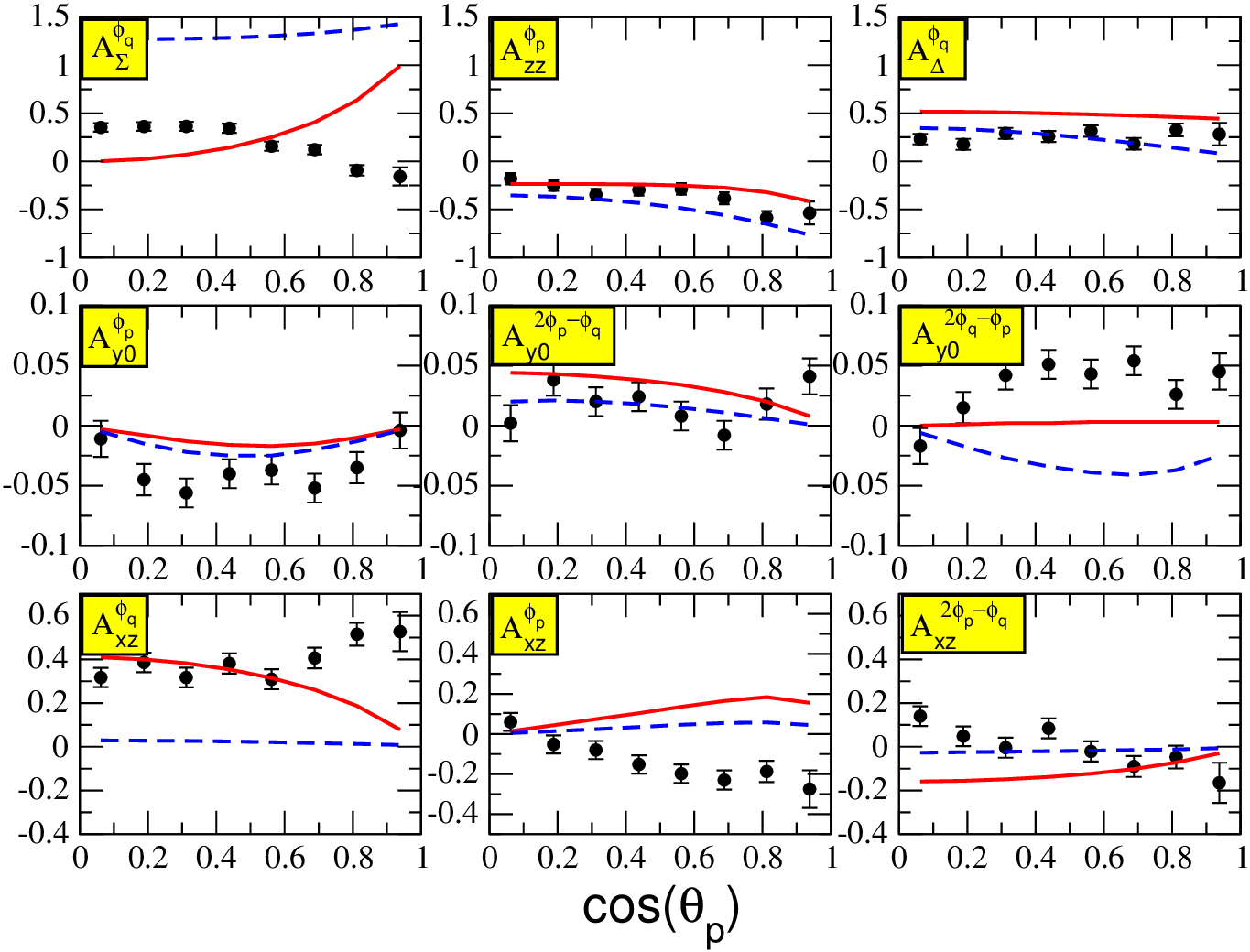} 
\caption{\it Some polarization observables reported
in  Ref. \protect{ \cite{meyerpol}} for the reaction $\vec p\vec p\to pp\pi^0$
at $\eta=0.83$
 as a function of
$\cos (\theta_p)$ compared to 
predictions of the model of Ref. \protect\cite{unserpol}. The solid lines
show the results for the full model whereas
contributions from the Delta were omitted for 
the dashed lines.}
\label{thNNpol}
\end{center}
\end{figure}

As was stressed above, the phase space for the production
reactions is of high dimension. To allow for a proper
presentation of the data as well as of calculations,
one either needs high dimensional plots (see discussion in the previous section)
or the dimensionality needs to be reduced to one dimensional
quantities\footnote{For the experimental side this is the far more demanding
  procedure, for the angular dependence of efficiency as well as acceptance
  needs to be known very well over the full angular range for those variables
that are integrated in order not to introduce false interferences.}, while, however,
still preserving the full complexity of the data. As can
be seen in Eq. (\ref{tdecompphi}), each polarization observable
is described by $2N+1$ functions  $g_{\rho,n}(\theta_{p'},\theta_{q'})$,
where the number of relevant terms is given by the number of partial waves.
In order to allow disentanglement of these functions,
in Ref. \cite{meyerpol} it was proposed to integrate
each observable over both azimuthal angles under a particular
constraint,
\begin{equation}
\Phi=m\phi_p+n\phi_q=c \ .
\label{constraint}
\end{equation}
This integration projects on those terms that depend on $\Phi$ or do not show
any azimuthal dependence at all \cite{meyerpol}.
To further reduce the dimensionality of the data, either
the relative proton angle or the meson angle can be integrated
to leave one
with a large number of observables that depend
on one parameter only. Those are then labeled as $A_{ij}^\Phi(\theta_k)$,
where $k$ is either $p$ or $q$. In Figs. \ref{thpipol} and \ref{thNNpol}
some observables reported in Ref. \cite{meyerpol}
 are shown for the energy with highest statistics, namely  $\eta = 0.83$.
 In the figures the data
is compared to the model predictions of Ref. \cite{unserpol}. The
solid lines are the results of the full model whereas for the dashed lines
the contribution from the Delta isobar was switched off. In section
\ref{pionprod}
this model will be discussed in more detail.

In the case of Ref. \cite{meyerpol} the complete set of polarization observables
for the reaction $\vec p \vec p \to pp\pi^0$ is given. Since the particles in
the initial state are identical there are 7 independent observables, all
functions of 5 independent parameters (c.f. table \ref{tensobs}).  In the
analysis of the data presented in the same reference it was assumed that only
partial waves up to $p$--waves in both the $NN$ as well as the $(NN)\pi$ system
were relevant.  Thus the various integrations described in the previous
paragraph lead to 32 independent integrated observables that depend only on a
single parameter.  On the other hand, if the
assumption about the maximum angular momenta holds, only 12 partial waves have to
be considered in the partial wave analysis. Since the amplitudes are complex
and two phases are not observable (c.f. discussion at the end of section
\ref{polobs}), a total of 22 degrees of freedom needs to be fixed from the
data. Thus a complete partial wave decomposition for the reaction $\vec p \vec
p \to pp\pi^0$ seems feasible.

\subsection{General structure of the amplitudes}
\label{generalstructure}

In this section we give the recipe for constructing the
most general transition amplitude for reactions
of the type $NN\to B_1B_2x$, where we focus on spin 1/2 baryons in the final
state. A
generalization to other reactions is straightforward.
For further applications we refer to a recent review \cite{tr}.

For simplicity let us restrict ourselves to those reactions in which
there is only one meson produced. The system is then characterized
by three vectors,
$$
\vec p \ , \ \ \vec q \, ' \ , \ \ \mbox{and} \ \vec p \, ' \ ,
$$
denoting the relative momentum of the two nucleons in the initial state,
the meson momentum, and the relative momentum of the two nucleons in the final
state, respectively---in the over all center of mass system. In addition, as
long as $x$ denotes a scalar or pseudoscalar meson, we find 6 axial vectors,
namely those that can be constructed from the above:
$$
i(\vec p \times \vec p \, ' ) \ , \ \ 
i(\vec p \times \vec q \, ' ) \ , \ \ \mbox{and} \ 
i(\vec p \, ' \times \vec q \, ' ) \ ;
$$
and those that contain the final or initial spin of the two nucleon 
system
$$
\vec S \ , \ \ \vec S \, ' \ , \ \ \mbox{and} \ i(\vec S \times \vec S \, ') 
\ ,
$$
where $\vec S = \chi^T_1\sigma_y\vec \sigma\chi_2$ and $\vec S \, ' =
 \chi^\dagger_3\vec \sigma \sigma_y (\chi^\dagger_4)^T$. Here
$\chi_i$ denotes the Pauli spinors for the incoming (1,2)
and outgoing (3,4) nucleons and $\vec \sigma$ denotes the
usual Pauli spin matrices.
If $x$ is a vector particle, an additional axial vector,
namely the polarization vector of the vector meson $\vec \epsilon \, ^*$
occurs. In addition, if instead of a two nucleon state in the 
continuum a deuteron occurs in the final state, its polarization
direction will be characterized by the same  $\vec \epsilon \, ^*$.
Since the energy available for the final state is
small (we focus on the close to threshold regime),
 we restrict ourselves to a non relativistic treatment of the
outgoing particles. This largely simplifies the formalism since
a common quantization axis can be used for the whole system.

In order to construct the most general transition amplitude that
satisfies parity conservation,
  we have to
combine the vectors and axial vectors given above so that
the final expression form a scalar or pseudoscalar for reactions
where the produced meson has positive or negative intrinsic parity,
respectively. 
The most general form of the transition matrix element may be written
as
\begin{equation}
{\cal M}=H({\cal I}\, {\cal I \, '})+ i\vec Q \cdot 
(\vec S \, {\cal I '})+i\vec A \cdot (\vec S \, '\, {\cal I})
 +(S_i \, S_j \, ')
B_{ij} \ ,
\label{ampdef}
\end{equation}
where 
${\cal I}=\left(\chi_2^T\sigma_y \chi_1\right)$ and
${\cal I \, '}=\left(\chi_4^\dagger \sigma_y (\chi_3^T)^\dagger\right)$.
In addition, the amplitudes have to satisfy the Pauli Principle
as well as invariance under time reversal. This imposes constraints
on the terms that are allowed to appear in the various coefficients.

 The 9
amplitudes
that contribute to $B$ may be further decomposed according to the total spin to
which $\vec S$ and $\vec S\, '$ may be coupled
$$
B_{ij} = b^s\delta_{ij}+b^v_k\epsilon_{ijk}+b^t_{ij} \ ,
$$
where the superscripts indicate if the combined spin of the initial and the
final state are coupled to 0 ($s$), 1 ($v$) or 2 ($t$), where $b^t_{ij}$ is to
be a symmetric, trace free tensor.

Once the amplitudes are identified the evaluation of the
various observables  is straightforward. In this
case the polarization comes in through
$$
\chi_i \chi_i^\dagger = \frac{1}{2}(1+\vec P_i \cdot \vec \sigma) \ ,
$$
where $\vec P_i$ denotes the polarization direction of particle $i$.
Using the formulas given in Appendix \ref{sptr} one easily derives
(summation over equal indices is implied):
\begin{eqnarray}
4\sigma_0 &=& |H|^2+|\vec Q|^2+|\vec A|^2+|B_{mn}|^2 \, , \label{si0def}
 \\
4A_{0i}\sigma_0 &=&
+i\epsilon_{ijk}\left( Q_j^*Q_k+ B_{jl}^*B_{kl}\right)+2\mbox{Im}
\left(B_{il}^*A_l-Q_i^*H\right) \, , \label{a0idef} \\
4D_{0i}\sigma_0 &=&
-i\epsilon_{ijk}\left( A_j^*A_k+ B_{lj}^*B_{lk}\right)-2\mbox{Im}
\left(B_{li}^*Q_l-A_i^*H\right) \, , \label{d0idef} \\
\nonumber
4A_{ij}\sigma_0 &=& \delta_{ij}\left(
-|H|^2+|Q|^2-|A|^2+|B_{mn}|^2\right) \\
 & &
\phantom{2\mbox{Re}}  \ \ \ \ \  
+2\mbox{Re}\left(\epsilon_{lij}(Q_l^*H-A_m^*B_{lm})-Q_i^*Q_j-B_{im}^*B_{jm}
\right)
 \, ,
\\
\nonumber
4D_{ij}\sigma_0 &=& 2\mbox{Re}\left(
Q^*_iA_j+\epsilon_{ilm}Q^*_lB_{mj}\phantom{\frac{1}{2}} \right. \\
 & &
\phantom{2\mbox{Re}}  \ \ \ \  \ \left.+ \epsilon_{jml}B_{il}A^*_m
+\frac{1}{2}\epsilon_{ilm}\epsilon_{jnk}B^*_{ln}B_{mk}+B_{ij}^*H\right) \ ,
\\
\nonumber
4A_{ij,k0}\sigma_0 &=& \mbox{Im}\left(\delta_{ij}(2H^*A_k
-2Q^*_lB_{lk}-\epsilon_{\alpha \beta k}(A_\alpha^*A_\beta
-B_{l\alpha}^*B_{l\beta}))\right.\\
\nonumber
 & &
\phantom{2\mbox{Re}}  \ \ \  \ \ +2\epsilon_{lij}(
Q^*_lA_k-\epsilon_{nmk}A_n^*B_{lm}+HB^*_{lk})\\
 & &
\phantom{2\mbox{Re}}  \ \ \  \ \ \left.+2(Q_i^*B_{jk}+Q_{j}^*B_{ik})
-\epsilon_{mnk}(B_{im}^*B_{jn}+B_{jm}^*B_{in})\right) \ .
 \label{aiidef}
\end{eqnarray}

For illustration we also give here the
explicit expressions for $A_\Sigma$ and $A_\Delta$ defined in 
the previous section:
\begin{eqnarray}
  \nonumber \sigma_0A_{\Sigma}&=& \frac12\left(
    -|H|^2+|Q_z|^2-|\vec A|^2+|B_{zn}|^2\right)\\
  \nonumber
  \sigma_0A_{\Delta}&=&-\frac12\left(Q_x^*Q_x-Q_y^*Q_y+B_{xm}^*B_{xm}
    -B_{ym}^*B_{ym} \right) \ .
\end{eqnarray}

As was stressed in the previous section, the method of spherical tensors
is very well suited to identifying equivalent observables.
This is significantly more difficult in the amplitude approach.
However the amplitude method becomes extremely powerful
if---due to physical arguments or appropriate kinematical cuts---one of the
subsystems can be assumed in an $s$--wave, for then the number of available
vectors is reduced significantly and rather general arguments become
possible (c.f. sec. \ref{hnpol}).

Since any amplitude can be made successively more complex by multiplying
it by an arbitrary scalar, in most of the cases an ordering
scheme is demanded  in order to make the approach useful.
 In the near--threshold regime this 
is given by
the power of final momenta occurring---in analogy with the
partial wave expansion. Actually, the amplitude approach presented here
and the partial wave expansion presented in
the previous subsection are completely equivalent. However, 
in the near--threshold regime the amplitude method is more transparent.
As one goes away from the threshold the number of partial waves contributing
as well as the number of the corresponding terms in the amplitude expansion
increases rapidly. As a consequence the construction
of the most general transition amplitude is rather involved.
The partial wave expansion, on the other hand,
can be easily extended to an arbitrary
number of partial waves.

\begin{figure}
\begin{center}
\vskip 7cm          
\includegraphics{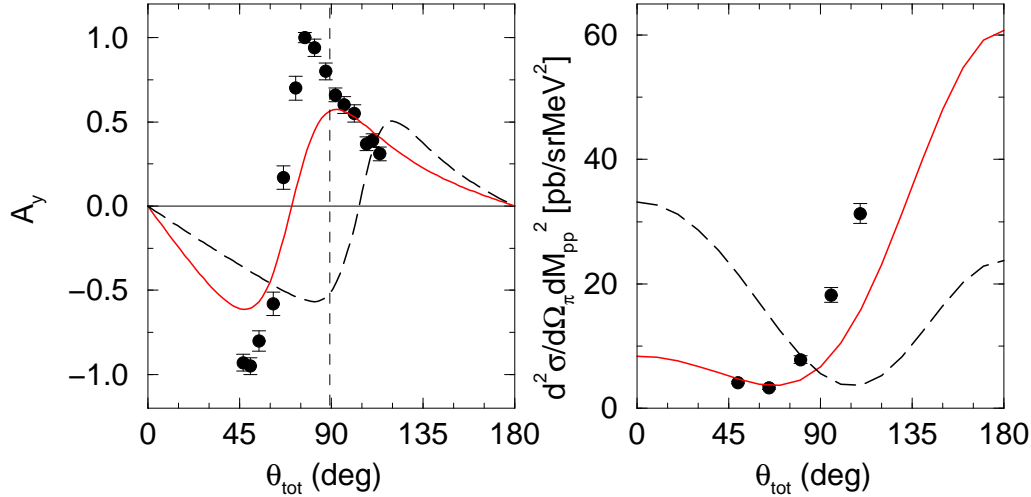} 
\caption{\it Sensitivity of the analyzing power
as well as the differential cross section for the
reaction $pn\to pp\pi^-$ to the sign of the
$^3P_0\to ^1S_0s$ amplitude $a_1$. The lines corresponds
to the model of Ref. \protect{\cite{unserD}}: the solid
line is the model prediction whereas for the dashed line
the sign of $a_1$ was reversed. The experimental
data are from Ref. \protect{\cite{triumf1}} and \protect{\cite{triumf2}}
at T$_{Lab}$ = 353 MeV ($\eta = 0.65$).
  }
\label{pimisigndep}
\end{center}
\end{figure}

\begin{table}[t]
\begin{center}
\caption{\it List of a possible set of the independent amplitude structures
  that contribute to the reaction $pp\to pp+$(pseudoscalar).  In the last
  column shows the lowest partial waves for the final state that contribute to the
  given amplitude (using the notation of section \ref{sr}).
To keep the expressions simple we omitted to give the symmetric, trace free expressions
for the terms listed in the last two lines.  }  \vskip 0.2cm
\begin{tabular}{|l|l c r|l|}
\hline
Amplitudes & Structures & & & Lowest pw \\
\hline \hline
$H$ & \multicolumn{2}{l}{$(\vec p\times \vec p\, ')\vec q\, ' (\vec p\cdot
  \vec p\, ')$} & 
 & Ds \\
\hline
$\vec Q$ & $\vec p$ & $\vec p\, '(\vec p\cdot \vec p\, ')$ &
$\vec q\, '(\vec p\cdot \vec q\, ')$ & Ss, Ds, Sd \\
\hline
$\vec A$ & $\vec p\, '$ & $\vec p(\vec p\cdot \vec p\, ')$ &
$\vec q\, '(\vec p\, '\cdot \vec q\, ')$ & Ps, Pd \\
\hline
$B_{ij}$ & $\epsilon_{ijk}p_k (\vec p\, '\cdot \vec q\, ')$
& $\epsilon_{ijk}p_k' (\vec p \cdot \vec q\, ')$ &
$\epsilon_{ijk}q_k' (\vec p\, '\cdot \vec p)$ & Pp \\
         & $\delta_{ij}(\vec p\times \vec p\, ')\vec q\, '$ & $(\vec p\times
         \vec p\, ')_iq'_j$ & $(\vec p\times \vec q\, ')_ip'_j$ & \\
         &   $p_ip_j(\vec p\times \vec p\, ')\vec q\, '$ & 
  $(\vec p\times \vec p\, ')_ip_j(\vec p\cdot \vec q\, ')$ & 
$(\vec p\times \vec q\, ')_ip_j(\vec p\cdot \vec p\, ')$ & \\
\hline
\end{tabular}
\label{pppi0ampslist}
\end{center}
\end{table}

As follows directly from Eq. (\ref{ampdef}), in the general case the matrix
element ${\cal M}$ is described by 16 complex valued scalar functions.  One
can show, e.g. by explicit construction, that for general kinematics of the
reaction $NN\to NNx$ all 16 amplitudes are independent. A possible choice is
given explicitly in Table \ref{pppi0ampslist} for the reaction $pp\to
pp+$(pseudoscalar). However, for particular reaction channels or an
appropriate kinematical situation their number sometimes reduces drastically.
For example in collinear kinematics (where $ \vec p \ , \ \vec q \, ' \ , \ 
\mbox{and} \ \vec p \, ' $ are all parallel) the number of amplitudes that
fully describes the reactions $pp\to pp+$(pseudoscalar) is equal to 3 (this
special case is discussed in Ref. \cite{pakrekalo}). This can be directly read
off Table  \ref{pppi0ampslist}, for under collinear
conditions
all cross products vanish and all structures of one group that are given by
the different
vectors
of the system collapse to one structure ($\vec Q \to \alpha \vec p \ , \ \
\vec A \to \beta p\ , \ \ B_{ij}\to \gamma \epsilon_{ijk}p_k$).

 Another interesting
example is that of elastic $pp$ scattering.  The Pauli Principle demands (c.f.
section \ref{sr}) that odd (even) parity states are in a spin triplet
(singlet). Time reversal invariance requires the amplitude to be invariant
under the interchange of the final and the initial state.  When parity
conservation is considered in addition, one also finds that the total spin is
conserved. Thus, only $H$ and $B_{ij}$ will contribute to elastic $pp$
scattering. In addition, from the two vectors available in the system ($\vec
p$ and $\vec p\, '$) one can construct only 4 structures that contribute to
the latter, namely $\delta_{ij}(\vec p\cdot \vec p\, ')b_1$,
$\epsilon_{ijk}(\vec p\times \vec p \, ')_kb_2$,
$(p_ip'_j+p_jp'_i-(2/3)\delta_{ij})(\vec p\cdot \vec p\, ')b_3$, and
$(p_ip_j+p'_jp'_i-(2/3)\vec p\, ^2\delta_{ij})(\vec p\cdot \vec p\,
')b_4$.\footnote{Note, this choice of structures in not unique; we could also
  have used $(\vec p\times \vec p\, ')_i(\vec p\times \vec p\, ')_j$ as a
  replacement of any other structure in $b^t_{ij}$.}  Thus, $pp$ scattering is
completely characterized by 5 scalar functions.

As a further example and 
to illustrate how the formalism simplifies in the vicinity of the production threshold,
we will now discuss in detail the production of pions in $NN$ collisions.
Throughout this report, however, the formalism will be applied to various
reactions.  In our example there are three reaction channels experimentally
accessible, namely $pp\to pp\pi^0$, $pp\to pn\pi^+$, and $pn\to pp\pi^-$, that
can be expressed in terms of the three independent transition amplitudes
${\cal A}_{T_iT_f}$ (as long as we assume isospin to be conserved),
where $T_i$($T_f$) denote the total isospin of the initial (final) $NN$ system
\cite{rosenfeld} (c.f. section \ref{isospinrole}).  As in section \ref{sr}, we
will restrict ourselves to those final states that contain at most one
$p$--wave. We may then write for the amplitudes of ${\cal A}_{11}$,
\begin{eqnarray}
\nonumber
H^{11}&=&0 \, , \\
\nonumber
\vec Q^{11}&=&a_1\hat p \,  , \\
\nonumber
\vec A^{11}&=&a_2\vec p \, '
+a_3\left[(\hat p \cdot \vec p \, ') \hat p
-\frac{1}{3}\vec p \, '\right] \, , \\
B_{ij}^{11}&=& 0 \, ;
\label{a11amps}
\end{eqnarray} 
for the amplitudes of ${\cal A}_{10}$,
\begin{eqnarray}
\nonumber
H^{10}&=&0 \, , \\
\nonumber
\vec Q^{10}&=&0 \,  , \\
\nonumber
\vec A^{10}&=&b_2\vec q \, '
+b_3\left[(\vec q\, '\cdot \hat p)\hat p-\frac{1}{3}
\vec q \, '\right] \, , \\
\vec B_{ij}^{10}&=&\epsilon_{ijk}b_1\hat p_k \, ;
\end{eqnarray}
for the amplitudes of ${\cal A}_{01}$,
\begin{eqnarray}
\nonumber
H^{01}&=&0 \, , \\
\nonumber
\vec Q^{01}&=&c_1\vec q \, '+
c_2\left[\hat p(\hat p\cdot \vec q\, ')
-\frac{1}{3}\vec q \, '\right]  \,  , \\
\nonumber
\vec A^{01}&=&0 \, , \\
\nonumber
B_{ij}^{01}&=&\epsilon_{nmk}\vec p_k \, '\left(c_3\delta_{in}\delta_{jm}+
c_4\delta_{jm}\left(\hat p_i\hat
  p_n-\frac{1}{3}\delta_{in}\right)\right.\\
& & \qquad \qquad \qquad \qquad \qquad \qquad
 +\left.c_5\delta_{in}\left( \hat p_j \hat p_m
-\frac{1}{3}\delta_{jm}\right)\right) \ ,
\end{eqnarray}
where $\hat p$ denotes the initial $NN$ momentum, normalized to $1$,
and $\vec p\, '$ and $\vec q\, '$ denote the final nucleon and pion
relative momentum, respectively.
The coefficients given are directly proportional to the corresponding
partial wave amplitudes as listed in Table \ref{tab_secr}; e.g., $a_3$ is proportional to 
the transition amplitude $^1D_2\to ^3P_2s$.
When constructing amplitude structures for higher partial waves care
has to be taken not to list dependent structures.
In order to remove dependent structures the reduction formula Eq. \eqref{reduc} proved useful.
In addition, one should take care that the number of coefficients appearing
exactly matches the number of partial waves allowed.
For example, the partial waves that contribute to $B_{ij}^{01}$ are
$^3S_1\to ^3P_1s$, $^3D_1\to ^3P_1s$ and $^3D_2\to ^3P_2s$.

The large number of zeros appearing in the above list of amplitude
structures reflects the strong selection rules discussed in section \ref{sr}.

As an example we will calculate the beam analyzing power and the differential
cross section for the reaction $\vec p n\to pp\pi^-$. These observables were
measured at TRIUMF \cite{triumf1,triumf2} and later at PSI
\cite{daum1,daum2}---here, however, with a polarized neutron beam). 
In accordance with the TRIUMF
experiment, where the relative $NN$ energy in the final
  state was restricted to at most 7 MeV,  we assume that the outgoing $NN$
system is in a relative $S$--wave. This largely simplifies the expressions. We then find
\begin{eqnarray}
\nonumber
\sigma_0 
&=& \frac14|a_1|^2+\frac12q'\mbox{Re}
\left(a_1^*\left(c_1+\frac{2}{3}c_2\right)\right)\cos (\theta ) \ , \\ 
\nonumber
\sigma_0 A_y
 &=& 
\left(\frac14q'{}^2\mbox{Im}(c_1^*c_2)\sin (2\theta )\right.
 \\
& & \qquad \qquad 
\left.-\frac12q'\mbox{Im}\left(a_1^*\left(c_1-\frac{1}{3}c_2\right)\right)\sin
  (\theta )\right)\cos (\phi)
\ ,
\label{interference}
\end{eqnarray}
where we used the definitions $(\vec q\, ' \times \hat p)_y =
-q'\sin(\theta)\cos (\phi )$ and $(\vec q\, ' \cdot \hat p)=q'\cos
(\theta)$ (c.f. Appendix \ref{vecdef}).  Thus, the forward--backward asymmetry in $\sigma_0$ as well
as the shift of the zero in $A_y$ directly measure the relative phase
of the $^3P_0\to ^1S_0s$ transition in $A_{11}$ ($a_1$) in the pion
$p$--wave transitions of $A_{01}$ ($c_1$ and $c_2$)), as was first
pointed out in Ref. \cite{maeda}. This issue will be discussed below
(c.f. sec. \ref{pionprod}).
Note: As before we neglected here pion $d$--waves, since
they are kinematically suppressed close to the threshold (c.f.
Eq. (\ref{edep})).

\subsubsection{Example I: Polarization observables for a baryon pair in the
  $^1S_0$
final state}
\label{hnpol}

As an example of the efficiency of the amplitude method, in this subsection
we will present an analysis of the angular pattern of some polarization
observables for the reaction $pp \to B_1B_2x$ under the constraint that the
outgoing two baryon state ($ B_1B_2$) is in the $^1S_0$ partial wave and $x$
is a pseudoscalar, as  is relevant for the reaction $pp \to pK\Lambda$.
 The dependence of the
observables on the meson emission angle is largely constrained under these
circumstances, as was stressed in Ref. \cite{vigdor}.
In contrast to the discussion in sec. \ref{sr}, here we will not
assume $B_1$ and $B_2$ to be identical particles. 
The results of this subsection will show how to disentangle in a model
independent way the two spin components of the hyperon--nucleon interaction
\cite{ynfsi}.

The analysis starts with identifying the tensors that are to be considered in
the matrix element of Eq. \eqref{ampdef}\footnote{Note, here we could as well
  refer to table \ref{pppi0ampslist} to come to the same conclusion as in this
  section, for the $^1S_0$ state is allowed for the $pp$ system. However,
   the argument given is quite general and instructive.}. For this we go
through a chain of arguments similar to those in sec. \ref{sr}. Given that we
restrict ourselves to a spin--zero final state, only $H$ and $\vec Q$ can be
non--zero. In addition, the quantum numbers of the final state are fixed by
$l_x$, the angular momentum of the pseudoscalar with respect to the two baryon
system, since \be J = l_x \qquad \mbox{and} \qquad
\pi_{tot}=(-)^{\left(l_x+1\right)} \ee for the total angular momentum and the
parity, respectively.  Conservation of parity and angular momentum therefore
gives \be (-)^L=(-)^{l_x+1}=(-)^{J+1} \qquad \longrightarrow \qquad S=1 \ ,
\ee and consequently we get $H=0$.  In addition, for odd values of $l_x$ we
see from the former equation that $L$ must be even. In $pp$ systems, however,
even values of $L$ correspond to $S=0$ states, not allowed in our case.
Therefore $l_x$ must be even. We may thus make the following ansatz: \be \vec
Q = \alpha \vec p + \beta \vec q\, '(\vec q \, '\cdot \vec p) \ , \ee where
$\alpha$ and $\beta$ are even functions of $p$, $p'$ and $(\vec p\cdot \vec
p\, ')$. All other coefficients appearing in Eq. \eqref{ampdef} vanish.  This
has serious consequences for the angular dependences of the various
observables. For example, the expression for the analyzing power collapses to
\be A_{0i}\sigma_0 = \frac i4\epsilon_{ijk}\left( Q_j^*Q_k\right) =
\frac12\mbox{Im}(\beta^*\alpha)(\vec q \, '\cdot \vec p)(\vec q \, '\times
\vec p)_i \ .  \ee Therefore, independently of the partial wave of the
pseudoscalar emitted, for a two--baryon pair in the $^1S_0$ state the
analyzing power $A_{0y}$ vanishes if the pseudoscalar is emitted either in the
$xy$--plane or in the $zx$--plane. In Ref. \cite{ynfsi} this observation was
used to disentangle the different spin states of the $\Lambda N$ interaction.

\subsubsection{Example II: Amplitude analysis for  $pp\to d\bar K^0 K^+$ close to threshold}
\label{a0amp}

In sec. \ref{a0dat} we discussed in some detail the data of Ref. \cite{a0exp}
for the reaction  $pp\to d\bar K^0 K^+$ based on rather general arguments on
the cross section level. In this subsection we will present the
corresponding production amplitude based on the amplitude method presented
above. This study will allow us at the same time to extract information on the
relative importance of the $a_0^+$ and the $\Lambda (1405)$ in the reaction
dynamics.

As in sec. \ref{a0dat}, we assume that either the $\bar KK$ system or
the deuteron with respect to the $\bar KK$ system is in a $p$--wave, whereas
the other system is in an $s$--wave, calling for an amplitude linear in $p \,
'$ or $q \, '$, respectively. We use the same notation as in sec. \ref{a0dat}.
Therefore the final state has odd parity and thus also the amplitude needs to
be odd in the initial momentum $\vec p$. An odd parity isovector $NN$ state
has to be $S=1$ and thus has to be linear in $\vec S$, defined in section
\ref{generalstructure}. In addition, the deuteron  in the
final state demands that each term is linear in the deuteron polarization
vector $\epsilon$.
 We therefore get for the full transition amplitude, in slight variation to
 Eq. \eqref{ampdef} due to the presence of the deuteron in the final state,
\be
{\cal M}   = B_{ij}\vec S_i\vec \epsilon_j \, ^* \ ,
\ee
where $\vec \epsilon$ appears
as complex conjugate, since the deuteron is in the final state, and
\be
\nonumber
 B_{ij}  &=&
\phantom{ \ + \ }
 a_{Sp}\, \hat p_i\vec q_j\, '
    \ + \ b_{Sp}\, (\hat p\cdot\vec q\, ')\, \delta_{ij} 
    \ + \ c_{Sp} \vec q_i\, '\hat p_j
     \ + \ d_{Sp} \hat p_i\, \hat p_j(\vec q\,
    '\cdot\hat p)\, \\
& & \ + \ a_{Ps}\, \hat p_i\vec p_j\, '
    \ + \ b_{Ps}\, (\hat p\cdot\vec p\, ')\delta_{ij}
 \ + \    c_{Ps}\, \vec p_i\, ' \hat p_j
   \ + \ d_{Ps}\, \hat p_i\hat p_j(\vec q\, '\cdot\hat p)\, ,
\label{8}\ee
where capital letters in the amplitude label indicate the partial wave of the
$\bar KK$ system and small letters that of the deuteron with respect to the
$\bar KK$ system.

Once the individual terms in the amplitude are identified, it is
straightforward to express the $C_i$ defined in
sec. \ref{a0dat} in terms of them. We find, for example,
\be
\nonumber
C_0^{q'}&=&\frac{1}{2}\left(| a_{Sp}|^2+| c_{Sp}|^2    \right) \ , \\
\nonumber
C_1^{q'}&=&| b_{Sp}|^2+\frac{1}{2}| b_{Sp}+d_{Sp}|^2
+\mbox{Re}\left[a_{Sp}^*c_{Sp}+(a_{Sp}+c_{Sp})^*(b_{Sp}+d_{Sp})\right] \ ,  \\
\nonumber
C_2&=&a_{Sp}^*a_{Ps}+c_{Sp}^*c_{Ps} \ .
\ee
A fit to the experimental data revealed that, within the experimental
uncertainty, $C_0^{p'}$ is compatible with zero. Thus, given the previous
formulas, both $a_{Ps}$ and $ c_{Ps}$ have to vanish individually.

\begin{figure}[t]
\begin{center}
\epsfig{file=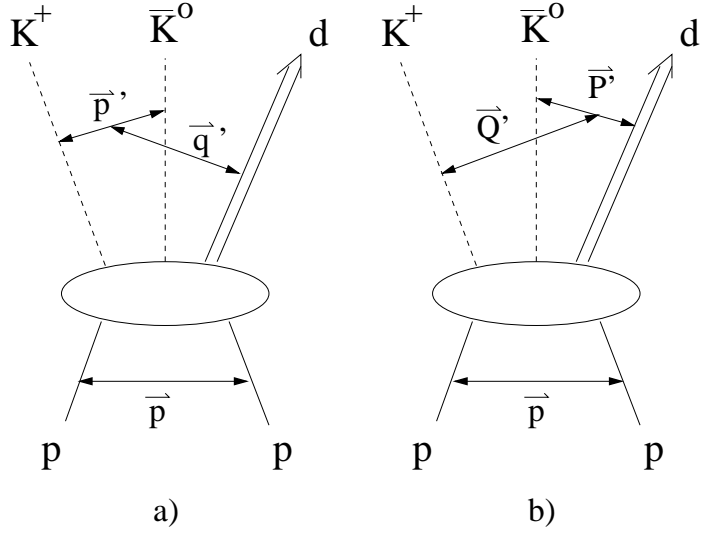, height=7cm}
  \caption{\it Illustration of the coordinate system used in the analysis
for the reaction $pp\to d\bar K^0 K^+$. }
\label{koor}
\end{center}
\end{figure}

Based on the strongly populated $\bar KK$ $s$--waves in Ref. \cite{a0exp} it
was argued that the  reaction $pp\to d\bar K^0 K^+$ is governed by the
production of the $a_0^+$. In Ref. \cite{oseta0lam}, however, it was argued
that the strong $\bar K d$ interaction caused by the proximity of the $\Lambda
(1405)$ resonance should play an important role as well. This FSI
should enhance the $\bar K d$ $s$--wave. We now want to calculate the
contribution of this partial wave relative to the $\bar K d$ $p$--wave
based on the amplitudes given in table \ref{cpara}. This will illustrate a
further strength of the amplitude method, for within this scheme changing the
coordinate system is trivial. The coordinate system suited to study resonances
in the $\bar KK$ system and in the $\bar K d$ system are illustrated in the
left and right panels of Fig. \ref{koor}. All we need to do now is express
the vectors that appear in Eq. \eqref{mform} in terms of $\vec Q\, '$ and $\vec
P\, '$.
We find
$$
\vec q\, ' = \vec P\, ' - \alpha \vec Q\, ' \qquad \mbox{and} \qquad \vec p\, '
=\frac{1}{2}((2-\alpha)\vec Q\, ' + \vec P\, ') \ ,
$$
where $\alpha = m_d/(m_d+m_{\bar K})$. Obviously, the squared amplitude
expressed in the new coordinates reveals the same structure as
Eq. \eqref{mform}:
\be \nonumber
\bar{|{\cal M}|^2} &=& B_0^{Q'}{Q'}^2+B_0^{P'}{P'}^2
+B_1^{Q'}(\vec {Q'}\cdot \hat p)^2+B_1^{p'}(\vec {P'}\cdot \hat p)^2 \\
& & \qquad \qquad \qquad \qquad \qquad +B_2(\vec {P'}\cdot \vec
{Q'})+B_3(\vec {P'}\cdot \hat p)(\vec {Q'}\cdot \hat p) \ ,
\label{mformnew}
\ee
where the coefficients appearing can be expressed in terms of the $C$
coefficients of Eq. \eqref{mform} so that, for example,
\be
\nonumber
B_0^{Q'}&=&\frac{(2-\alpha)^2}{4}C_0^{q'}+\alpha^2C_0^{k'}-\frac{\alpha(2-\alpha)}{2}\frac{1}{2}C_2
\ ,\\
\nonumber
B_1^{Q'}&=&\frac{(2-\alpha)^2}{4}C_1^{q'}+\alpha^2C_1^{k'}-\frac{\alpha(2-\alpha)}{2}\frac{1}{2}C_3
\ . \\
\ee
With these expressions at hand it is easy to verify, that
$K\bar K$ $s$--waves contribute to 83  \% to the total cross section,
whereas
$\bar K d$ $s$--waves contribute to  54  \% only \cite{utica}.
Here we used the total $s$--wave strength for $\bar K K$,
$(C_0^{q'}+(1/3)C_1^{q'})$, and the total $s$--wave strength for $\bar K d$,
$(B_0^{Q'}+(1/3)B_1^{Q'})$, as a measure of the strength of the partial waves.

As we do not see a significant population of the $\bar K d$ $s$--wave, it
appears that the $\Lambda (1405)$ does not play an essential role in the
reaction dynamics of $pp\to d\bar K^0 K^+$ close to threshold in contrast to
the $a_0^+$.

\subsection{Spin Cross Sections}

As early as 1963, Bilenky and Ryndin showed  \cite{br}, that from the spin correlation coefficients
that can be extracted from measurements with polarized beam and target,
the cross section can be separated into pieces that stem from
different initial spin states. Their results were
recently re-derived  \cite{pia} and the formalism was
generalized to the differential level in Ref. \cite{deepak}.
With these so--called spin cross sections it can easily be
demonstrated how the use of spin observables enable one to filter out particular
aspects of a reaction. We begin this subsection with
a derivation of the spin cross sections using the amplitude method
of the previous subsection and then use the
reaction $\vec p\vec p \to pn\pi^+$ as an illustrative example.

Since we have the amplitude decomposition of the individual
observables given in Eqs. (\ref{si0def})--(\ref{aiidef}),
one easily finds
\begin{eqnarray}
\sigma_0(1-A_{xx}-A_{yy}-A_{zz})=|H|^2+|A|^2&=:&^1\sigma_0 \, , \\
\sigma_0(1+A_{xx}+A_{yy}-A_{zz})=|Q_z|^2+|B_{zn}|^2&=:&^3\sigma_0
\,  ,\\
\sigma_0(1+A_{zz})=\frac12\left(|Q_x|^2+|Q_y|^2
+|B_{xn}|^2+|B_{yn}|^2\right)&=:&^3\sigma_1 \, ,
\label{spinwqs}
\end{eqnarray}
where the assignment of the various spin cross sections
$^{(2S+1)}\sigma_{M_S}$, with $S$ ($M_S$) the total spin (projection of the
total spin on the beam axis) of the initial state can be easily
confirmed from the definition of the amplitudes in Eq. (\ref{ampdef}).

\begin{table}[t]
\begin{center}
\caption{\it List of the lowest partial waves in the final state that contribute
to the individual spin cross sections. Capital letters denote the
baryon--baryon
partial waves whereas small letters that of the meson with respect to the
baryon-baryon system.}
\vskip 0.2cm 
\begin{tabular}{|c|c|c|}
\hline
$^{2S+1}\sigma_m$ & \multicolumn{2}{|c|}{possible final states
for $pp\to bb'x$} \\
                  &  $Sl_x$ & $Pl_x$ \\
\hline
$^1\sigma_0$ & $^3S_1p$ & $^3P_js, \ ^1P_1s$ \\
$^3\sigma_0$ & $^1S_0s, \ ^1S_0d , \ ^3S_1d$ & $ ^3P_jp, \ ^1P_1p$  \\
$^3\sigma_1$ & $^3S_1s, \ ^1S_0d , \ ^3S_1d$ & $^3P_jp, \ ^1P_1p$  \\
\hline
\end{tabular}
\label{pws}
\end{center}
\end{table}

As was shown in section \ref{sr}, in case of two nucleon initial or final
states restrictive selection rules apply. For example, for $pp$ final states
the isospin of the final $NN$ system is 1 and therefore states with even (odd)
angular momentum have total spin 0 (1). From Table \ref{pws} it thus follows,
that the $pp$ $S$--wave in connection with a meson $s$--wave contributes only
to $^3\sigma_0$. This example of how spin observables can be used to
filter out particular final states was used previously in sec.  \ref{fsisec}
(c.f. discussion to Fig. \ref{partxs}).

\begin{figure}[t]
\begin{center}
\epsfig{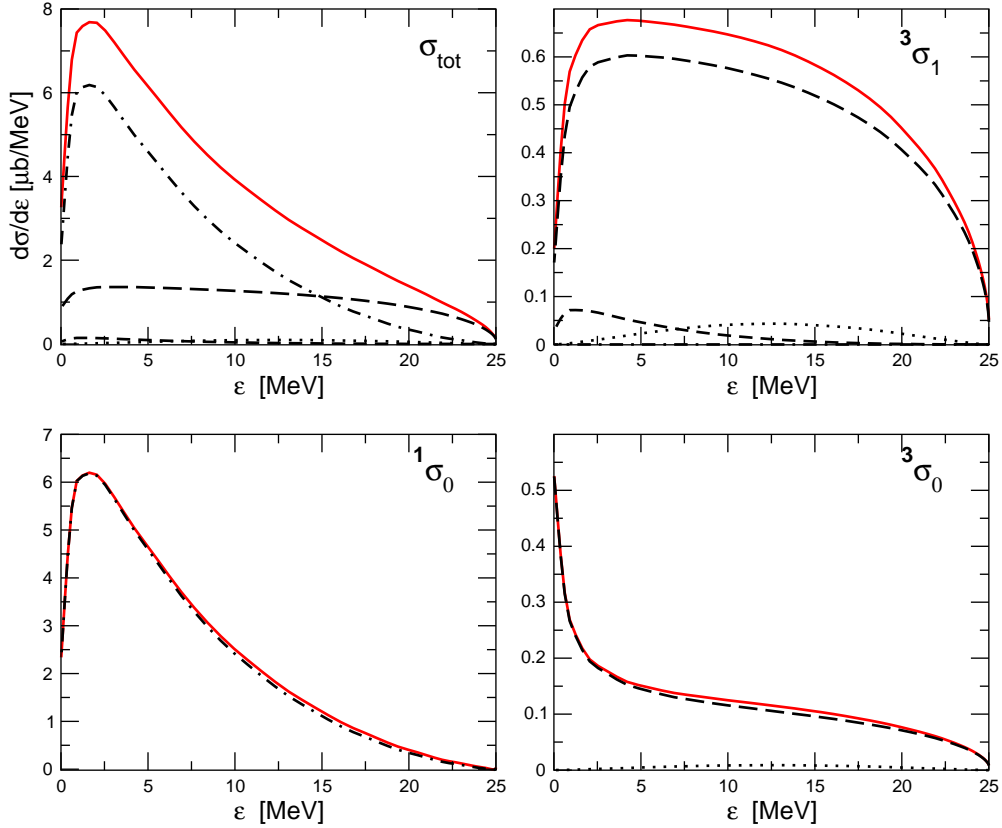}
\caption{\it Demonstration of the selectivity of
the spin cross sections. Shown are the 
spectra of the reaction $\vec p\vec p\to pn\pi^+$ as
a function of $\epsilon$ for an excess
energy of 25 MeV. In each panel the solid line shows
the full result for the corresponding cross section,
the dot--dashed, long--dashed, dashed, and dotted
lines show the $Sp$, $Ss$, $Sd$, and $Pp$ contribution
respectively. The curves are from the model of Ref. \protect{\cite{unserpol}}.
}
\label{spwqs}
\end{center}
\end{figure}

For illustrative purposes we show in Fig. \ref{spwqs} 
the spectra for the various cross sections
for the reaction $\vec p\vec p \to pn\pi^+$, as a function
of the relative energy of the nucleon pair in the final state. 
The curves correspond to the model of Ref. \cite{unserpol} that
very well describes the available data in the $\pi^+$ production
channel. The model is described in detail in sec. \ref{pionprod}.
In the upper left
panel the unpolarized cross section is shown. It is dominated by
the $Sp$ final state (the dominant transition is $^1D_2\to ^3S_1p$), 
and from this spectrum alone it would be 
a hard task to extract information on final states other than
the $^3S_1$ $NN$ state. This can be clearly seen by the similarity
of the shapes of the dot--dashed line and the solid line.
The spin cross sections, however, allow separation of the spin
singlet from the spin triplet initial states. Naturally
$^1\sigma_0$ (lower left panel of Fig. \ref{spwqs}) is now saturated by the $Sp$ final state, but in
$^3\sigma_0$ and $^3\sigma_1$ other structures appear: the former
is now dominated by the transition $^3P_0\to ^1S_0s$ and the
latter by $^3P_1\to ^3S_1s$.

\begin{table}[t]
\begin{center}
\caption{\it List of observables measured
for various $NN\to NNx$ channels for excess 
energies up to $Q=40$ MeV.}
\begin{tabular}{|c|l|l|l|l|l|}
\hline
channel                  &  $\sigma_{tot}$ & ${d\sigma}/{d\Omega}$ & ${d\sigma}/{dm}$
& $A_{oi}$ & $A_{ij}$ \\
\hline
$pp\to pp\pi^0$ & \cite{meyer1,meyer2,ups,ups_pi0,tof_pi0,pppi02} & 
 \cite{pppi0sat,tof_pi0,ups_pi0} & \cite{ups_pi0,tof_pi0,pppi02}
& \cite{pppi02,meyerpol}& \cite{meyerpol} \\
$pp\to pn\pi^+$ & \cite{daelett,pnpipl,kaipipl} & \cite{pnpipl} &
& \cite{pnpipl}& \cite{polpnpipl} \\
$pn\to pp\pi^-$ & \cite{daum1} & \cite{daum1,triumf1} & \cite{daum1}
& \cite{daum2,triumf2}&  \\
$pp\to d\pi^+$ &
\cite{dpi1,dpi2,dpi3,dpi4,dpi5,dpi6,dpi7,dpi8,dpi9,dpi10}
 & \cite{dpi1,dpi9,dpi10} & ---
& \cite{dpipl_ay,dpi11,dpi12}& \cite{poldpipl} \\
\hline
$pp\to pp\eta$ & \cite{calen1,calen4,smyrski,paweleta,eduard_eta,calen6,etatot1,etatot2,etatot3} &
\cite{eduard_eta,paweleta,calen5} & \cite{eduard_eta,paweleta}
& \cite{etapol}&  \\
$pn\to pn\eta$ & \cite{calen2} & &
&  &  \\
$pn\to d\eta$ & \cite{calen3,calen6} &  & ---
& &  \\
\hline
$pp\to pp\eta'$ & \cite{etatot3,pawel_etap,pawel_etap2} & &
&  &  \\
\hline
$pp\to pK^+\Lambda$ & \cite{jan,jan2,jan3} & & \cite{jan3}
& &  \\
$pp\to pK^+\Sigma^0$ & \cite{sig0} & &
& &  \\
\hline
$pp\to pp\omega$ & \cite{ppomega} & &
& &  \\
$pn\to d\omega$ & \cite{inti} & & ---
& &  \\
\hline
$pp\to pp\phi$ & \cite{ppphi} & \cite{ppphi} & 
& &  \\
\hline
$pp\to ppf_0/a_0$ & \cite{pawelf0a0} &  & 
& &  \\
\hline
$pp\to pp\pi^+\pi^-$ & \cite{piplpim1,piplpim2,twopion} &  \cite{piplpim1,piplpim2,twopion}
& \cite{piplpim1,piplpim2,twopion} & &  \\
$pp\to pn\pi^+\pi^0$ & \cite{twopion} &  \cite{twopion} & \cite{twopion}
& &  \\
$pp\to pp\pi^0\pi^0$ & \cite{twopion} &   \cite{twopion} & \cite{twopion}
& &  \\
\hline
$pp\to ppK^+K^-$ & \cite{ppkpkm} & &
& &  \\
$pp\to dK^+\bar K^0$ & \cite{vera} &  \cite{vera} & \cite{vera}
& &  \\
\hline
\end{tabular}
\label{measurements}
\end{center}
\end{table}

\subsection{Status of Experiment}
In this presentation we will be rather brief on details
about current as well as planned experiments, as this
subject was already covered in recent reviews \cite{machnerrep,oelertrep}.
Here we only wish to give a brief list of observables  and reactions
that are measured already or are planned to be measured
in nucleon--nucleon and nucleon--nucleus induced reactions.

\begin{figure}[t!!]
\begin{center}
\epsfig{file=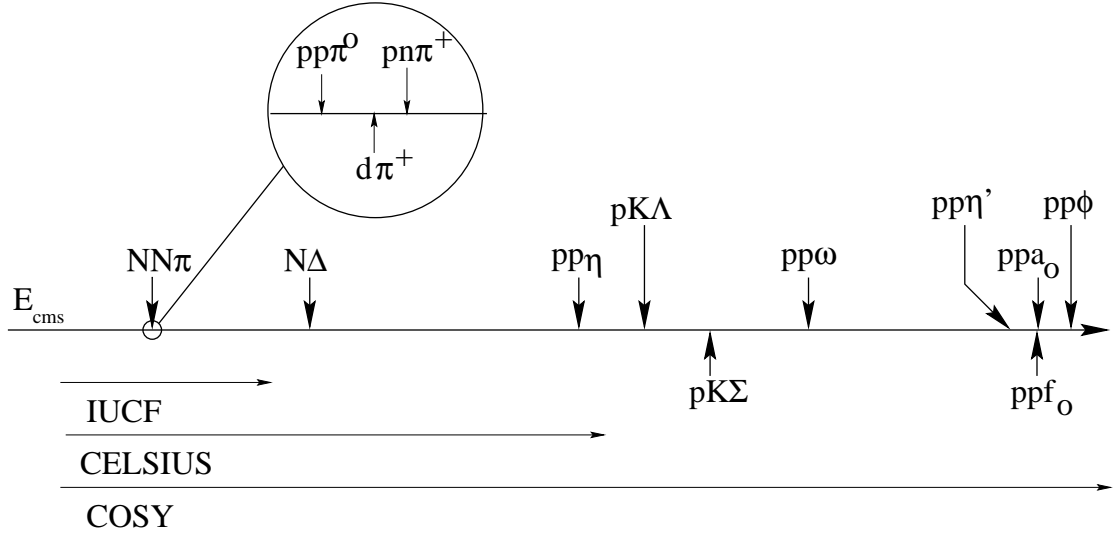, height=7cm}
\caption{\it The lowest meson production thresholds for
single meson production in proton--proton collisions
together with the corresponding energy ranges of
the modern cooler synchrotrons.}
\label{expfig}
\end{center}
\end{figure}

 In the case of pion production, measurements with vector--
and tensor--polarized deuteron and vector--polarized proton targets
and polarized proton beams
have been carried out at IUCF \cite{stori99}.

  Because of good $4\pi$ detection of
photons and charged particles, CELSIUS, at least in the near future,
is well equipped for studies involving $\eta$ mesons. For the
production of heavier mesons, COSY, due to its higher beam energies
and intensities, but due most importantly to the possibility to use
polarization, is in a
position to dominate the field during the years to come.
The energy range of the different cooler synchrotrons is
illustrated in Fig. \ref{expfig}.

In table \ref{measurements} a list is given for the various $NN$ induced
production reactions measured in recent years at SACLAY, TRIUMF, PSI, COSY,
IUCF, and CELSIUS in the near--threshold regime ($Q < 40$ MeV). The
corresponding references are listed as well.  In the subsections to come we
will discuss some examples of the kind of physics that can be studied with the
various observables in the many reaction channels.


\section{Symmetries and their violation}
\label{syms}

As was already mentioned in several places in this article,
symmetries strongly restrict the allowed pattern for
various observables. This leads to observable
consequences. Naturally, it is therefore also straight forward to investigate
the breaking of these symmetries by looking at a violation of 
these symmetry predictions.

Probably the most prominent example of an experiment for a storage ring is
that proposed by the TRI collaboration to be performed at the COSY
accelerator. The goal is to do a null experiment in order to put an upper
limit on the strength of T--odd P--even interactions via measuring $A_{y,xz}$
in polarized proton--deuteron scattering, which should vanish if time reversal
invariance holds \cite{trith}. For details we refer to Ref. \cite{tri}.

\subsection{Investigation of charge symmetry breaking (CSB)}

If the masses of the up and down quark were equal, the QCD Lagrangian were
invariant under the exchange of the two quark flavors. In reality these
masses are not equal and also the presence of electro magnetic effects leads
to small but measurable charge symmetry breaking effects. Note, the mass
difference of a few MeV \cite{leutwyler} is small compared to the typical
hadronic scale of 1 GeV.  Quantifying CSB effects therefore allows to extract
information on the light quark mass differences from hadronic observables.

As should be clear from the previous paragraph, CSB is closely linked to the
isospin symmetry. However, does isospin symmetry demand an independence of the
interaction under an arbitrary rotation in isospin space, a system is charge
symmetric, if the interaction does not change under a 180 degree rotation in
isospin space. Therefore isospin symmetry or charge independence is the
stronger symmetry than charge symmetry. For an introduction into the subject
we refer to Ref. \cite{csbrep}.

The advantage of reactions with only nucleons or nuclei in the initial state
is, that one can prepare initial states with well--defined isospin. This is
in contrast to photon induced reactions, since a photon has both isoscalar
and isovector components.  Therefore, in the case of meson photo-- or
electro--production, CSB signals can only be observed as deviations from some
expected signal. (As an example of this reasoning see Ref. \cite{tabakin}.)
In the
case of hadron induced reactions on the other hand, experiments can be
prepared that give a non--vanishing result only in the presence of CSB. This
makes the unambiguous identification of the effect significantly easier.

One complication that occurs if a CSB effect is to be extracted from the
comparison of two cross sections with different charges in the final state is
that of the proper choice of energy variable. Obviously, the quantity that
changes most quickly close to the threshold is the phase space, so that it
appears natural to compare two reaction channels that have the same
phase space volume. However, due to the differences in the particle masses,
this calls for different initial energies. In the case of a resonant production
mechanism this might lead to effects of the same order as the effect of
interest. In Ref. \cite{jounikin} this is discussed in detail for the
reactions $pp\to d\pi^+$ and $pn\to d\pi^0$.

 In this report we will concentrate on the
implications
of CSB on observables in $NN$ and $dd$ collisions. For
details of the mechanisms of CSB we refer to Ref. \cite{csbrep}. 
In the corresponding class of experiments
the deuteron as well as the alpha particle play an exceptional role
since as isoscalars they can act as isospin filters.

The most transparent example of a CSB
reaction is 
$$
dd\to \alpha x \ ,
$$
where $x$ is some arbitrary isovector. The reaction with $x=\pi^o$ was
recently measured at IUCF for the first time close to the threshold \cite{dd2alphapi0}. The initial
state as well as the $\alpha$ are  pure isoscalars. Thus the final state as
an isovector can not be reached as long as isospin is conserved.

In addition
$pn$ reactions can be used for producing clean signals of CSB. More
generally, whenever a $pn$ pair has a well--defined isospin they behave as
identical particles. Specifically, the differential cross section needs to be
forward--backward symmetric because nothing should change if
beam and target are interchanged. Any deviation from this symmetry is an unambiguous signal of
CSB \cite{csbrep}. An experiment performed at TRIUMF
recently claimed for the first time a non--vanishing forward--backward asymmetry in $pn\to
d\pi^0$ \cite{allenapi0}.  Note that not every forward--backward asymmetry in
$pn$ reactions stems from isospin violation. A counter example was given
at the end of section \ref{generalstructure}, where the differential cross
section for $pn\to pp\pi^-$ is discussed in detail. There, in contrast to
the previous example, $T=0$ and $T=1$ initial states interfere.

In case of pion production in nucleon--nucleon collisions it is
possible to define a convergent effective field theory (see section
\ref{pionprod}). Within this theory it is possible to relate effects
of CSB in these reactions directly to the up--down quark mass
difference \cite{csbrep,jchiral}. Preliminary studies show, that the relative
importance of different CSB mechanisms in the reaction $pn\to d\pi^0$
and $dd\to \alpha \pi^0$ are very different \cite{csbprep}, and thus it should be possible to
extract valuable information on the leading CSB operators from a combined
analysis of the two reactions.

\begin{figure}[t]
\begin{center}
\epsfig{file=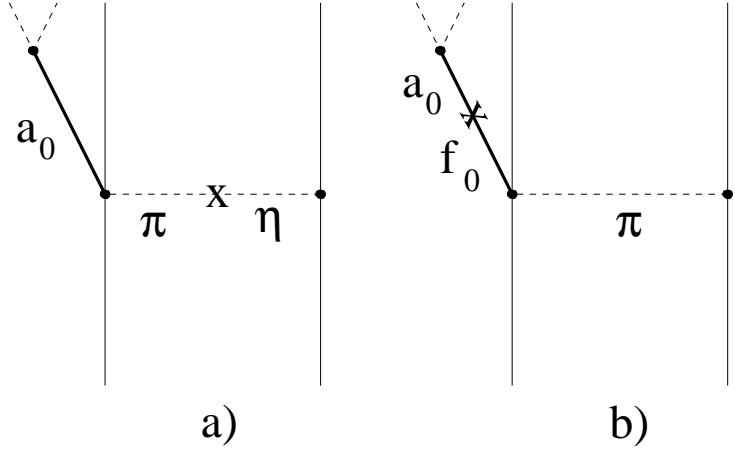, height=6cm}
  \caption{\it Illustration of different sources of charge symmetry breaking:
    diagram a) shows CSB in the production operator through $\pi-\eta$ mixing
    and diagram b) shows CSB in the propagation of the scalars. Thin solid
    lines denote nucleons, thick solid lines scalar mesons and dashed ones
    pseudoscalar mesons. The $X$ indicates the occurrence of a CSB
    matrix element.}
\label{diffcsb}
\end{center}
\end{figure}
In the arguments given all that was used was the isovector character of the
meson produced. Thus, the same experimental signals will be seen also in
$pn\to d(\pi^0\eta)$ \cite{saschafb} and $dd\to \alpha (\pi^0\eta)$
\cite{leoniddd}. In Ref. \cite{unsera0f0} also the analysing power was
identified as a useful quantity for the extraction of the $a_0-f_0$
mixing matrix element (see also discussion in Ref. \cite{ayfrompwaves}). 
The $(\pi^0\eta)_{s-wave}$  is interesting especially
close to the $\bar KK$ threshold, since it should give insight into the nature of the
light scalar mesons $a_0(980)$ and $f_0(980)$ (c.f. sec. \ref{a0f0}). Note that
the $\pi\eta$ channel is the dominant decay channel of the $a_0$, which is an
isovector--scalar particle.  It should be stressed that the charge symmetry
breaking signal in case of the scalar mesons is significantly easier to
interpret in comparison to the case of pion production. The reason is that the
two scalar resonances of interest overlap and therefore the effect of CSB as
it occurs in the propagation of the scalar mesons is enhanced compared to
mixing in the production operator \cite{utica}.  To make this statement more
quantitative we compare the impact of $f_0-a_0$ mixing in the propagation of
the scalar mesons (Fig. \ref{diffcsb}b) to that of $\pi-\eta$ mixing in the
production operator (Fig. \ref{diffcsb}a). We regard the latter as a typical
CSB effect and thus as a reasonable order--of--magnitude estimate for CSB in the
production operator.  Observe that the relevant dimensionless quantity for this
comparison is the mixing matrix element times a propagator (c.f. Fig.
\ref{diffcsb}). In the production operator the momentum transfer---at least
close to the production threshold---is given by $t=-M_Nm_R$, where $m_R$
denotes the invariant mass of the meson system produced (or equivalently the
mass of the resonance) and $M_N$ denotes the nucleon mass. Thus, the
appearance of the $\eta$ propagator introduces a factor of about $1/t$ into
the amplitude, since $t\gg m_\eta^2$.  On the other hand, the resonance
propagator is given by $1/(m_R\Gamma_R)$, as long as we concentrate on
invariant masses of the outgoing meson system close to the resonance position.
Here $\Gamma_R$ denotes the width of the scalar resonance. Thus we find using
$\Gamma_R= 50$ MeV \cite{pdb} that the CSB in the production operator is
kinematically suppressed by a factor of more than $\Gamma_R/M_N \sim 1/20$ as
compared to CSB in the propagation of the scalars.  In addition the mixing
matrix element is enhanced in the case of $f_0-a_0$ mixing (c.f. sec. \ref{a0f0})
and therefore it should be possible to extract the $f_0-a_0$ mixing matrix
element from $NN$ and $dd$ induced reactions.


\section{The reaction $NN\to NN\pi$}
\label{pionprod}

The production of pions in nucleon--nucleon collisions has
a rather special role. First of all, it is the lowest hadronic inelasticity
for the nucleon--nucleon interaction  and
thus an important test of our understanding of the 
phenomenology of the $NN$ interaction. Secondly, since
pions are the Goldstone bosons of chiral symmetry,
it is possible to study this reaction using chiral perturbation
theory. This provides the opportunity to improve the phenomenological
approaches via matching to the chiral expansion as well
as to constrain the chiral contact terms via resonance saturation.
Last but not least, a large number of (un)polarized data is
available (c.f. table \ref{measurements}).

After a brief history,
we continue this chapter with a discussion of a particular phenomenological
model for pion production near the threshold, followed by
a presentation of recent results from chiral perturbation
theory. 

\subsection{Some History}
\label{hist}

In section \ref{thapp} it was argued, that for the
near--threshold regime the distorted wave born approximation is
appropriate, and here we will concentrate on those models
that work within this scheme\footnote{In
the next section a further argument in favor of the 
distorted wave born approximation will be given.}. 

Pioneering work on pion production was done by Woodruf~\cite{woodruf}
as well as by Koltun and Reitan~\cite{KuR} in the 1960s. The diagrams
included are shown in Figs. \ref{kur}a and \ref{kur}b where,
 in these early approaches, the $\pi N \to \pi N$ transition
amplitudes (denoted by $T$ in diagram \ref{kur}b) were parameterized by the scattering
lengths.

When the first data on the reaction $pp\to pp\pi^0$ close
to the threshold were published \cite{meyer1,meyer2},
it came as a big surprise that the model of Koltun and Reitan \cite{KuR}
underestimated
the data by a factor of 5--10. This is
in vast contrast to the reaction $pp\to pn\pi^+$ reported
in Ref. \cite{daelett},
where the discrepancy was less than a factor of 2. On the other hand, it was shown
that the energy dependence of the total cross section 
can be understood from that of the $NN$ FSI once the
Coulomb interaction is properly included \cite{MuS} (c.f. sec. \ref{fsisec}).

Niskanen investigated whether the inclusion of the Delta isobar,
as well as keeping the rather strong on--shell energy dependence
of the $\pi N$ interaction,
could help to improve the theoretical results for neutral pion production \cite{Nis1}.
Although these improvements lead to some enhancement, the cross section was
still missed by more than a factor of 3.

The first publication that reported a quantitative understanding of the $pp\to
pp\pi^0$ data was that by Lee and Riska \cite{LuR} and later confirmed by
Horowitz et al. \cite{HMG}, where it was demonstrated that short range
mechanisms as depicted in Fig. \ref{kur}c, can give a sizable contribution.
However, shortly after this discovery Hern\'andez and Oset demonstrated, using
various parameterizations for the $\pi N\to \pi N$ transition amplitude and
qualitatively reproducing earlier work by Hachenberg and Pirner \cite{HuP}
that the strong off--shell dependence of that amplitude can also be sufficient
to remove the discrepancy between the Koltun and Reitan model and the data.
Gedalin et al. came to the same conclusion within a relativistic one boson
exchange model \cite{moalemOBE}.  In Ref. \cite{unsers} the $\pi N$ amplitude
needed as input for the evaluation of diagram \ref{kur}b was extracted from a
microscopic model. Also there a significant although smaller contribution from
the pion rescattering
was found. Thus, in this model still some
additional short range mechanism is needed.

In the succeeding years many theoretical works presented 
calculations for  the $pp\to pp\pi^0$ cross section.
 In Refs. \cite{shyampiprod,ulfnovel}
covariant one boson exchange models were used in combination
with an approximate treatment of the nucleon--nucleon interaction.
Both models turned out to be dominated by heavy meson exchanges and
thus give further support to the picture proposed in Ref. \cite{LuR}.
However, in Ref. \cite{jiri} the way that the anti nucleons were treated
in Refs. \cite{LuR,HMG,shyampiprod} was heavily criticized: the authors
argued that the anti  nucleon contributions get significantly suppressed
once they are included non perturbatively. It is interesting to note,
that also in Bremsstrahlung the contribution from anti--nucleons in a
non--perturbative treatment is significantly reduced compared to a
perturbative inclusion \cite{brems1,brems2}.
Additional short range contributions were also suggested, namely
the $\rho-\omega$ meson exchange current \cite{RKM}, resonance
contributions \cite{resonancespi0} and loops that contain
resonances \cite{osetroperloop}, all those,  however, turned out to
be smaller compared to the heavy meson exchanges and the off--shell
pion rescattering, respectively.

At that time the hope was that chiral perturbation theory 
might resolve the true ratio of rescattering and short
range contributions. It came as a big surprise, however,
that the first results for the reaction $pp\to pp\pi^0$\cite{chiral1,chiral2}
found a rescattering contribution that interfered destructively
with the direct contribution (diagram a in Fig. \ref{kur}),
making the discrepancy with the data even more severe.
In addition, the same isoscalar rescattering amplitude also
worsened the discrepancy in the $\pi^+$ channel \cite{mitulf,derocha}.
 Some
authors interpreted this finding as a proof for the 
failure of chiral perturbation theory in these
large momentum transfer reactions \cite{ulfnovel,moalemHHF}.
Only recently was it demonstrated,
that it is possible to appropriately modify the
chiral expansion in order to make it capable of analysing
meson production in nucleon--nucleon collisions. We
will report on those
studies in sec. \ref{cpt} that in the future will
certainly prove useful for improving the phenomenological approaches
(c.f. subsec. \ref{pionlessons}).

Before we close this section a few remarks on the 
off--shell $\pi N$ amplitude are necessary.
For this purpose we write the relevant piece of the $\pi N$ interaction
$T$--matrix in the following
form 
\begin{equation}
T_{\pi N} = -t^{(+)}\frac{1}{2}N^\dagger{\mathbf \pi}\cdot {\mathbf \pi}N
+t^{(-)}\frac{1}{2m_\pi}N^\dagger{\bf \tau}\cdot 
({\mathbf \pi}\times \dot {\mathbf \pi})N \ ,
\label{pinham}
\end{equation}
where $t^{(+)}$ ($t^{(-)}$) denote the isoscalar (isovector) component.
Note that it is only the former that can contribute
to the reaction $pp\to pp\pi^0$\footnote{This is 
only true if we do not include the
$\Delta$ isobar explicitly, as will be discussed in the
next section.}, for the isospin structure of the 
latter changes the total isospin of the two nucleon system.

As long as we neglect the distortions due to the final and
initial state interactions, what is relevant for the
discussion in this paragraph is the half off--shell 
$\pi N$ amplitude. We thus may write $t=t(s,k^2)$,
where $s$ denotes the invariant energy of the $\pi N$ system 
and $k^2$ is the square of the four momentum of the incoming pion.
At the threshold for elastic $\pi N$ scattering ($s=(m_\pi+M_N)^2$, $k^2=m_\pi^2$)
we may write
\begin{eqnarray}
\nonumber
 t^{(+)}(s_0,m_\pi^2) &:=& \frac{4\pi}{3}\left(1+\frac{m_\pi}{M_N}\right)(a_1+2a_3)
= (-0.05 \pm 0.01) \,  m_\pi^{-1} \ , \\
t^{(-)}(s_0,m_\pi^2)&:=& \frac{4\pi}{3}\left(1+\frac{m_\pi}{M_N}\right)(a_1-a_3)\phantom{2}
= (1.32 \pm 0.02)  \, m_\pi^{-1} \ ,
\label{lamthr}
\end{eqnarray}
where the $a_{2I}$ denote the scattering lengths in the corresponding
isospin channels $I$. 
The corresponding values were extracted from data on $\pi^-d$ atoms in Ref. \cite{ulfpid}.
Note that the dominance of the isovector interaction is a consequence of the
chiral symmetry: the leading isoscalar rescattering is suppressed by a factor $m_\pi/M_N$
compared to the leading iso--vector contribution---the so called
Weinberg--Tomozawa term \cite{wein3,tom}. However, it is still remarkable,
that also the higher order chiral corrections are small leaving
a value consistent with zero for the isoscalar scattering length. 
A detailed study showed, that this smallness is a consequence of
a very efficient cancellation of several individually large terms
that are accompanied with different kinematical factors \cite{ulfpin}.
To be concrete: to order ${\cal O}(p^2)$ one finds
\begin{equation}
 t^{(+)} = \frac{2}{f_\pi^2}\left(-2m_\pi^2c_1+q_0'k_0
\left(c_2-\frac{g_A^2}{8M_N}\right)+(q'k)c_3
\right) \ ,
\label{lam1}
\end{equation}
where $k$ and $q'$ denote the four momentum of the initial and final pion
respectively. The values for the various $c_i$ are given in table \ref{citable}.
For on--shell scattering at the threshold ($q'=k=(m_\pi,\vec 0)$) one gets
$
 t^{(+)}(s_0,m_\pi^2) = -0.24 \, m_\pi^{-1}$
using the values of Ref. \cite{ulf1}\footnote{Note, this value is inconsistent
  with the empirical value given in Eq. \eqref{lamthr}. To come to a
  consistent value one has to go to one loop order as discussed in Ref. \cite{ulf2}.}.
 Please note, that the linear combination
of the $c_i$ appearing above turns out to be an order of magnitude smaller than
the individual values.
As a consequence, the on--shell isoscalar amplitude
shows a rather strong energy dependence above threshold.
It should not then come as a surprise, that the transition
amplitude corresponding to Eq. (\ref{pinham}), when
evaluated in the kinematics relevant for pion production 
in $NN$ collisions, 
within chiral perturbation theory turns out to be rather large numerically
\cite{chiral1,chiral2}. For the non--covariant expression
given in Eq. (\ref{lam1}) this translates into $q'=(m_\pi, \vec 0)$
and $k=(m_\pi/2,\vec k)$, leading to $
t^{(+)}((m_\pi+M_N)^2,-M_Nm_\pi) = 0.5 \, m_\pi^{-1} 
$.
This is why the first calculations using chiral perturbation theory
found a big effect from pion rescattering---unfortunately
increasing the discrepancy with the data. 
In Ref. \cite{chiralimp} the tree level chiral perturbation theory
calculations where repeated using a different prescription
for the energy of the exchange pion\footnote{This prescription was
later criticized in Ref. \cite{toy1}.}. The authors found 
agreement with the data, but with a sign of the full amplitude opposite to the one
of the direct term.

One year earlier it was shown, that
 within phenomenological approaches
the isoscalar transition amplitude evaluated in off--shell kinematics
is also significantly different from its on--shell value.
For example, in the J\"ulich meson exchange
model it is the contribution from the iterated 
$\rho$ $t$--channel exchange and the $\sigma$ exchange
that are individually large but basically cancel in threshold
kinematics in the isoscalar channel \cite{unsers}. 
This cancellation gets weaker away from
the threshold point. This rescattering contribution, however, turned out
to interfere constructively with the direct term.

Since chiral perturbation theory as the effective field theory for low energy
strong interactions is believed to be the appropriate tool to study pion
reactions close to threshold, it seemed at this stage as if there were a severe
problem with the phenomenology.  However, as was shown in section
\ref{generalstructure}, there are observables that are sensitive to the sign
of the $s$--wave $pp\to pp\pi^0$ amplitude---relative to a $p$--wave amplitude
that is believed to be under control---and the experimental results
\cite{triumf1,triumf2,daum1,daum2} agree with the sign as given by the
phenomenological model (c.f. Fig. \ref{pimisigndep}) .

Does this mean that chiral perturbation theory is wrong or not applicable?
No. As we 
will discuss in the subsequent sections, it was demonstrated recently
that the chiral counting scheme needs to be modified in the case of large momentum
transfer reactions. No complete calculation has been carried out up to now, but
intermediate results look promising for a consistent picture to emerge
in the years to come. The insights gained so far from the effective field theory
studies call also for a modification of the phenomenological treatment.
This will be discussed in detail in section \ref{pionlessons}.

\subsection{Phenomenological approaches}
\label{secphen}

As  stressed in the previous section, the number of phenomenological
models for pion production is large. 
For definiteness in this section we will focus on one particular
model, namely that presented in Refs. \cite{unserpol,unserD,unsers},
mainly because it incorporates most of the mechanisms
proposed in the literature for pion production in nucleon--nucleon 
collisions, its ingredients are consistent
with the data on $\pi N$ scattering, and it is the only model so far
whose results have been compared to the polarization data recently
measured at IUCF \cite{meyerpol,polpnpipl,poldpipl}.

\begin{figure}
\begin{center}
\vskip 10cm          
\includegraphics{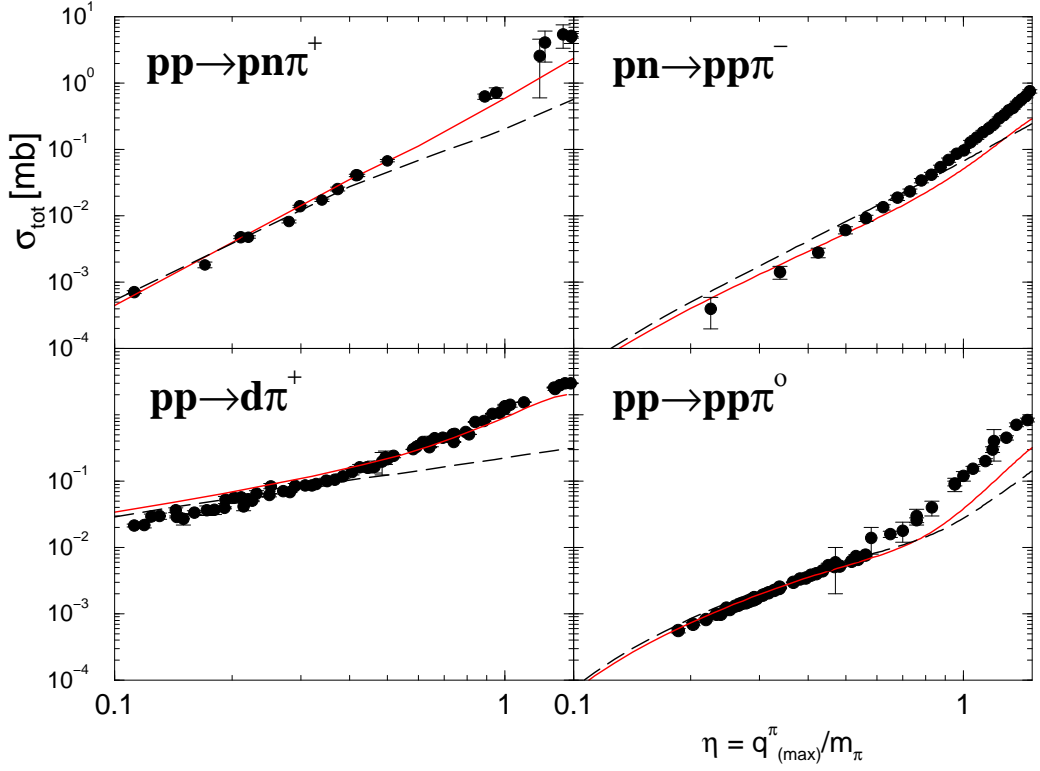} 
\caption{\it Comparison of the model predictions of
Ref. \protect\cite{unserD} to the data. 
The references for the experimental data can be found in table
\protect\ref{measurements}. The solid lines show
the results of the full model; the dashed
line shows the results without the $\Delta$ contributions.}
\label{cserg}
\end{center}
\end{figure}

The model is the first attempt to treat consistently the $NN$ as well as the
$\pi N$ interaction for meson production reactions close to the threshold:
both were taken from microscopic models (described in Refs. \cite{ccf} and
\cite{SHH} for the $NN$ and the $\pi N$ interaction, respectively). These were constructed
from the same effective Lagrangians consistent with the symmetries
of the strong interaction and are solutions of a Lippman--Schwinger equation
based on  time--ordered perturbation theory.
Although not all parameters and approximations used in the two systems are the
same, this model should still be viewed as a benchmark calculation for pion
production in $NN$ collisions within the distorted wave Born approximation.
We will start this section with a description of the various ingredients of
the model and then present some results.

\subsubsection{The $N N$ interaction}
\label{nnmod}

A typical example of a so--called realistic model for $NN$ scattering
is the  Bonn potential \cite{MHE}.
The model used for the $NN$ distortions in the final and initial states is based
on this model, where a pseudo--potential is constructed
based on the $t$--channel exchanges of all established mesons below one GeV in
mass between both nucleons and Delta isobars. 

\begin{figure}[ht!]
\begin{center}
\epsfig{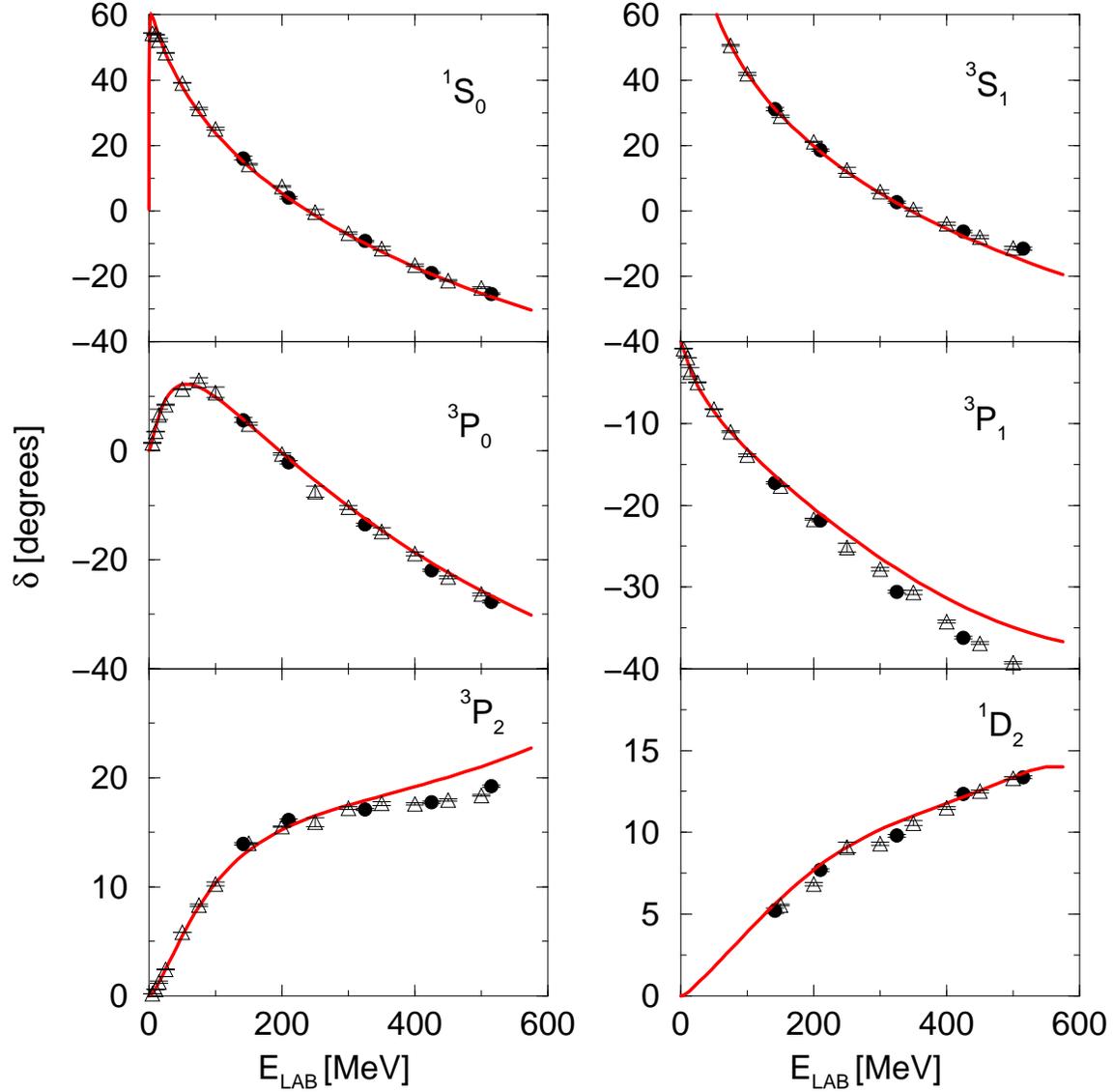}
\caption{\it The $NN$ phase shifts for the model of Ref.  \protect{\cite{ccf}}
in the energy range relevant for pion production. The
pion production threshold is at $E_{LAB}=286$ MeV. The
experimental data are from  Refs. \protect{\cite{bugg}} (triangles) and
\protect{\cite{said}} (circles).}
\label{nnphases}
\end{center}
\end{figure}
The interaction amongst the various dynamical fields in the model derived from
the following Lagrange densities:
\begin{eqnarray}
\label{nnpv}
{\cal L}_{NN\pi} &=& \frac{f_{NN\pi}}{m_\pi}\bar  \psi \gamma_5 \gamma_\mu
  \boldtau \cdot  \partial^\mu \boldpi \psi \\
{\cal L}_{NN\rho} &=& g_{NN\rho}\bar  \psi \gamma_\mu
  \boldtau \cdot \boldrho^\mu \psi + 
\frac{f_{NN\rho}}{4M_N}\bar  \psi \sigma_{\mu \nu} \boldtau \cdot 
(\partial^\mu \boldrho^\nu - \partial^\nu \boldrho^\mu )\psi \\
{\cal L}_{NN\omega} &=& g_{NN\omega}\bar  \psi \gamma_\mu
 \boldomega^\mu \psi \\ 
{\cal L}_{NN\sigma} &=& g_{NN\sigma}\bar \psi  \boldsigma \psi \\ 
{\cal L}_{NNa_0} &=& g_{NNa_0}\bar   \psi   \boldtau \cdot {\boldaz} \psi \\
\label{ndpv}
{\cal L}_{N\Delta \pi} &=& \frac{f_{N\Delta\pi}}{m_\pi}\bar  \psi 
  \vec{T} \cdot  \partial^\mu \boldpi \psi_\mu \ +  \ \mbox{h.c.} \\
{\cal L}_{N\Delta \rho} &=& i\frac{f_{N\Delta \rho}}{m_\pi}\bar  \psi \gamma_5 \gamma_\mu
 \vec{T} \cdot 
(\partial^\mu \boldrho^\nu - \partial^\nu \boldrho^\mu ) \psi_\nu \ +  \ \mbox{h.c.} \ .
\end{eqnarray}

Note: The particle called $\delta$ in the original Bonn publication
\cite{MHE} is nowadays called $a_0$.
The operator $ \vec{T}$ as well as the fields are defined in Ref. \cite{MHE}.
Note that there is no tensor coupling for the $\omega NN$ vertex given for the
fit to the elastic $NN$ scattering data did not need any such coupling.

There is a difference between
the nucleon--nucleon model used \cite{ccf} and the Bonn potential
\cite{MHE}. The original Bonn model
has an energy dependent interaction, for it keeps the full meson retardation
in the intermediate state. This, however, leads to technical problems, once
the model is to be evaluated above the pion production threshold due to the
occurrence of three body singularities. In Ref. \cite{EHSM} those singularities
were handled by solving the dynamical equations in the complex plane. Unfortunately,
this method is not useful for the application in a distorted wave Born approximation.
Instead we used a model based on the so--called folded diagram formalism
developed in Ref. \cite{mikkel}. This formalism, worked out to infinite order,
is fully equivalent to time--ordered perturbation theory. When truncated at
low order, however, it leads to energy independent potentials that can formally be
evaluated even above the pion production threshold. 
In addition, the model of Ref. \cite{ccf} is constructed as a coupled--channel
model including the $NN$ as well as the $N\Delta$ and $\Delta \Delta$
channels. This enables us to treat the $\Delta$ isobar on equal footing
with the nucleons.

The resulting phase
shifts are shown in Fig. \ref{nnphases}. Note that the model parameters were
adjusted to the phase shifts below the pion production threshold only, which is
located at $E_{LAB}=286$ MeV. Fig. \ref{nnphases}
thus clearly illustrates that using this model for the $NN$ interaction is indeed
justified and we may conclude that---at least up to laboratory energies of 600
MeV--the $NN$ phenomenology is well understood.

\subsubsection{The $\pi N$ model}

The $\pi N$ interaction that enters in the pion rescattering diagrams can be
taken from a meson exchange model as well \cite{schuetz}. This allows a
consistent treatment of the meson and nucleon dynamics. It should be stressed
that this is the precondition for comparing the results of the phenomenological
model to those of chiral perturbation theory, as we will do below.
In addition, since we also want to include the rescattering diagram in
partial waves higher than the $s$--wave, after a fit to the $\pi N$
data the pole contributions (nucleon and Delta) need to be removed
from the amplitudes in order to avoid double counting with the direct
production. This is possible only within a microscopic model.

\begin{figure}[t!]
\begin{center}
\epsfig{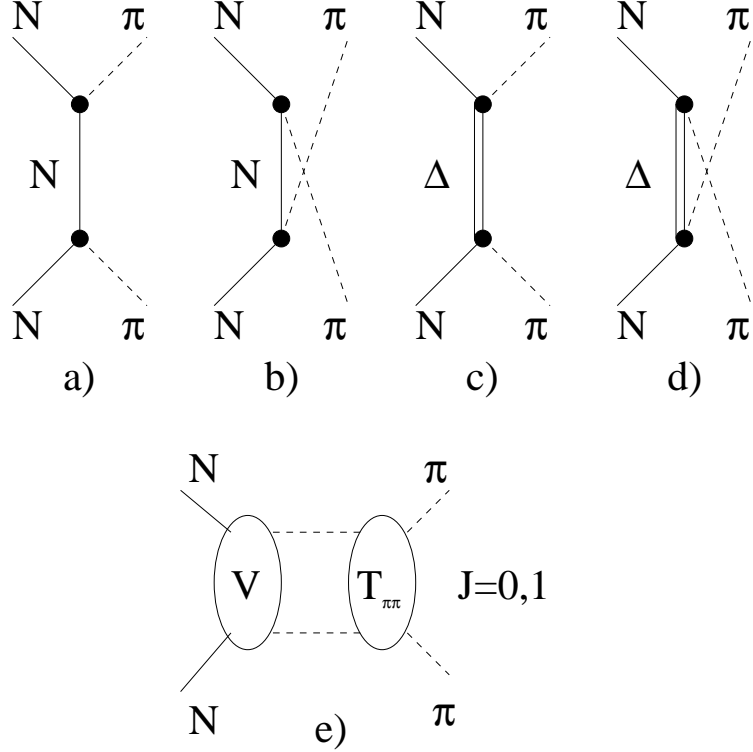}
\caption{\it Contributions to the potential
of the model of Ref. \protect\cite{schuetz}.}
\label{pinpot}
\end{center}
\end{figure}

The main features of the $\pi N$ model of Ref. \cite{schuetz} are that
it is based on an effective Lagrangian consistent with chiral symmetry to
leading order and that the $t$--channel exchanges in the isovector ($\rho$)
and isoscalar ($\sigma$) channel are constructed from dispersion 
integrals\footnote{There are ambiguities in how to extrapolate
the results of the dispersion integrals to off--shell kinematics. This
issue is discussed in detail in Ref. \cite{wdrm}.}.
For details on how the $t$--channel exchanges are included we refer to
Ref. \cite{schuetz}. The diagrams that enter the potential are
displayed in Fig. \ref{pinpot}. This potential is then unitarized
with a relativistic Lippmann--Schwinger equation---in complete analogy
to the nucleon--nucleon interaction.

Within this model, at tree level the isoscalar and isovector $\pi N$ interaction are
give by the corresponding $t$--channel exchanges. In the latter case the
unitarization does not have a big influence in the near--threshold regime, and
thus also the isovector scattering length is governed by the tree level
$\rho$--exchange. The famous KFSR relation \cite{kfsr1,kfsr2},
that relates the couplings of the $\rho$--meson to pions and nucleons to the
coupling strength of the Weinberg--Tomozawa term, is a consequence of this.
On the other hand, for the isoscalar $\pi N$ interaction the effects
of the unitarization are large and lead to an almost complete cancellation of
the isoscalar potential with the iterated $\rho$--exchange in the near
threshold regime. As one moves away from the threshold value this cancellation
gets weaker leading to the strong variation of the off--shell isoscalar $\pi N$
$T$--matrix mentioned at the end of sec. \ref{hist}.

\begin{figure}[t!]
\begin{center}
\epsfig{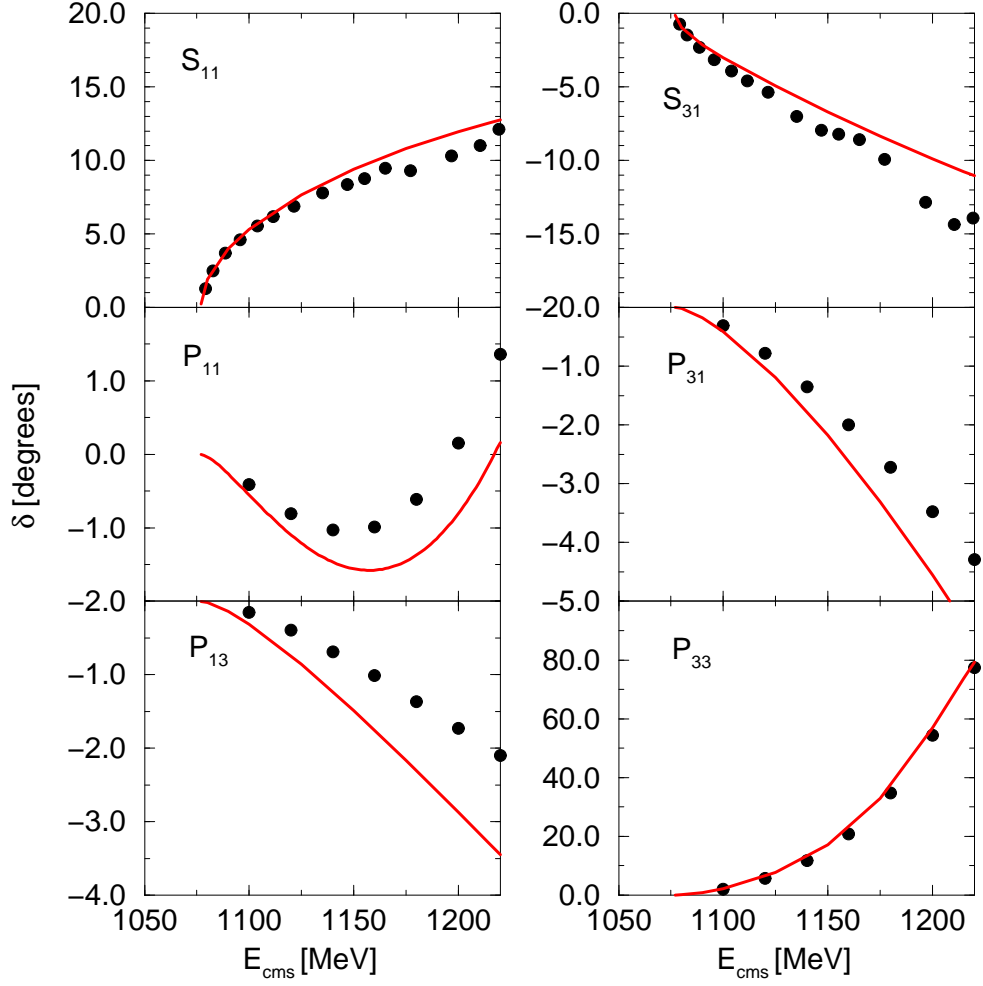}
\caption{\it The $\pi N$ phase shifts for the model of Ref.  \protect{\cite{schuetz}}
  in the energy range relevant for pion production. The experimental data are
  from Refs. \protect{\cite{hoehler,arndt}}. Note the different scales of the
  various panels.}
\label{pinphases}
\end{center}
\end{figure}

In Figure \ref{pinphases} the results for the model of Ref. \cite{schuetz}
are compared to the data of Refs. \cite{hoehler,arndt}. As one can see the
model describes the data well, especially in
the most relevant partial waves: $S_{11}, \ S_{31}$ and $P_{33}$.

\subsubsection{Additional short range contributions and model parameters}
\label{shortrange}

As was stressed above, a large class of additional
mechanisms was suggested in the literature to contribute
significantly to pion production in nucleon--nucleon collisions. 
Since they are all of rather short range and mainly influence the
production of $s$--wave pions, in this work only a single diagram
was included (heavy meson exchange through the $\omega$---Fig. \ref{kur}c) to parameterize
these various effects. Consequently, the strength of this contribution
was adjusted to reproduce the total cross section of the 
reaction $pp\to pp\pi^0$ close to the production threshold.
The short--range contributions turn out to contribute about 20\% to the
amplitude.
After this is done, all parameters of the model are fixed.

\subsubsection{Results}

\begin{figure}
\begin{center}
\vskip 10cm          
\includegraphics{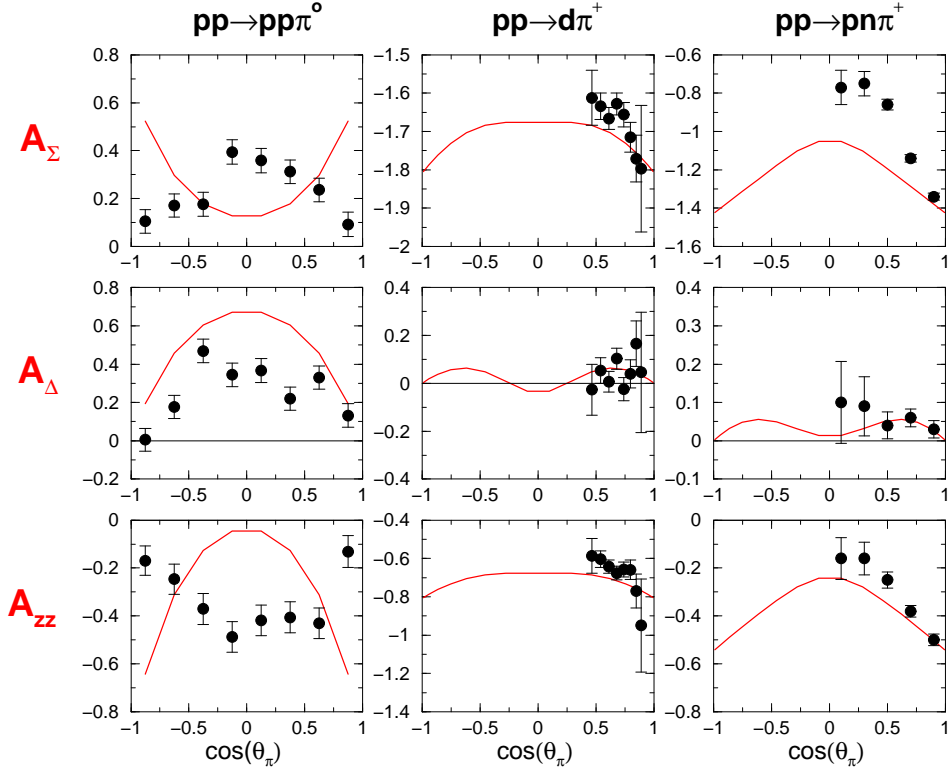} 
\caption{\it Comparison of the model predictions to the data taken
from Ref. \protect{ \cite{meyerpol}} ($pp\to pp\pi^o$),
 Ref. \protect{ \cite{poldpipl}} ($pp\to d\pi^+$) and
 Ref. \protect{ \cite{polpnpipl}} ($pp\to pn\pi^+$).}
\label{erg}
\end{center}
\end{figure}

The results of the model presented have already appeared several times in this
report (Figs.
\ref{thpipol},\ref{thNNpol},\ref{pimisigndep},\ref{spwqs},\ref{cserg},\ref{erg})---mainly
for illustrative purposes---and they are discussed in detail in Refs.
\cite{unserD,unserpol}.

Overall the model is rather successful in describing the data, given that
only one parameter was adjusted to the total cross section for low energy
neutral pion production (see sec. \ref{shortrange}). One important finding is
that the sign of the $s$--wave neutral pion production seems to be in
accord with experiment, as is illustrated in Fig. \ref{pimisigndep} (c.f.
corresponding discussion in sec. \ref{generalstructure}), in contrast to the
early calculations using chiral perturbation theory.  As we will see in sec.
\ref{cpt}, today we know that those early calculations using effective field
theory were incomplete.

The most striking differences appear, however, for double polarization
observables in the neutral pion channel, as shown in Figs. \ref{thpipol} and
\ref{thNNpol}. As a general pattern the amplitudes seem to be of the right
order of magnitude, but show a wrong interference pattern.
To actually allow a detailed comparison of the model and data a partial wave
decomposition of both is necessary. Work in this direction is under way.

It is striking that for charged pion production the pattern is very different,
for here almost all observables are described satisfactorily. In Fig.
\ref{erg} the results of the model for charged and neutral pion production are
compared to the data for a few observables.  In contrast to the neutral
channel, the charged pion production is completely dominated by two
transitions, namely $^3P_1\to ^3S_1s$, which is dominated by the isovector
pion rescattering, and $^1D_2\to ^3S_1p$, which governs the cross section
especially in the regime of the Delta resonance.  The prominence of the Delta
resonance is a consequence of the strong transition $^1D_2(NN)\to
^5S_2(N\Delta)$ that even shows up as a bump in the $NN$ phase shifts (see
Fig. \ref{nnphases}). This effect was first observed long ago and is well
known (see, e.g., discussion in Ref. \cite{ew}). As a consequence, the $NN\to
N\Delta$ transition potential should be rather well constrained by the $NN$
scattering data and is not the case for all the many transitions relevant in
case of the neutral pion production, where the $Sp$ final state is not allowed
due to selection rules (see sec. \ref{sr}). One might therefore hope to learn
more about the $\Delta N$ interaction from the pion production data.

One can also ask how well we know the production
operator. Fortunately, at the pion production threshold it is still possible to
analyze meson production in $NN$ collisions within effective field
theory. This analysis will give deeper insight into the production dynamics,
as will be explained in the following section. The two approaches are then compared
in sec. \ref{pionlessons}.

%
%
%
\subsection{Chiral perturbation theory}
\label{cpt}

The phenomenological approaches, such as the one described in the previous
section, lack a systematic expansion. Thus it is neither possible to
estimate the associated model uncertainties nor to systematically
improve the models. On the other hand, for various meson production
reactions the phenomenological approaches proved to be quite successful.
One might therefore hope that effective field theories will give insights
into why the phenomenology works, as was first stressed
by Weinberg \cite{wein2}.

A first attempt to construct---model independently---the transition amplitude
$NN\to NN\pi$ was carried out almost 40 years ago
\cite{ssy,current2,current3}. The authors tried to relate what is known about
nucleon--nucleon scattering to the production amplitude via the low energy
theorem of Adler and Dothan \cite{aandd}, which is a generalization of the
famous Low theorem for conserved currents \cite{low} to partially conserved
currents via the PCAC relation (see, e.g., Ref. \cite{josebook}).  However,
it soon was realized that the extrapolation from the chiral limit
($m_\pi=0$) to the physical point can change the hierarchy of different
diagrams.  The reason for the non--applicability of soft radiation theorems to
meson production close to the threshold is easy to see: a necessary condition
for the soft radiation theorem to be applicable is that the energy emitted is
significantly smaller than the typical energy scale, that characterizes 
variations of the nuclear wave function. In the near--threshold regime,
however, the scale of variation is set by the inverse of the $NN$ scattering
lengths---thus a soft radiation theorem of the type of Low or that of Adler
and Dothan could only be applicable to meson production if $m_\pi \ll
1/(M_Na^2)$. In reality, however, the pion mass exceeds the energy scale
introduced by inverse $NN$ scattering length by more than two orders of
magnitude.  The range of applicability of soft radiation theorems is discussed
in Ref.  \cite{axion} in a different context.

More recent analyses, however, show that in the case of
pion production it is indeed possible to define a convergent effective
field theory that allows a systematic study of the structure
of the production operator.
As we will show, most of the diagrams included in the Koltun and Reitan 
model \cite{KuR} are indeed the leading operators in pion production.
In addition this study will show
\begin{itemize}
\item that the use of the distorted wave born approximation is justified
\item why neutral pion production is the more problematic case
\item the importance of loop contributions 
\item that there is a close connection between pion production in
nucleon--nucleon collisions and the three nucleon problem.
\end{itemize}

The problem with strong interaction phenomena is their non--perturbative
nature with respect to the coupling constants.
 To construct a controlled expansion it is necessary, to
identify a small expansion parameter. In general this is only
possible for a limited energy range. 
The conditio sine qua non for constructing an effective field theory
for any system is the separation of scales characteristic for the
system. Once the---in this context light---scales are identified, one treats
them dynamically, while all the dynamics that are controlled by the heavy
scales are absorbed in contact interactions. As long as the relevant external
momenta and energies are such that structures of the size of the inverse of
the heavy scale can not be resolved, this procedure should always work.
Weinberg \cite{weinchpt} as well as
Gasser and Leutwyler  \cite{Gasser} have shown that this general
idea works even when loops need to be included. 

In case of low energy pion physics it is the chiral symmetry that provides
both preconditions for the construction of an effective field theory, in
that it forces not only the mass of the pion $m_\pi$, as the Goldstone boson of the chiral
symmetry breaking, to be low, but also the interactions to be weak, for 
the pion needs to be free of interactions in the chiral limit for vanishing
momenta. The corresponding effective field theory is called chiral perturbation
theory ($\chi PT$) and was successfully applied to meson--meson  \cite{pipi}
scattering. Treating baryons as heavy allows straightforward extension of
the scheme to 
 meson--baryon  \cite{ulfbible}
as well as baryon--baryon  \cite{vk94,vk96,evgeni1,evgeni2} systems. 
In all these references the expansion parameter used
was $p/\Lambda_\chi \sim p/(4\pi f_\pi) \sim p/M_N$, where $\Lambda_\chi$ denotes
the chiral symmetry breaking scale, $f_\pi$ the pion decay constant and $M_N$ the
nucleon mass. Recently it was shown that also the Delta isobar can be included
consistently in the effective field theory  \cite{thomas}. The authors treated the
new scale, namely the Delta nucleon mass splitting $\Delta = M_\Delta-M_N$,
to be of the order of $p$ that is taken to be of the order of $m_\pi$.

An additional new scale occurs for meson production in nucleon--nucleon
collisions, namely the initial momentum $p_i \sim \sqrt{m_\pi M_N}$.
Note that, although larger than the pion mass, this momentum is still smaller than
the chiral symmetry breaking scale and thus the expansion should still
converge, but slowly. A priori there are now two options to construct an effective
field theory for pion production. The first option, called Weinberg scheme in
what follows, treats all light scales to be of order of $m_\pi$. Thus, in
this case, there is one expansion parameter, namely $\chi_W=m_\pi/M_N$.
 The other
option is to expand in two scales simultaneously, namely $m_\pi$ and
$p_i$. In this case the expansion parameter is
$$
\chi = \sqrt{\frac{m_\pi}{M_N}} \sim 0.4 \ .
$$
This scheme was advocated in Refs. \cite{chiral2,ourpwaves} and
applied in Ref. \cite{withnorbert}. 
The two additional scales, namely $\Delta$ and
$m_\pi$,  are identified with 
\be
\frac{\Delta}{\Lambda_\chi} \sim \frac{p_i}{\Lambda_\chi}=\chi \qquad \mbox{and} \qquad
\frac{m_\pi}{\Lambda_\chi}
 \sim \frac{p_i^2}{\Lambda_\chi^2}=\chi^2 \ ,
\label{count}
\ee
where the former assignment was made due to the numerical similarity of the
two numbers\footnote{Note that $\Delta$ stays finite in the chiral limit,
  whereas both $p_i$ as well as $m_\pi$ vanish.} ($\Delta = 2.1 m_\pi $ and $p_i = 2.6 m_\pi$).
Only explicit calculations can reveal which one is the more appropriate
approach. 

Within the Weinberg counting scheme, tree level calculations were performed
for $s$--wave pion production in the reactions $pp\to pp\pi^0$
\cite{chiral1,chiral2,chiralimp} as well as $pp\to pn\pi^+$
\cite{mitulf,derocha}. In addition, complete calculations to
next--to--next--to--leading order (NNLO), where in the Weinberg scheme for the
first time loops appear, are available for $pp\to pp\pi^0$ \cite{dmitra,ando}.
The authors found that some of the NNLO contributions exceeded significantly
the next--to--leading (NLO) terms leading them to the conclusion that the chiral expansion
converges only slowly, if at all.  This point was further stressed in Refs.
\cite{moalem,ulfnovel},  however, it was shown recently that as soon as the
scale induced by the initial momentum is taken into account properly
(expansion in $\chi$ and not in $\chi_W$), the series indeed converges
\cite{ourpwaves,withnorbert}. For illustration, in this section we
compare the order assignment of the Weinberg scheme to that of the modified
scheme.

In appendix
\ref{chped} the relevant counting rules of the new scheme are presented
as well as justified via
application to a representative example. Especially, it is not clear a priori
what scale to assign to the zeroth component $l_0$ of the four dimensional
integration volume $d^4l$ as it occurs in covariant loops. After all, the typical energy of the
system is given by $m_\pi$, but the momentum by $p_i$. As is shown in the
appendix by matching a covariant analysis to one carried out in
time--ordered perturbation theory, in loops $l_0 \sim p_i$. This assignment
was also confirmed in explicit calculations \cite{withnorbert}.

The starting point is an appropriate Lagrangian density, constructed to be consistent
with the symmetries of the underlying more fundamental theory (in this case
QCD) and ordered according to a particular counting scheme. 
Omitting terms that do not contribute to the order we will be considering here,
we therefore have for the leading order Lagrangian 
 \cite{vk943b,vk96,ulfbible}
\begin{eqnarray}
 {\cal L}^{(0)} & = & 
          \frac{1}{2}\partial_\mu{\boldpi}\partial^\mu{\boldpi}
          -\frac{1}{2}m_{\pi}^{2}\boldpi^{2}
          +\frac{1}{2f_\pi^2}\left[
(\boldpi \cdot \partial_\mu \boldpi)^2-\frac14m_\pi^ 2\left(\boldpi^2\right)^ 2 \right]
\nonumber   \\ \nonumber
    &   & +N^{\dagger}[i\partial_{0}-\frac{1}{4 f_{\pi}^{2}} \boldtau \cdot
         (\boldpi\times\dot{\boldpi})]N \\ 
    &   & +\frac{g_{A}}{2 f_{\pi}} 
         N^{\dagger}\boldtau\cdot\vec{\sigma}\cdot\left(\vec{\nabla}\boldpi
+\frac{1}{2f_\pi^2}\boldpi(\boldpi \cdot \vec \nabla \boldpi)
\right)N
                                               \nonumber \\
    &   & +\Psi_\Delta^{\dagger}[i\partial_{0}- \Delta]\Psi_\Delta 
          +\frac{h_{A}}{2 f_{\pi}}[N^{\dagger}(\boldT\cdot
          \vec{S}\cdot\vec{\nabla}\boldpi)\Psi_\Delta +h.c.] +\cdots \ .
\label{la0}
\end{eqnarray} 
The expressions for interactions with more than two pions depend on the
interpolating field used. The choice made here was the so called sigma
gauge---c.f. Appendix A of Ref. \cite{ulfbible}, where also the corresponding
vertex functions are given explicitly\footnote{As usual, all observables are
independent of the choice of the pion field.}.

For the next--to--leading order Lagrangian we get
\begin{eqnarray}
 {\cal L}^{(1)}& 
        =&\frac{1}{2m_{N}}[N^{\dagger}\vec{\nabla}^{2}N
+\Psi_\Delta^{\dagger}\vec{\nabla}^{2}\Psi_\Delta]                            
                                    \nonumber \\
  &   & 
        +\frac{1}{8M_N f_{\pi}^{2}}(iN^{\dagger}\boldtau\cdot
        (\boldpi\times\vec{\nabla}\boldpi)\cdot\vec{\nabla}N + h.c.)
                                       \nonumber \\ \nonumber
  &   & +\frac{1}{f_{\pi}^{2}}N^{\dagger}[(c_2 +c_3 - \frac{g_A ^2}{8 m_{N}})
        \dot{\boldpi}^{2} -c_3 (\vec{\nabla}\boldpi)^{2} 
        -2c_1 m_{\pi}^{2} \boldpi^{2} \\ \nonumber
  &   & -
        \frac{1}{2} (c_4 + \frac{1}{4m_{N}}) 
        \varepsilon_{ijk} \varepsilon_{abc} \sigma_{k} \tau_{c} 
        \partial_{i}\pi_{a}\partial_{j}\pi_{b}]N \\
  &   & -\frac{g_{A}}{4 m_{N} f_{\pi}}[iN^{\dagger}\boldtau\cdot\dot{\boldpi}
        \vec{\sigma}\cdot\vec{\nabla}N + h.c.]             
                                                               \nonumber \\
  &   &         -\frac{h_{A}}{
        2 m_{N} f_{\pi}}[
        iN^{\dagger}\boldT\cdot\dot{\boldpi}\vec{S}\cdot\vec{\nabla}
        \Psi_\Delta + h.c.]                       \nonumber \\
  &   & -\frac{d_1}{f_{\pi}} 
        N^{\dagger}(\boldtau\cdot\vec{\sigma}\cdot\vec{\nabla}\boldpi)N\,
        N^{\dagger}N   \nonumber \\
  &   &         -\frac{d_2}{2 f_{\pi}} \varepsilon_{ijk} \varepsilon_{abc} 
        \partial_{i}\pi_{a}  
        N^{\dagger}\sigma_{j}\tau_{b}N\, N^{\dagger}\sigma_{k}\tau_{c}N 
        +\cdots \ ,            \label{la1}
\end{eqnarray}
\noindent

where $f_\pi$ denotes the pion decay constant in the chiral limit, 
$g_A$ is the 
axial--vector coupling of the nucleon, $h_A$ is the $\Delta N \pi$ coupling, and
$\vec S$ and $\boldT$ are the transition spin and isospin
matrices, normalized such that
\begin{eqnarray}
S_iS_j^\dagger &=& 
\frac{1}{3}(2\delta_{ij}-i\epsilon_{ijk}\sigma_k) \ , \\
T_iT_j^\dagger &=&
 \frac{1}{3}(2\delta_{ij}-i\epsilon_{ijk}\tau_k) \ .
\end{eqnarray}
These definitions are in line with the ones introduced in sec. \ref{nnmod}.
The dots symbolize that what is shown are only those terms that are relevant
for the calculations presented.  As demanded by the heavy baryon formalism, the
baryon fields $N$ and $\Psi_\Delta$ are the velocity--projected pieces of the
relativistic fields appearing in the interactions discussed in section
\ref{nnmod}; e.g. $N=1/2(1 + \barr v )\psi$, where $v^\mu$ denotes the nucleon
4--velocity.

The terms in the Lagrangians given are ordered according to the conventional
counting ($p\simeq m_\pi$). A reordering on the basis of the new scheme does
not seem appropriate, for what order is to be assigned to the energies and
momenta occurring depends on the topology of a particular diagram
(see also Appendix \ref{chped}).

\begin{figure}
\begin{center}
\vskip 2cm          
\includegraphics{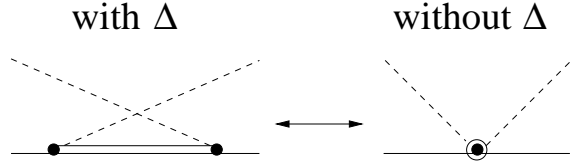} 
\caption{\it Illustration of resonance saturation.}
\label{ressat}
\end{center}
\end{figure}

\renewcommand{\arraystretch}{1.4}
\begin{table}[t!]
\begin{center}
\caption{\it The various low energy constants $c_i$ in units of GeV$^{-1}$. The
first two columns give the values from the original references (left column:
tree level calculation of Ref. \cite{ulf1}; right column: one loop calculation
of Ref. \cite{ulf2}), whereas
the last two columns give those values reduced by the Delta contribution
as described in the text. It is those numbers that where used in the
calculations presented here.}
\begin{tabular}{|c|c c |c c| }
\hline
$i$  & $c_i^{tree}$  & $c_i^{loop}$ & 
$c_i^{tree}(\Delta \!\!\!\! / )$ 
   & $c_i^{loop}(\Delta \!\!\!\! / )$ \\ 
\hline
1 & -0.64 & -0.93 & -0.64 & -0.93 \\
2 &  1.78 &  3.34 &  0.92 &  0.64 \\
3 & -3.90 & -5.29 & -1.20 & -2.59 \\
4 &  2.25 &  3.63 &  0.9 & 2.28 \\
\hline
\end{tabular}
\label{citable}
\end{center}
\end{table}

The constants $c_i$ can be extracted from a fit to elastic
$\pi N$ scattering. This was done in a series of papers with successively
improved methods
 \cite{ulf1,ulf2,piN3,piN4,piN5,piN6}. However, here we will focus on
the values extracted in Refs. \cite{ulf1,ulf2} for it is those that were used
in the calculations for pion production in nucleon--nucleon collisions. In the
former work the $c_i$ were extracted at
tree level and in the latter to one loop.
The corresponding values are given in table \ref{citable}.
In both papers the Delta isobar was not considered as explicit degree of freedom.
In an effective field theory the low energy constants appearing
depend on the dynamical content of the theory.
Thus, in a theory without explicit Deltas, their effect is
absorbed in the low energy constants \cite{ulf2}.  
This is illustrated graphically in Fig. \ref{ressat}.
Thus we need to subtract the Delta contribution from
the values given in the first columns of table \ref{citable}.
Analytical results for those contributions are given in Refs.
\cite{vanKolckcired,ulf2}\footnote{To match the
results of the two references the large $N_C$ value has to be used
for $h_A = 3g_A/\sqrt{2}\simeq 2.7$.}:
\begin{equation}
c_2^{\Delta} \ = \ -c_3^{\Delta} = 2c_4^{\Delta} = \frac{h_A^2}{9(M_\Delta-M_N)}
\ = \ 2.7 \  \mbox{GeV}^{-1} \ .
\end{equation}
Thus, the only undetermined parameters in the interaction Lagrangian
are $d_1$ and $d_2$. Since they are the strength parameters of
4--nucleon contact interactions that do not contain any derivatives
of the nucleon fields,
they only contribute to those amplitudes that have an $NN$ $S$--wave
to $NN$ $S$--wave plus pion $p$--wave transition. This automatically excludes
a transition from an isospin triplet to an isospin triplet state, for this
would demand to go from $^1S_0$ to $^1S_0$ accompanied by a $p$--wave pion,
which is forbidden by conservation of total angular momentum. Thus, only two
transitions are possible: $T=0\to T=1$, as it can be studied in $pn\to pp\pi^-$,
and $T=1\to T=0$, as it can be studied in $pp\to pn\pi^+$.
Even more
importantly, in both channels the $d_i$ appear with the 
same linear combination $d$ which is completely fixed by the corresponding isospin factors:
\begin{equation}
d = \frac{1}{3}(d_1+2d_2) \sim \frac{\delta}{f_\pi^2M_N} \ ,
\label{ddef}
\end{equation}
where we have introduced the dimensionless parameter $\delta$ . In order
for the counting scheme to work, $\delta = {\cal O}(1)$ has to hold.
As we will show below, this order of magnitude is indeed consistent with the data.
Please note, the four--nucleon--pion contact term with $p$--wave discussed here was to some
extend also investigated in Ref. \cite{lutz}, however, within a different
expansion scheme. Thus direct comparison is not possible.

\begin{figure}[t!]
\begin{center}
\epsfig{file=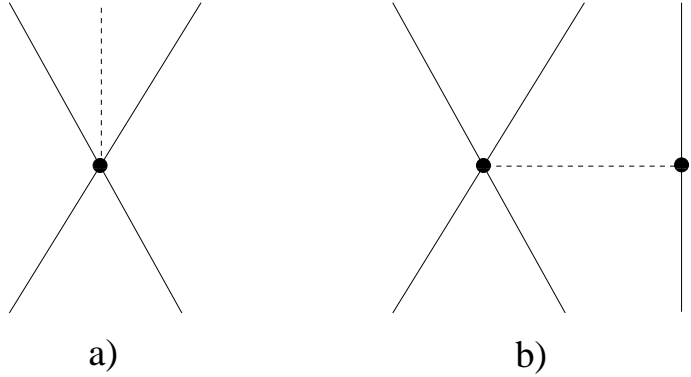, height=5cm}
\caption{\it Illustration of the 
role of the $4N \pi$ contact term in $NN\to NN\pi$ and
three nucleon scattering. Solid lines denote nucleons, dashed lines
denote pions.}
\label{dill}
\end{center}
\end{figure}
 
The parameter $d$ is very interesting, for it is at the same time the leading
short-range--long-range contribution\footnote{In this context, pion exchanges
  are called long--ranged, whereas any exchange of heavier mesons---absorbed
  in the contact terms---are called short--ranged.} to the three nucleon
force, as illustrated in Fig. \ref{dill}. Naturally, it can also be fixed from
$pd$ scattering data directly. This was done in Ref. \cite{evgeni3}.  We come
back to this point below.

The effect of the large scale on the vertices is best illustrated with an
example. In ${\cal L}^{(0)}$ the so--called Weinberg--Tomozawa (WT)
term ${1}/(4 f_{\pi}^{2})N^{\dagger} \boldtau
\cdot(\boldpi\times\dot{\boldpi})N$ appears. The corresponding recoil correction
$1/(8M_N f_{\pi}^{2})(iN^{\dagger}\boldtau\cdot
(\boldpi\times\vec{\nabla}\boldpi)\cdot\vec{\nabla}N + h.c.)$ appears in
${\cal L}^{(1)}$. Therefore the vertex function derived from the WT term is
proportional to $(q_0+k_0)/f_{\pi}^{2}$, where $q_0$ ($k_0$) denote the zero--th
component of the 4--vectors for the outgoing (incoming) pion. If the WT term
appears as the $\pi N$ vertex in the rescattering contribution (c.f. Fig.
\ref{kur}b), in threshold kinematics $k_0=m_\pi/2$ and $q_0=m_\pi$. Also in
threshold kinematics, where the 3--momentum transfer equals the initial
momentum, we find for the contribution from the recoil term
$p^{2}/(2M_Nf_{\pi}^{2})$.  Obviously, since $p^{2} = M_Nm_\pi$, the
contribution from the WT term and from its recoil term are of the same order.
This changes if the WT term appears inside a loop, for then the scale
for $k_0$ is also set by $p$---in this case the recoil term is suppressed by one
chiral order compared to the WT term itself. The assignments made are
confirmed by explicit calculation \cite{withnorbert} as well as by a toy model
study \cite{toy1}.

\renewcommand{\arraystretch}{1.4}
\begin{table}[t!]
\caption{\it Comparison of the corresponding chiral order in the
Weinberg scheme ($p \sim m_\pi$) and the new counting scheme ($p \sim \sqrt{m_\pi M_N}$)
for several nucleonic contributions for $p$--wave pion production.
Subleading vertices are marked as $\odot$.
Here $q$ denotes the external pion momentum. For simplicity we
assume the outgoing nucleons at rest.}
\begin{center}
\begin{tabular}{c c | c | c  }
$p$--wave diagrams & scale & $p \sim m_\pi $  & $p\sim$ \\
(nucleons only)& &\phantom{$\sqrt{m_\pi M_N}$} & $\sqrt{m_\pi M_N}$ \\
\hhline{~---}
\parbox[c]{7.3cm}{\epsfig{file=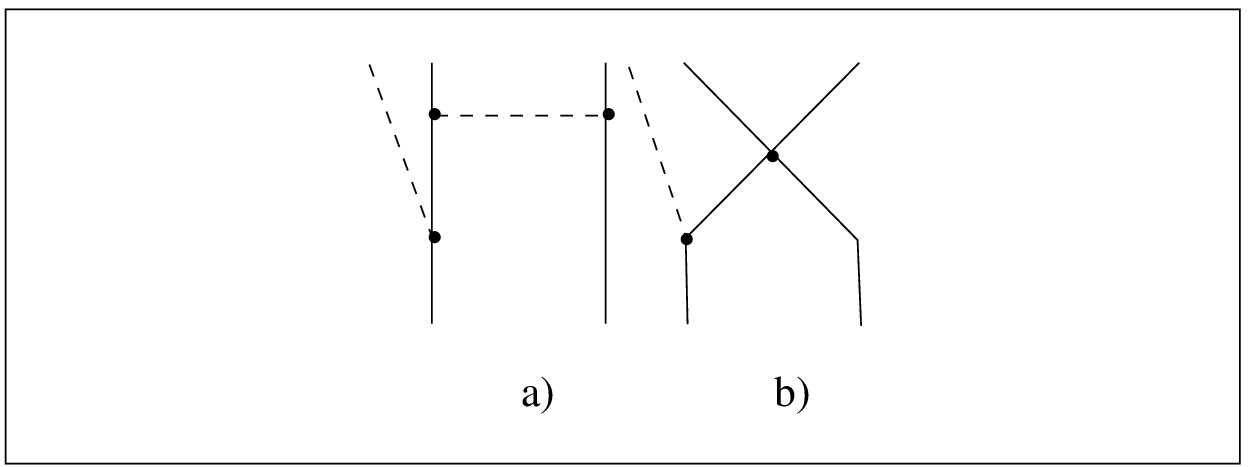, height=2.7cm}} & ${q}/{m_\pi}$ & LO & LO \\
\hhline{~---}
\parbox[c]{7.3cm}{\epsfig{file=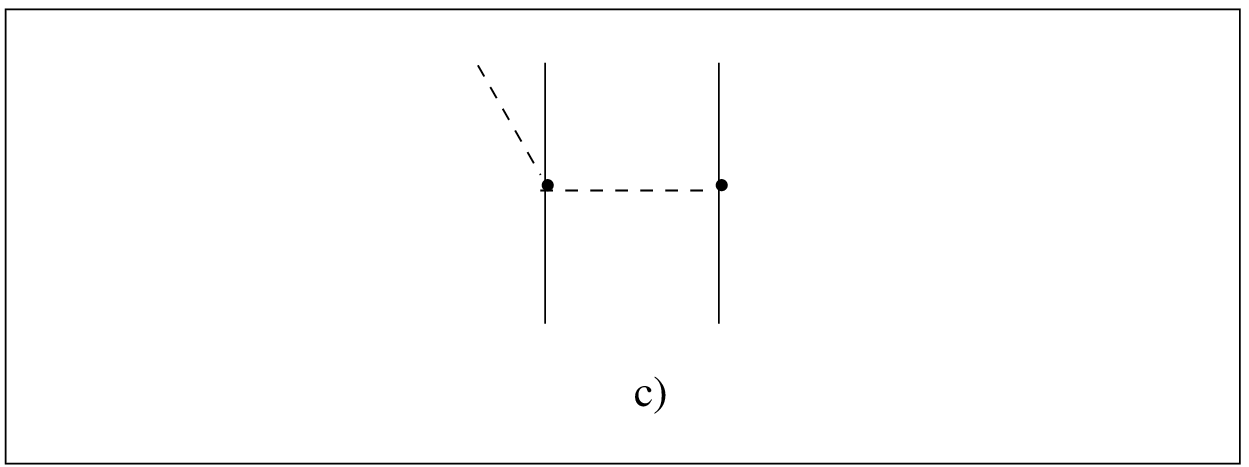, height=2.7cm}} & $qm_\pi/p^2$ & LO & N$^2$LO \\
\hhline{~---}
\parbox[c]{7.3cm}{\epsfig{file=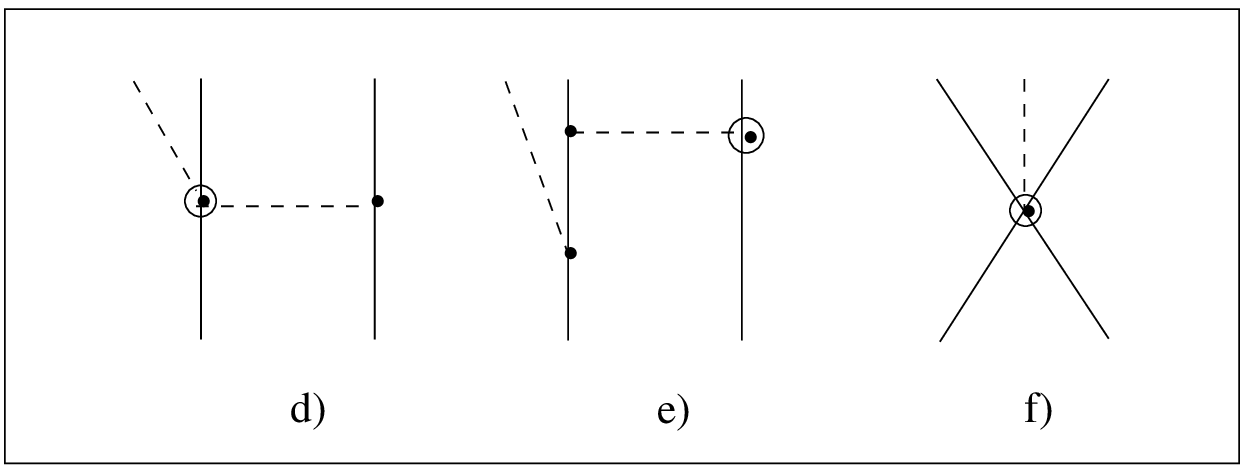, height=2.7cm}} & ${q}/{M_N}$ & NLO & N$^2$LO \\
\hhline{~---}
\parbox[c]{7.3cm}{\epsfig{file=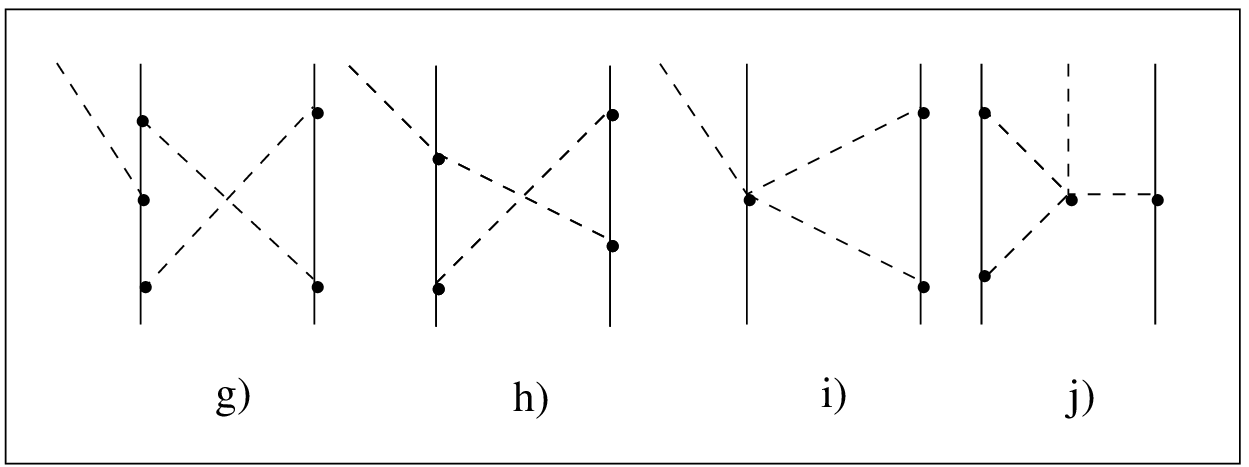, height=2.7cm}} & ${pq}/{M_N^2}$ & N$^2$LO & N$^3$LO  \\
\hhline{~---}
\end{tabular}
\end{center}
\label{diatabp}
\end{table}

Since we know now the interactions of pions and nucleons we
can investigate the relevance of subleading loops, where we
mean loops that can be constructed from a low order
diagram by inserting an additional pion line. Obviously
this procedure introduces at least one $\pi NN$ vertex $\sim p/f_\pi$,
a pion propagator $\sim 1/p^2$, an integral measure $p^4/(4\pi)^2$
and either an additional $\pi NN$ vertex together with
two nucleon propagators, or one additional nucleon propagator
together with a factor $1/f_\pi$.
To get the leading piece of the loops in the latter case,
the integration over the energy variable has to pick the
large scale and thus we are to count the nucleon propagator
as $1/p$, leading to an overall suppression factor for
this additional loop of $p^2/M_N^2$. In the former case,
however, a topology is possible that contains a two--nucleon
cut. This unitarity cut leads to an enhancement of that
intermediate state, for it pulls a large scale ($M_N$) into the
numerator. 
Based on this observation Weinberg established 
a counting scheme for  $NN$ scattering that strongly
differs from that in $\pi \pi$ as well as $\pi N$ scattering \cite{wein2}.
The corresponding factor introduced by this loop is $p/M_N$
(and in addition typically comes with a factor of $\pi$).
As was stressed by Weinberg, this suppression is compensated by
the size of the corresponding low energy constants of
the two--nucleon interaction and therefore all diagrams that
can be cut by crossing a two nucleon line only are called reducible
and the initial as well as final state $NN$ interaction is summed
to all orders. This is what is called distorted wave Born
approximation. There is one exception  to this rule: namely
when looking at the two--nucleon intermediate state close to the pion
production vertex in diagram a of Fig. \ref{kur}. Either the incoming or the outgoing nucleon
needs to be off--shell and thus this intermediate state does not 
allow for a two nucleon unitarity cut. Therefore those two--nucleon
intermediate states are classified as irreducible.

There is a special class of pion contributions not yet discussed, namely those
that contain radiative pions (on--shell pions in intermediate
states). We restrict ourselves to a kinematic regime close to the production
threshold. As soon as an intermediate pion goes on--shell, the
typical momentum in the corresponding  loop automatically needs to be of
order of the outgoing momenta. This leads to an effective suppression of
radiative pions; e.g., for $s$--wave pions the effects of
pion retardation become relevant at N$^5$LO.

Note that each loop necessarily contributes at many orders simultaneously. The
reason for this is that the two scales inherent to the pion production
problem can be combined to a dimensionless number smaller than one:
$m_\pi/p_i=\chi$ \ .  However, what can always be done on very general grounds is to
assign the minimal order at which a diagram can start to contribute and this is
sufficient for an efficient use of the effective field theory.

The next step is  the consistent inclusion of the nuclear wave functions.
However, the $NN$ potentials constructed consistently
with chiral perturbation theory \cite{vk94,vk96,evgeni1,evgeni2}
are not applicable at the pion production threshold.
Therefore we use the so--called hybrid approach originally introduced by
Weinberg \cite{hybrid},
where we convolute the production operator, constructed within
chiral perturbation theory, with a phenomenological $NN-N\Delta$ 
wavefunction. We use the CCF model described in the
previous section \cite{ccf}. 

Let us start with a closer look at the production of 
$p$--wave pions, for those turned out 
easier to handle than $s$--wave pions. The reason for
this pattern lies in the nature of pions as Goldstone
bosons of the chiral symmetry: since in the chiral limit
for vanishing momenta the interaction of pions with matter
has to vanish, the coupling of pions naturally
occurs in the company of either a derivative or an even power
of the pion mass\footnote{Only even powers of the pion
mass are allowed to occur in the interaction, since, due to
the Gell-Mann--Oakes--Renner relation \cite{gmor}, $m_\pi^2 \propto m_q$,
where $m_q$ is the current quark mass and in the interaction
no terms non--analytic in the quark masses are allowed.}.
As a consequence, the leading piece of the $\pi NN$ vertex
is of $p$--wave type, whereas the corresponding $s$--wave
piece is suppressed as $\omega_\pi/M_N$, where $\omega_\pi$
denotes the pion energy. Note that also in
the case of neutral pion photoproduction the
$s$--wave amplitude is dominated by pion loops, whereas
the $p$--wave amplitude is dominated by tree level diagrams \cite{pi0photo,ulfbible}.

\begin{figure}
\begin{center}
\vskip 8cm          
\includegraphics{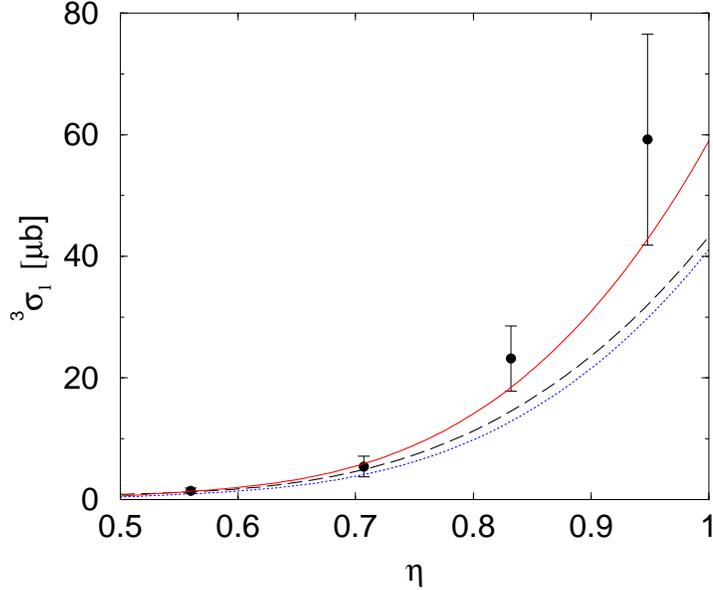} 
\caption{\it Comparison of the predictions from
effective field theory at LO (dashed line) and NLO
(with the $\pi N$ parameters from a NLO (dotted line) and an
NNLO (solid line) analysis) with the $^3\sigma_1$ cross
section for $pp\to pp\pi^0$ \protect{ \cite{meyerpol}}.}
\label{3p1}
\end{center}
\end{figure}

The corresponding diagrams are shown in table \ref{diatabp} up to N$^3$LO.
So far in the literature calculations have been carried out only up to N$^2$LO 
\cite{ourpwaves}. An important test of the approach is to show its
convergence. For that we need an observable to which $s$--wave pions do
not contribute. Such an observable is given by the spin cross section
$^3\sigma_1$ recently measured 
at IUCF for the reaction $pp\to pp\pi^0$ \cite{meyerpol} (c.f. table
\ref{pws} in section \ref{polobs}).
The parameter--free prediction of chiral perturbation theory compared to the data is shown
in Fig. \ref{3p1}. As one can see, the total amplitude is clearly dominated by
the
leading order suggesting a convenient rate of convergence for the series. In
addition, the prediction agrees with the data.

\begin{figure}[tb]
\vspace{6.2cm}
\includegraphics{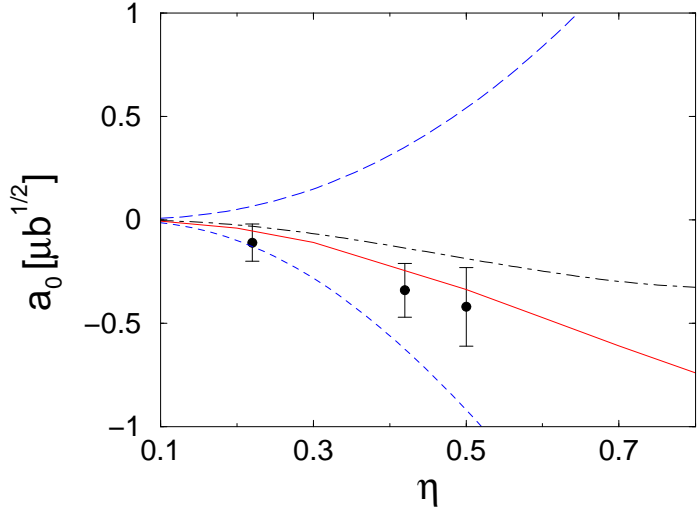}
\caption{\it  $a_0$
of  $pp \to np\pi^+$ in chiral perturbation theory.  
The different lines correspond to 
values of the parameter related to the three-nucleon force:
$\delta=1$ (long dashed line).
$\delta=0$  (dot-dashed line),
$\delta=-0.2$ (solid line), and
$\delta=-1$ (short dashed line).
Data are  from Ref. \protect\cite{flammang}. }
\label{a0pn} 
\end{figure}

Thus we are now prepared to extract the parameter $d$ from 
data on the reaction $pp\to pn\pi^+$ \cite{flammang}.
As it was argued above, only the amplitude corresponding to the transition
$^1S_0\to ^3S_1p$, called $a_0$, is influenced
by the corresponding contact interactions. The results
of the chiral perturbation theory calculations are shown
in Fig. \ref{a0pn}. The figure shows four curves
for different values of the parameter $\delta$ defined in Eq. (\ref{ddef}), namely
the result for $\delta = 0$ (dot--dashed line), for $\delta = -0.2$---the
authors of
Ref. \cite{huebner} claim this value to yield an important contribution to $A_y$
in $Nd$ scattering at energies of a few MeV\footnote{The calculation of
  Ref. \cite{huebner} suffers from numerical problems.}---as well as the results
we get when $\delta$ is varied within its natural range $\delta = +1$ and
$\delta = -1$ shown as the long--dashed and the short--dashed curve, respectively.
Thus we find that the results for $a_0$ are indeed rather sensitive to
the strength of the contact interaction. This might be surprising at first glance,
since we are talking about a subleading operator, however it turned out
that the leading (diagrams a) and b) in table \ref{diatabp})
 and subleading contributions  (the same diagrams with a $\Delta$ intermediate state)
 largely
cancel. Thus, to draw solid conclusions the $p$--wave calculations should be
improved by one chiral order, the corresponding diagrams of which are shown in 
table \ref{diatabp}g)-j).

Please note that the rescattering contribution involving $c_4$ that occurs at
the same order as the $d$ contribution turned out to be sensitive to the
regulator used for the evaluation of the convolution integral with the nuclear
wavefunction. This cutoff dependence can be absorbed in $d$, which in turn is
now cutoff--dependent. Therefore, in order to compare the results for $d$ from
the pion production reaction to those extracted from the three--nucleon system
\cite{evgeni3} a consistent calculation that is not
possible at present has to be performed. Note that the $d$ parameter as fixed in Ref.
\cite{evgeni3} (there it is called $E$) also turned out to be sensitive to the
regulator.  On the long run, however, a consistent description of pion
production and three--nucleon scattering should be a rather stringent test of
chiral effective field theories at low and intermediate energies. As should be
clear from the discussion above, to yield values for the low energy constant
$\delta$ that are compatible both calculations $pd$ scattering as well as
$\pi$--production have to be performed using the same dynamical fields---at
present the $\Delta$--isobar is not considered as explicit degree of freedom
in Ref. \cite{evgeni3}, but plays a numerically important role in the
extraction of $\delta$ from $NN\to NN\pi$.

In section \ref{generalstructure} it was shown that the differential
cross section as well as the analysing power for the reaction
$pn\to pp\pi^-$ is sensitive to an interference term of
the $s$--wave pion production amplitude of the $A_{11}$ amplitude ($^3P_0\to ^1S_0s$)
and the $p$ amplitudes of $A_{01}$: $^3S_1\to ^1S_0p$ and $^3D_1\to ^1S_0p$.
Obviously, the 4--nucleon contact interaction contributes to the former.
Thus, once a proper chiral perturbation theory calculation is available
for the $s$--wave pion production---whose status will be discussed
in the subsequent paragraphs---the reaction $pn\to pp\pi^-$ close
to the production threshold might well be the most sensitive reaction
to extract the parameter $d$.

\renewcommand{\arraystretch}{1.4}
\begin{table}[ht!]
\caption{\it Comparison of the corresponding chiral order in the
Weinberg scheme ($p \sim m_\pi$) and the new counting scheme ($p \sim \sqrt{m_\pi M_N}$)
for several nucleonic contributions for $s$--wave pion production.
Subleading vertices are marked as $\odot$.
 Not shown explicitly are the
recoil corrections for low--order diagrams. For example, recoil corrections
to diagram b) appear at order $pm_\pi / M_N^2$.}
\begin{center}
\begin{tabular}{c c | c | c  }
$s$--wave diagrams & scale & $p \sim m_\pi $  & $p\sim$ \\
(nucleons only)& &\phantom{$\sqrt{m_\pi M_N}$} & $\sqrt{m_\pi M_N}$ \\
\hhline{~---}
\parbox[c]{7.3cm}{\epsfig{file=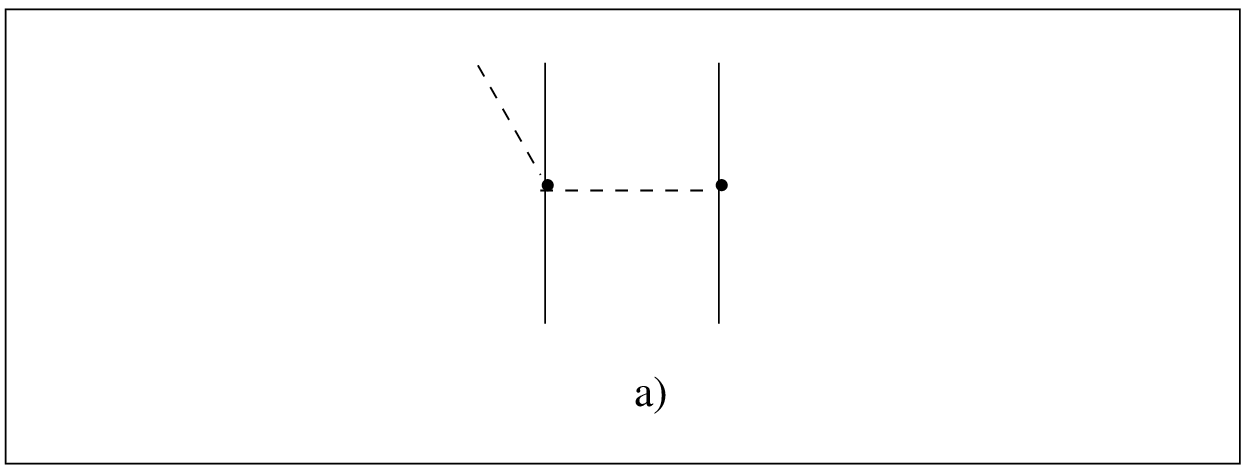, height=2.7cm}} & ${m_\pi}/{p}$ & LO & LO \\
\hhline{~---}
\parbox[c]{7.3cm}{\epsfig{file=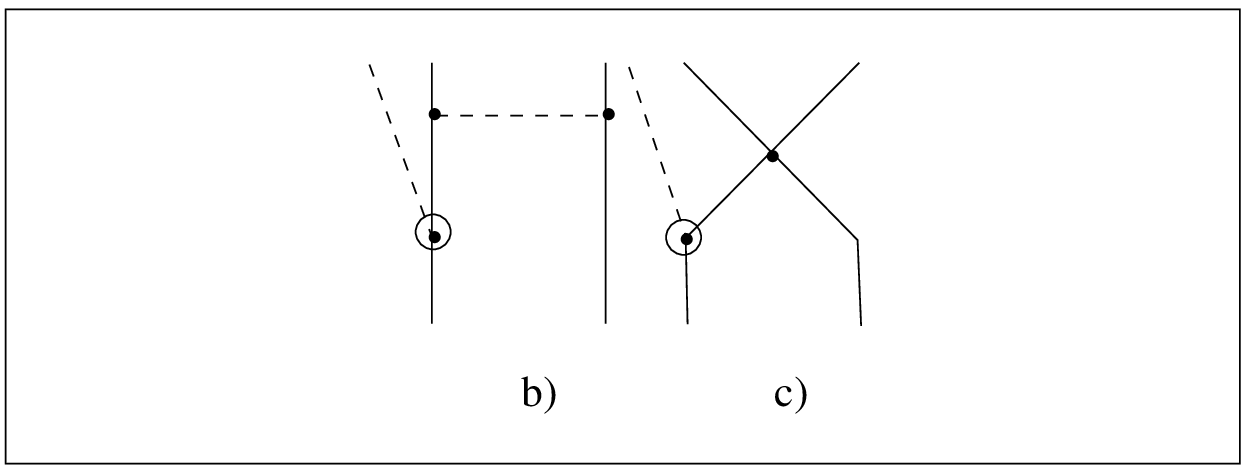, height=2.7cm}} & $p/M$ & NLO & LO \\
\hhline{~---}
\parbox[c]{7.3cm}{\epsfig{file=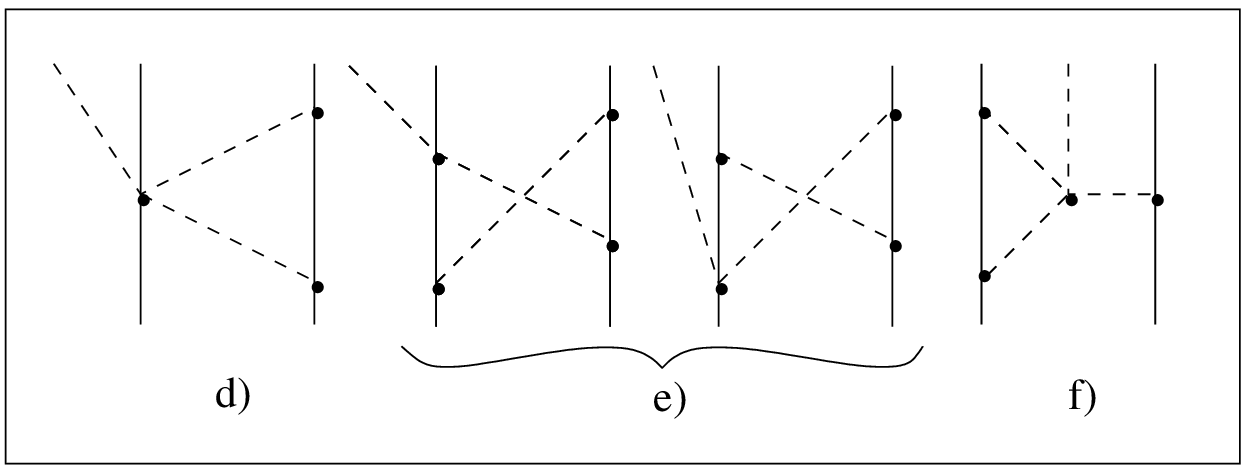, height=2.7cm}} & ${p^2}/{M_N^2}$ & N$^2$LO & NLO \\
\hhline{~---}
\parbox[c]{7.3cm}{\epsfig{file=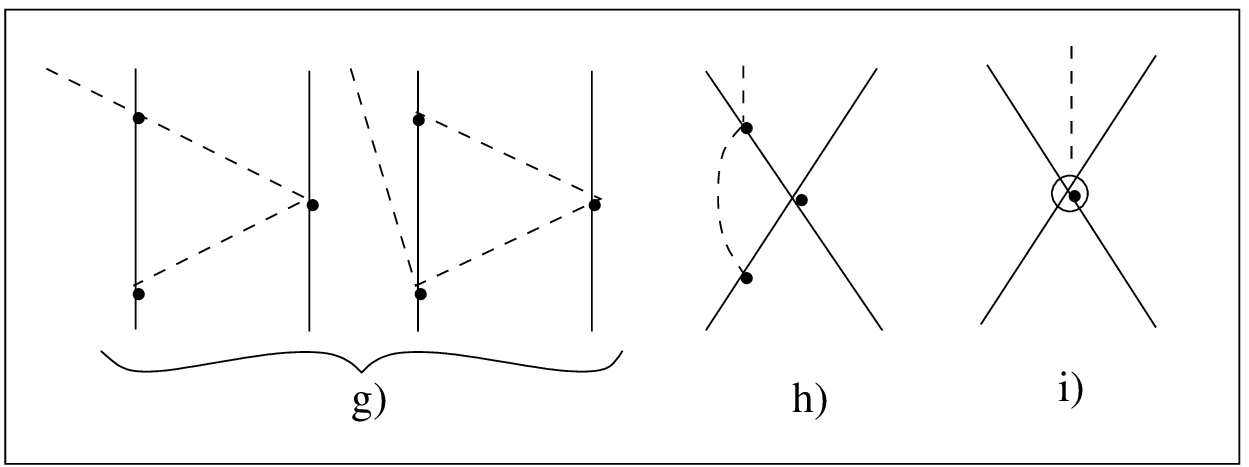, height=2.7cm}} & ${pm_\pi}/{M_N^2}$ & N$^2$LO & N$^2$LO  \\
\hhline{~---}
\parbox[c]{7.3cm}{\epsfig{file=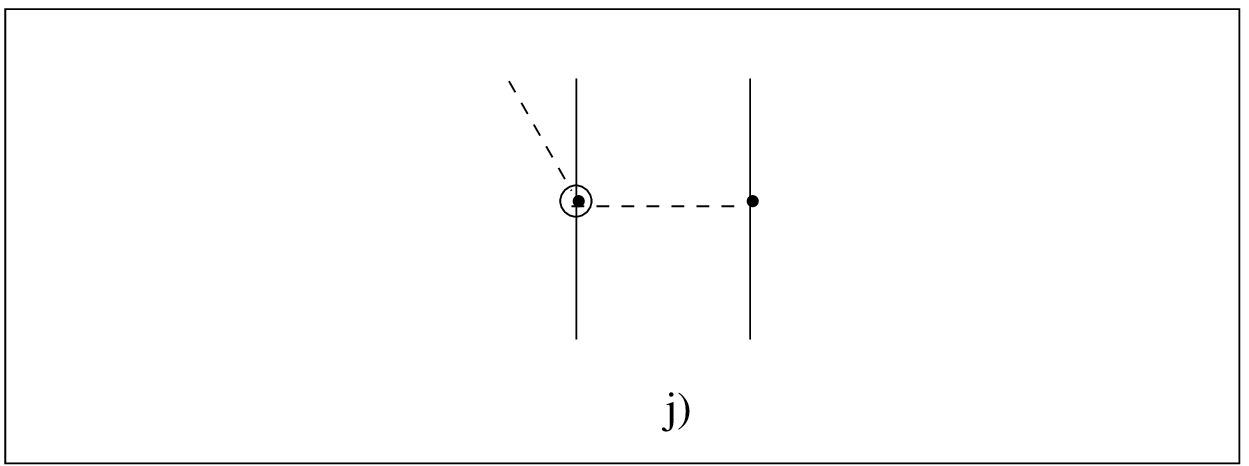, height=2.7cm}} & ${m_\pi^2}/{(pM_N)}$ & NLO & N$^2$LO  \\
\hhline{~---}
\parbox[c]{7.3cm}{\epsfig{file=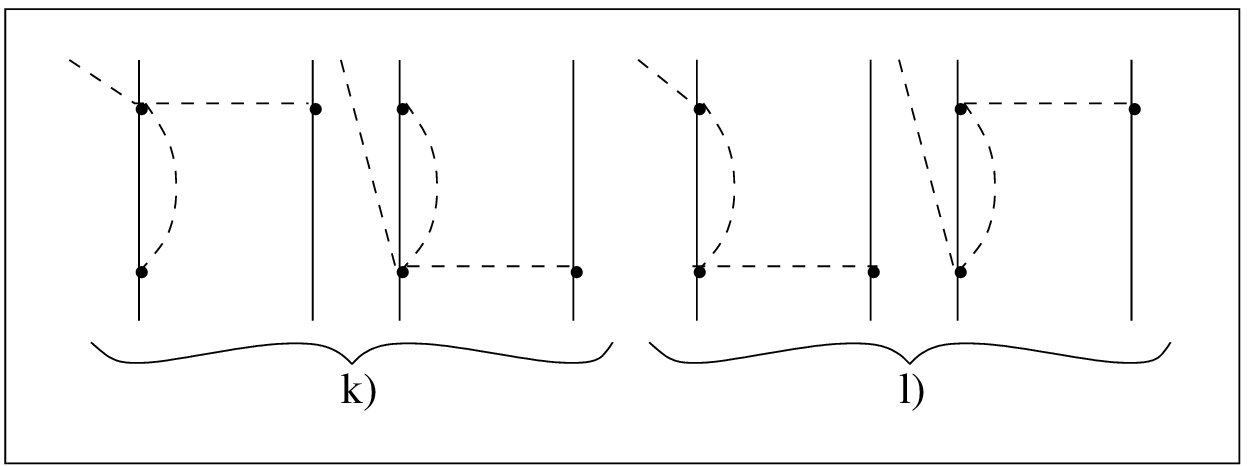, height=2.7cm}} &
${m_\pi^3}/{pM_N^2}$ & N$^2$LO & N$^4$LO  \\
\hhline{~---}
\end{tabular}
\end{center}
\label{diatabs}
\end{table}

Let us now turn to the $s$--wave contributions. The leading diagrams
containing nucleons only
are shown in table \ref{diatabs};
those that contain the $\Delta$ are shown in table \ref{diatabs_D}.
Again, the Weinberg scheme and the new scheme are compared.
The list is complete up to NNLO in both schemes. Please note, however,
that one class of diagrams ($k$ and $l$) that is of NLO in the Weinberg
counting in the new scheme is pushed to N$^4$LO!
 
\renewcommand{\arraystretch}{1.4}
\begin{table}[ht!]
\caption{\it Comparison of the corresponding chiral order in the
Weinberg scheme ($p \sim m_\pi$) and the new counting scheme ($p \sim \sqrt{m_\pi M_N}$)
for the leading and next--to--leading $\Delta$
 contributions for $s$--wave pion production.
Subleading vertices are marked as $\odot$.
 Not shown explicitly are the
recoil corrections for low--order diagrams.}
\label{diatabs_D}
\begin{center}
\begin{tabular}{c c | c | c  }
$s$--wave diagrams & scale & $\Delta \sim m_\pi $  & $\Delta \sim$ \\
(Deltas only)& &\phantom{$\sqrt{m_\pi M_N}$} & $\sqrt{m_\pi M_N}$ \\
\hhline{~---}
\parbox[c]{7.3cm}{\epsfig{file=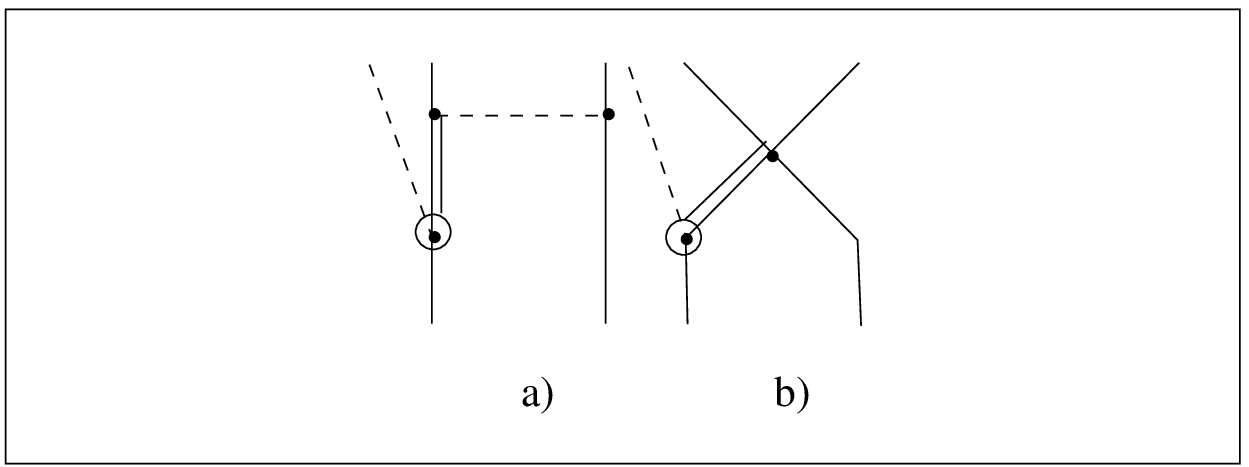, height=2.7cm}} & ${pm_\pi}/{\Delta M_N}$
 & NLO & NLO \\
\hhline{~---}
\parbox[c]{7.3cm}{\epsfig{file=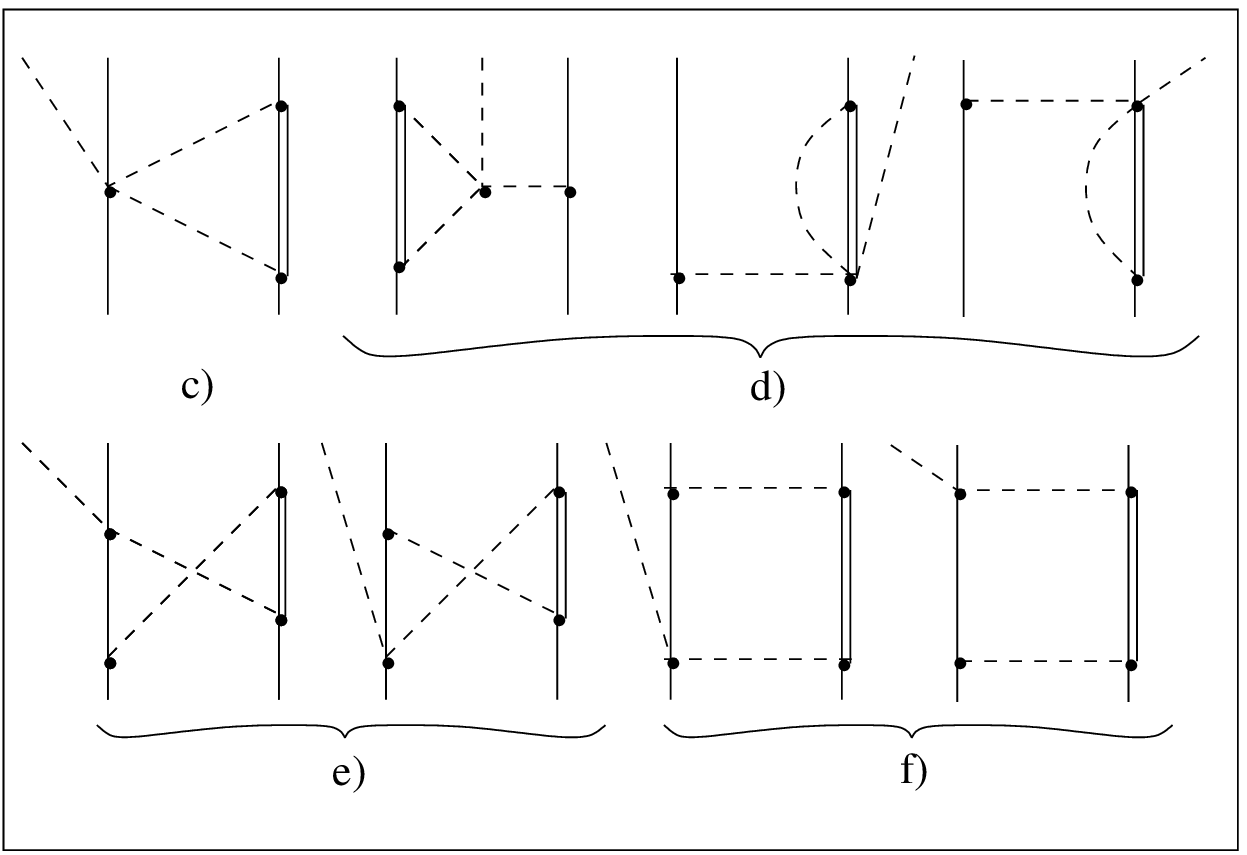, height=5.cm}} & $p^3/(\Delta M_N^2)$
 & N$^2$LO & NLO \\
\hhline{~---}
\end{tabular}
\end{center}
\end{table}
Thus in the new counting scheme---contrary to the Weinberg scheme---the leading
pieces of some loops appear one order lower than the tree level isoscalar
rescattering amplitudes.  As can also be read from the table, the corresponding
order is $m_\pi/M_N$.  If we consider, in addition, that due to the odd parity
of the pion the initial state has to be a $p$--wave, the loops themselves have
to scale as $\sqrt{m_\pi}$. Since no counter term non--analytic in $m_\pi$ is
allowed, the loops have to be finite or to cancel exactly.  This
requirement is an important consistency check of the new counting scheme.

The details of the loops calculations in threshold kinematics can
be found in Ref. \cite{withnorbert}. Here we only give the results.
Note, we only evaluate the leading order pieces of the 
 integrals corresponding to the diagrams of table 
(\ref{diatabs}). For example, in the integrals we
drop terms of order $m_\pi$ compared to $l_0$.

After these simplifications, straightforward evaluation gives for the production amplitude from
 the loops with nucleons only
\begin{equation}
A_n=\frac{i}{F_\pi^3}g_A^3
\left(\vec \sigma_1 \cdot \frac{\vec k}{|\vec k {}|}\right)
\left(\frac{M_Nm_\pi}{128F_\pi^2}\right)\iota_n \ ,
\end{equation}
where $n$ denotes the diagram (labels as in the figure).
For the isospin functions we find
\begin{eqnarray}
\iota_d  =   -\tau_1^c \ , \qquad
\iota_e  =   -\frac{1}{4}(\ \tau_1^c+\tau_2^c \ ) \ , \qquad
\iota_f  =   \frac{3}{2}\tau_1^c \ .  
\end{eqnarray}
Those can be easily evaluated in the different isospin channels. We find
for $\iota(TT_3)=\langle TT_3|\iota_d+\iota_e+\iota_f|11 \rangle$ (it is sufficient here
to look at the $pp$ initial state only)
\begin{eqnarray}
\iota(11)=-1-\frac{1}{2}+\frac{3}{2}=0 \ , \qquad 
\iota(00)=-1+0+\frac{3}{2}=-\frac{1}{2} \ .
\label{iotavalues}
\end{eqnarray}

The first observation is that the NLO contributions are of the order of
magnitude expected by the power counting, since
$$
\left(\frac{M_Nm_\pi}{128F_\pi^2}\right) \ = \ 0.8\chi^2 \ ,
$$
where we used $F_\pi = 93$ MeV.
The power counting proposed in Refs. \cite{chiral2,ourpwaves} thus is indeed
capable of treating properly the large scale inherent to the the $NN \to NN\pi$ reaction.

We checked that our results for the individual diagrams agree with the leading
non--vanishing pieces from the calculations of Ref. \cite{dmitra}\footnote{In
  this reference the same choice for the pion field is made and thus a
  comparison of individual diagrams makes sense. Note that there is a sign error
  in the formula for diagram $f$ in Ref. \cite{dmitra}. We are grateful to F.
  Myhrer for helping us to resolve this discrepancy. }.  In addition, for
almost all diagrams given in table \ref{diatabs}, Ref. \cite{dmitra} gives
explicit numbers for the amplitudes in threshold kinematics. It is intriguing
to compare those to what one expects from the different counting schemes. This
is done in table \ref{expecvscalc}, where the first line specifies the
particular diagram according to table \ref{diatabs} and the second gives the
result of the analytical calculation of Ref. \cite{dmitra}\footnote{In Ref.
  \cite{dmitra} the full result for the particular loops are given. Thus, any
  loop contains higher order contributions.} (normalized to the first column).
In the following two lines those numbers are compared to the expectations
based on the counting schemes---first showing those for the Weinberg scheme
and then those for the new counting scheme. As can clearly be seen, the latter
does an impressive job of predicting properly the hierarchy of diagrams.
Thus, at least when ISI and FSI are neglected, the counting scheme proposed is
capable of dealing with these large momentum transfer reactions.

\renewcommand{\arraystretch}{1.4}
\begin{table}[t!]
\caption{\it Comparison of the results of the analytic calculation of
Ref. \protect\cite{dmitra} with the expectations based on the two counting
schemes as discussed in the text. The diagrams are labeled as in Fig.
\protect\ref{diatabs} (the label $j_R$ shows that here the recoil term of
diagram $j$ is calculated; it appears at $NNLO$ in the Weinberg scheme as
well as in the new counting scheme).}
\begin{center}
\begin{tabular}{c |c |c|c |c |c |c | }
Diagram & d & e & g & $j_R$ & k & l \\
\hline
Ref. \cite{dmitra} & 1.0\phantom{0} & -1.0\phantom{0} & 
0.1\phantom{0} & 0.4\phantom{0} & 0.03 & 0.02 \\
$p \sim m_\pi $ & 1.0\phantom{0} &  1.0\phantom{0} & 1.0\phantom{0} &
 1.0\phantom{0} & 1.0\phantom{0} & 1.0\phantom{0} \\ 
 $p\sim  \sqrt{m_\pi M_N}$     & 1.0\phantom{0} &  1.0\phantom{0} &
 0.4\phantom{0} & 0.4\phantom{0} & 0.06 & 0.06
\end{tabular}
\end{center}
\label{expecvscalc}
\end{table}

It is striking that the sum of loops in the case of the neutral pion production
vanishes ($\iota (11)=0$ in Eq. \eqref{iotavalues}). In addition, for 
neutral pion production there is no meson exchange current at leading order
and the nucleonic current (diagrams b) and c) in table \ref{diatabs}) gets
suppressed by the poor overlap of the initial and final state wave functions
(see discussion in sec. \ref{remprodop})---an effect not captured by the
counting---and interferes destructively with the direct production off the
Delta (diagrams a) and b) in table \ref{diatabs_D}). Thus the first
significant contributions to the neutral pion production appear at NNLO. This
is the reason why many authors found many different mechanisms, all of similar
importance and capable of removing the discrepancy between the Koltun and
Reitan result and the data, simply because there is a large number of diagrams
at NNLO. The situation is very different for the charged pions. Here there is
a meson exchange current at leading order and there are non--vanishing loop
contributions. We therefore expect charged pion production to be significantly
better under control than  neutral pion production and this is indeed what
we found in the phenomenological model described in the previous section.

Next let us have a look at the loops that contain Delta isobars (c.f. Fig.
\ref{diatabs_D}).  In Ref. \cite{withnorbert} it was shown that the individual
loops diverge already at leading order, because the Delta--Nucleon mass
difference introduces a new scale. Therefore, as was argued above, the sum of
the diagrams has to cancel, for at NLO there is no counter term. Is is an
important check of the counting scheme that this does indeed happen. We take this
as a strong indication that expanding in $\chi$ is  consistent with the
chiral expansion.

The observation that we have now established a counting scheme for reactions
of the type $NN\to NN\pi$ also has implications for the understanding of other
reactions. One example is the analysis elastic $\pi d$ scattering, which is
commonly used to extract the isoscalar $\pi N$ scattering length that is
difficult to get at otherwise \cite{ulfpid}.  As mentioned above, due to
chiral symmetry constraints the isoscalar scattering length does not get a
contribution at leading order and thus, for an accurate extraction of this
important quantity from data on $\pi d$ scattering or on $\pi d$ bound states,
a controlled calculation of the few body--corrections is compulsory. One of
these corrections are the so--called dispersive corrections (see Ref.
\cite{piddisp} and references therein): loop contributions to the elastic
scattering that have intermediate two--nucleon states. Obviously, the
imaginary part of those loops gives the essential contribution to the
imaginary part of the $\pi d$ scattering length\footnote{About one third of
  the imaginary part was found to be related to the reaction $\pi d\to \gamma
  NN$ \cite{elt}.}, however there is also a real part to these loops that
needs to be calculated within a scheme consistent with that used for the
calculation of the other contributions.  Within chiral perturbation theory
that has not been done up to now.  Given the progress reported here, such a
calculation is now feasible.  The same technique can then also be used to
calculate the corresponding corrections for $\pi ^3He$ scattering, recently
calculated in chiral perturbation theory for the first time \cite{he3}.

\subsection{On the significance of off--shell effects}
\label{offshell}

A few years a ago there existed a strong program at
several hadron facilities to measure bremsstrahlung
in $NN$ collisions with the goal to identify the
{\it true} off--shell behavior of the $NN$ interaction.
Indeed it was found that the predictions for several highly differential
observables are significantly different
when different $NN$ potentials are employed.

On the other hand, a change in the off--shell behavior of any $T$--matrix can be realized on 
the level of the effective interaction by a field redefinition and it is known since long
that $S$ matrix elements do not change under those 
transformations (as long
as one works within a well defined field theory)\cite{haag,lsz}. Thus, for any given set
of fields the off--shell amplitudes might well have a significant impact
on the values of various observables, but it is an intrinsic feature of
quantum field theory that they can not be separated from the short range
interactions constructed within the same model space.
In a very pedagogical way those results were presented later in Refs. \cite{fear1,fear2}.

\begin{figure}
\begin{center}
\vskip 3cm          
\includegraphics{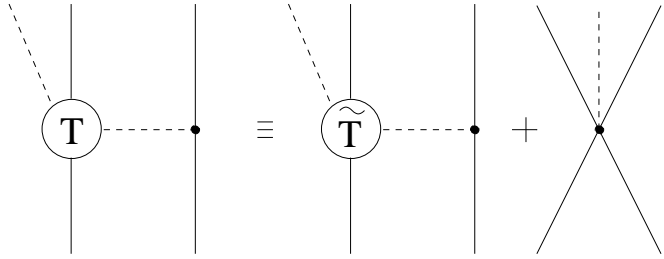} 
\caption{\it Visualisation of Eq. (\protect{\ref{os}}).}
\label{offshellfig}
\end{center}
\end{figure}

How can one understand this seemingly contradictory
situation: on the one hand off--shell amplitudes
enter the evaluation of matrix elements and in some
cases influence significantly the result (c.f. discussion
in previous sections), on the other hand they are claimed
not to have any physical significance? 
Following Ref. \cite{fear1} 
to start the discussion, let us consider some general half--off 
shell $\pi N$ $T$--matrix. In a covariant form the list 
of its arguments contains the standard Mandelstam variables
$s$, $t$ as well as $q^2$---the four momentum squared of the off--shell
particle(s). For our discussion let this be the incoming pion and
for simplicity  omit all spin indices. We now define $T_R (s,t,q^2)$ via
$$
T(s,t,q^2) = \tilde T(s,t,q^2)+(q^2-m_\pi^2)T_R(s,t,q^2) \ ,
$$
where $\tilde T(s,t,q^2)$ is arbitrary up to the
condition that it has to agree with the on--shell amplitude for  $q^2=m_\pi^2$.
Obviously, this is always possible and we can assume $T_R$ to be
smooth around the on--shell point. Then, when introduced into
the $NN\to NN\pi$ transition amplitude, we get
\begin{equation}
A = T\frac{1}{q^2-m_\pi^2}W_{\pi NN} = \tilde T\frac{1}{q^2-m_\pi^2}W_{\pi NN}
+T_RW_{\pi NN} \ , 
\label{os}
\end{equation}
where $W_{\pi NN}$ denotes the $\pi NN$ vertex function.
This example highlights two important facts: 
i) a change in the off--shell dependence of a particular amplitude can always be 
compensated via an appropriate additional contact term. The only quantities 
that are physically accessible are $S$--matrix elements and those are by definition
on--shell; and
ii) the contribution stemming from the off--shell $\pi N$ $T$--matrix is a short
range contribution. Therefore the distinction between short range and off--shell
rescattering is artificial.

\subsection{Lessons and outlook}
\label{pionlessons}

The previous section especially  made clear that it
is rather difficult to construct a model that gives
quantitatively satisfactory results for the reaction
$pp\to pp\pi^0$. On the other hand, it turned out that
for the reaction $pp\to pn\pi^+$ current models as well
as the effective field theory approach do well. In
the previous sections this difference was traced back to a suppression
of 
meson exchange currents in the neutral pion production. What
is the lesson to be learned from this? First of all, 
any model for meson production close to the threshold should
contain the most prominent meson exchange currents;
as long as these are not too strongly suppressed, one can
expect them to dominate the total production cross
section close to the threshold. However if there is
no dominant meson exchange current, then there is 
a large number of sub--leading operators that compete
with each other and make a quantitative understanding
of the cross section difficult. The optimistic conclusion
to be drawn from these observations is that (also
for heavier mesons)  the leading meson exchange
currents should give a reasonable description of the
data, while the contributions from irreducible loops 
largely cancel. Obviously, in heavy meson production there is
no reason anymore to consider pions only as the exchange particles. 
In this sense it is  $\pi^+$ production that is the
typical case, whereas the $\pi^0$ is exceptional due to
the particular constraints from chiral symmetry in that channel.
Note that also in the case of pion photoproduction close to
the threshold
the $\pi^0$ plays a special role, in that the $s$--wave
amplitude is dominated by loops \cite{pi0photo}.

\begin{figure}[t]
\begin{center}
\epsfig{file=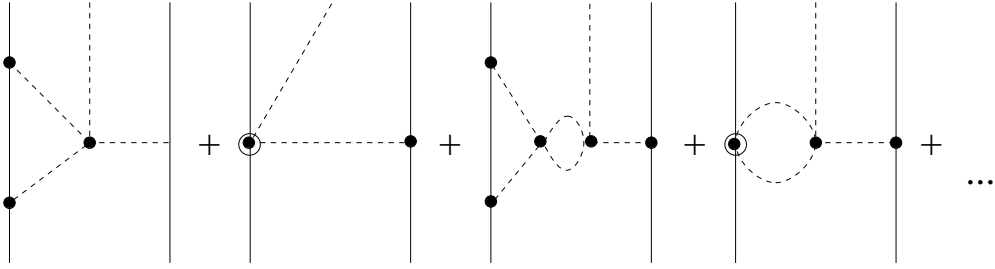, height=3.5cm}
\caption{\it The one sigma exchange as it is perturbatively 
built up in the effective field theory, starting from the left
with the lowest order diagram (NLO). The chiral order increases by one power
in $\chi$ between each diagram from the left to the right.}
\label{chsig}
\end{center}
\end{figure}

Let us look in somewhat more detail at the production operator for neutral
pion production. Within the model described above, the most prominent diagram
for neutral pion production close to the production threshold is pion
rescattering via the isoscalar pion--nucleon $T$--matrix that, for the kinematics
given,
is dominated by a one--sigma exchange
(diagram b) of Fig. \ref{kur}, where the $T$--matrix is replaced by the
isoscalar potential given by diagram e) of Fig. \ref{pinpot}). 
Within the effective field theory the isoscalar potential is built up
perturbatively. This is illustrated in Fig. \ref{chsig}. As was shown above,
the leading piece of the one--sigma exchange gets canceled by other loops
that cannot be interpreted as a rescattering diagram (the sum of diagrams d),
e) and f) of table \ref{diatabs} vanishes) and are therefore not included in
the phenomenological approach. 
This is an indication that in order to improve the phenomenological approach,
at least in case of neutral pion production, pion loops should be considered as
well\footnote{Note that within the effective field theory approach the
  convergence of the series shown in Fig. \ref{chsig} should also be checked, as
pointed out in a different context in Ref. \cite{osetnnsig}.}.

There is one more important conclusion to be drawn from the 
insights reported in the two previous sections: We
can now identify what was missing in the straightforward
extension of the Bonn potential reported in Ref. \cite{EHSM}.
This is especially relevant, since we will see that the
failure to describe the low energy pion production data
in that approach does not point at a missing degree
of freedom in the nucleon--nucleon phenomenology, but
at an incomplete treatment of the cut structure.

\begin{figure}
\begin{center}
\vskip 10cm          
\includegraphics{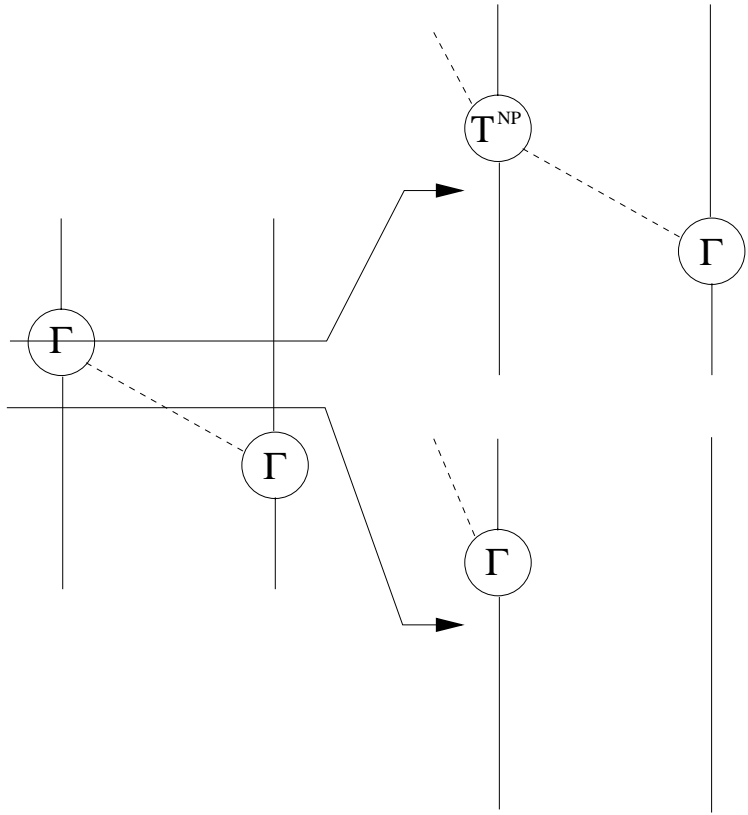} 
\caption{\it Leading cut structure of the one pion exchange potential
in nucleon--nucleon scattering.}
\label{cuts}
\end{center}
\end{figure}

Diagrammatically, the dressed $\pi NN$ vertex function is given by

\begin{center}
\parbox[c]{4.3cm}{\epsfig{file=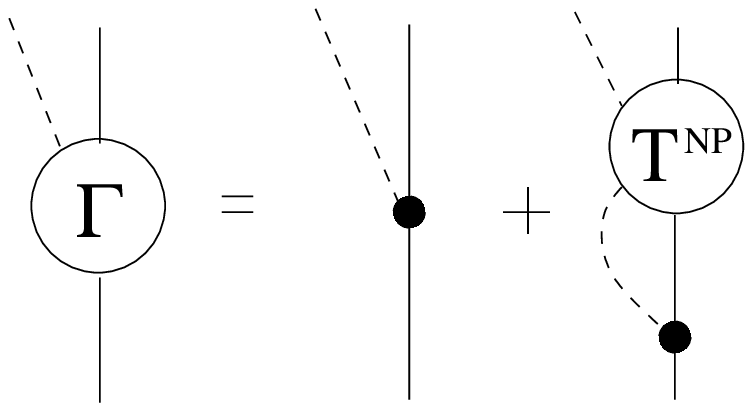, height=2.7cm}} 
\end{center}

where the solid dot indicates the so--called bare vertex
(in this picture it is to be understood as a parameterization
of both the extended structure of the nucleon as well
as meson dynamics not included in the non--pole $\pi N$ $T$--matrix $T^{NP}$,
such as $\pi \rho$ \cite{georgff} or $\pi \sigma$ \cite{rainerff}  correlations)
as well as the dressing due to the $\pi N$ interactions parameterized in
 $T^{NP}$. The latter structure, however, contains a $\pi N$ cut not
considered in Ref. \cite{EHSM}. It is straightforward to map
the different three--body cuts in the one pion exchange potential
to those that must occur in the pion production operator in order to allow for a consistent
description of scattering and production. This is indicated in Fig.  \ref{cuts}:
the additional cut due to the pion dynamics in the vertex function 
is related to the pion rescattering diagram in the production operator.
Therefore, at least close to threshold in Ref. \cite{EHSM}, the
most relevant mechanism that leads to inelasticities was missed.

As we saw, however, as we move away from the near--threshold region
the contribution from the Delta isobar becomes more and more significant. 
Consequently, the results of Ref. \cite{EHSM} look significantly better
at higher energies.


\section{Production of heavier mesons}
\label{heavyprod}
We start this section with some general remarks and then discuss some
special examples. The phenomenology will be discussed only very briefly; for
details on this we refer to the original references or the recent reviews
\cite{machnerrep,oelertrep}.  The list of reactions discussed in detail is by
no means complete. For example, we will not discuss here $\eta '$ production, for
this was already discussed in great detail in Ref. \cite{oelertrep}.  Neither will the
production of vector mesons be discussed in detail. Here we refer to
the recent contributions to the literature (Ref. \cite{kanzonew} and
references therein).

As can
be seen from the headlines already, special emphasis will now be put on the
physics that can be extracted from the particular reactions.

\subsection{Generalities}
As the mass of the produced meson increases, the initial energy
needs to increase as well and consequently also the typical
momentum transfer at threshold. This has various implications:
\begin{itemize}
\item the $NN$ interaction needed for a proper 
evaluation of the initial state interaction is less under control
theoretically; for the $T=0$ channel there is not even a partial wave analysis
available above the $\eta$--production threshold, due to a lack of data
\item a larger momentum transfer makes it more difficult
to construct the production operator; at least one should expect
that the exchanges of  heavier mesons become even more relevant
compared to the pion case
\item it is not known, how to formulate a convergent effective field
  theory\footnote{In principle one might expect an approach like that
    presented in sec. \ref{cpt} to work for the production of all Goldstone
    bosons. However, the next threshold after pion production is that for
     eta production and it already corresponds to an initial momentum of $p_i$=770 MeV.
    Thus the expansion parameter would be
    $\sqrt{m_\eta / M_N}=0.8$.}
\item less is known about the interactions of the subsystems
\item the treatment of three--body singularities due to the
exchange of light particles requires more care.
\end{itemize}
In the upcoming section we will not discuss in detail the role
of the three--body singularities. Only recently was a method developed
that in the future will allow inclusion of those singularities in calculations
within
the distorted wave born approximation \cite{schwamb}. This
method was already applied in a toy model calculation \cite{andreas},
where its usefulness was demonstrated.

In calculations these singularities can occur only if the initial state
interaction is included. For the production of mesons heavier than the $\eta$,
however, no reliable model is available at present.  Up to now no model for the
production of heavy mesons takes three--body singularities into account, and
thus their role is yet unclear.

\subsection{Remarks on the production operator for heavy meson production}
\label{remheavy}

As in sec. \ref{pionprod} we will only look at analyses, that work within the
distorted wave born approximation. For the two--baryon states nothing changes,
other than that the nucleon--nucleon interaction needed for the initial state
interaction gets less reliable from the phenomenological point of
view as we go up in energy. The construction of the production operator,
however, is now even more demanding, for with increasing momentum transfer
heavier exchange mesons can play a significant role. Therefore, in
order to get anything useful out of a model calculation, as many reaction
channels as possible should be studied simultaneously. Only in this way can the
phenomenological model parameters can be fixed and useful information be
extracted. In this context also the simultaneous analysis of $pp$ and $pn$
induced reactions plays an important role, for different isospin structures in
the production operator will lead to very different relative strengths of the
two channels (c.f. discussion in sec. \ref{isospinrole}). Then, once a basic
model is constructed that is consistent with most of the data, deviations in
particular reaction channels can be studied.

\begin{figure}
\begin{center}
\hspace*{-0.2cm}
\epsfig{file=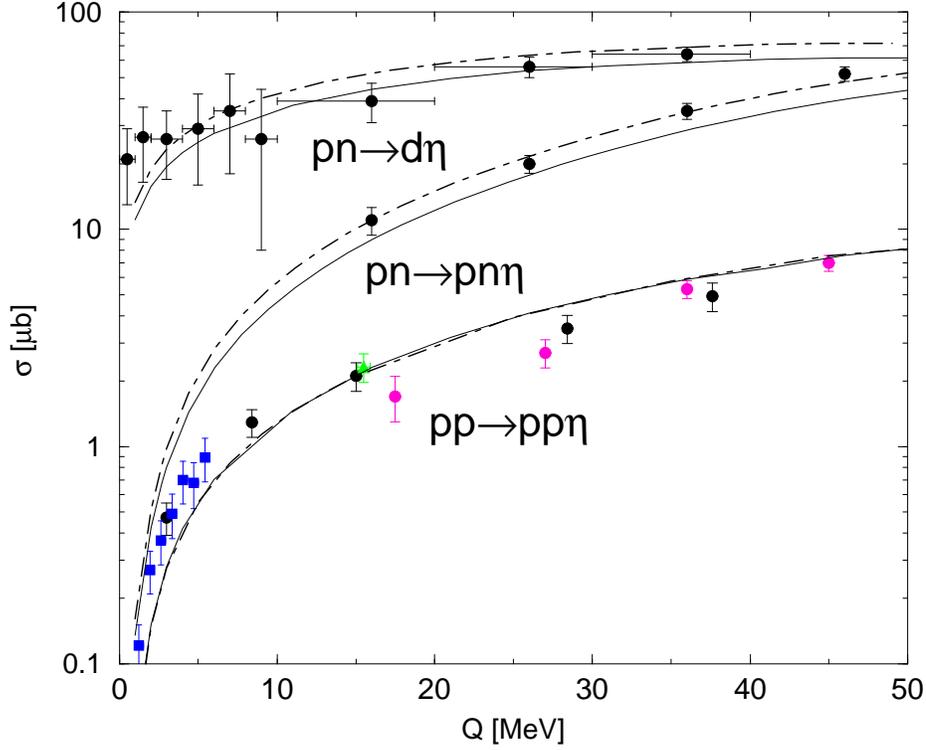, width=12.3cm}
\end{center}
\caption{\it Results for the total cross sections of 
  the reactions $pp\to pp\eta$, $pn\to pn\eta$, and $pn\to d\eta$ from Ref.
  \cite{vadim} employing different $NN$ models for the final state
  interaction.  The solid lines represent the results with the CCF $NN$ model
  \protect\cite{ccf} whereas the dashed-dotted lines were obtained for the
  Bonn B model \protect\cite{OBEPQB}. Data are from Refs.
  \protect\cite{calen2,calen1,calen4,smyrski,paweleta,calen3}.  }
\label{CCFBonnB}
\end{figure}

In Ref. \cite{vadim} the model described in sec. \ref{secphen} was extended to
$\eta$ production in nucleon--nucleon collisions, where the single channel
$\pi N$ $T$--matrix used in the pion production calculations was replaced by
the multi channel meson--baryon model of Ref. \cite{unserroper} to properly
account for the exchange of heavy mesons. So far only $s$--waves were
considered in the calculation. Results for the various total cross sections
using two different models for the final state $NN$ interaction are shown in
Fig. \ref{CCFBonnB}. The studies of Ref. \cite{vadim} indicate, that a
complete calculation for $NN\to NN\eta$ will put additional constraints on the
relative phases of the various meson--baryon$\to \eta N$ transition amplitudes.
Unfortunately, up to now the $\eta$ is the heaviest meson
that can be investigated using this kind of microscopic approach, for there
is no reliable model for the $NN$ interaction for energies significantly above
the $\eta$ production threshold.

The strategy that therefore needs to be followed is to study many reaction
channels consistently. This was done, for example, in a series of papers by Kaiser and others
for $\pi$, $\eta$, $\eta'$ \cite{ulfnovel}, $\omega$ \cite{norbertomega} as
well as strangeness production \cite{norbertkaon} and by Nakayama and
coworkers for $\omega$ \cite{kanzonew,kanzo_omega,kanzodv}, $\phi$
\cite{kanzodv,kanzo_phi}, $\eta$ \cite{kanzo_eta,unsereta} and $\eta'$
\cite{kanzo_etap} production.  Both groups construct the production operator
as relativistic meson exchange currents, where the parameters are constrained
by other data such as decay ratios.  The striking difference between the two
approaches is that the former studies $s$--waves only, does not consider
effects of the initial state interaction and treats the final state
interaction in an approximate fashion. The latter group includes the ISI
through the procedure of Ref. \cite{withkanzo} (c.f. sec. \ref{secisi}),
treats the FSI microscopically and includes higher partial waves as well. It
is important to stress that, where compatible, the two approaches give
qualitatively similar results, as stressed in Ref. \cite{norbertomega}.

Recently it was observed that the onset of higher partial waves can strongly
constrain the production operator \cite{unsereta}. In the same reference it
was demonstrated that to unambiguously disentangle effects from FSI or higher
partial waves, polarization observables are necessary. Thus one should expect
that once a large amount of polarized data is available on the
production of heavy mesons, the production operators and thus the relevant
short--range physics for the production of a particular meson, can be largely
fixed. As an example, this particular issue is discussed in detail in the next
section.

\subsection{The reaction $NN \to NN\eta$ $or$ properties
of the $S_{11}(1535)$}
\label{eta}

The $\eta$ meson is a close relative of the pion, since it is also
a member of the nonet of the lightest pseudo scalar mesons.
It is an isoscalar with a mass of 547.3 MeV---the difference in mass 
from the pion can be understood from its content of hidden strangeness
through the Gell-Mann--Okubo mass formula \cite{go1,go2}.

As for the pion, the $\eta$ couples strongly to a resonance, but a
resonance with different quantum numbers: is it the
positive parity $\Delta (1232)$
for the pion that in the $\Delta$ rest frame 
leads to $p$ wave pion production, the $\eta$ production
is dominated by the negative parity $S_{11}(1535)$\footnote{The quantum
  numbers are chosen in accordance with the partial wave in which the
  resonance would appear in $\pi N$ scattering. Therefore positive parity
  resonances appear in odd partial waves.},
which leads to $s$ wave $\eta$ production in its rest frame. 
As a consequence near--threshold $\eta$ production 
is completely dominated by the $S_{11}(1535)$ resonance.
This statement can well be reversed: studying $\eta$ production
in various reaction channels close to the production threshold
allows selective study of the properties of the $S_{11}(1535)$
resonance in various environments. In this context it is interesting
to note that the nature of this particular resonance is 
under discussion: within the so--called chiral unitary
approach the resonance turns out to be dynamically generated \cite{s11norbert,s11oset},
while on the other hand, detailed studies within the meson exchange 
approach \cite{christian2}, as described in sec. \ref{pionprod}, as well as
quark models (see Ref. \cite{bonnerbaryons} and references therein) call for a genuine
quark resonance. 
One expects that a molecule behaves differently in the
presence of other baryons than a three--quark state due to the naturally
enhanced affinity to meson baryon states of the former.
In this sense data
on $\eta$ production in few--baryon systems should provide valuable information
about this lowest negative parity excitation of the nucleon.

The reaction $NN\to NN\eta$ was studied intensively both theoretically 
\cite{vadim,ulfnovel,kanzo_eta,unsereta,garpen_eta,penaeta,eta_moalem1,eta_moalem2,eta_moalem3,eta_santra,eta_wilkin1,eta_wilkin2,eta_mosel,eta_instanton,eta_vera,eta_batinic} as well
as experimentally (c.f. table \ref{measurements}) in recent years.  The
outstanding feature of $\eta$ production in $NN$ collisions is the strong
effect of the $\eta N$ FSI, most visible in the reaction $pn\to d\eta$, where
in a range of 10 MeV the amplitude grows by about an order of magnitude
\cite{calen3}. This phenomenon was analyzed by means of Faddeev calculations,
showing this enhancement to be consistent with a real part of the $\eta N$ scattering length
of 0.4 fm \cite{garpen_eta}. It is interesting that the same value of the
$\eta N$ scattering length was recently extracted from an analyses of the
reaction $\gamma d\to \eta pn$ \cite{sonja} and that this value is consistent
with that stemming from the microscopic model for meson--baryon scattering of
Ref. \cite{unserroper}. First three body calculations for 
 the reaction with a two--nucleon pair in the
continuum show only a minor impact of the $\eta N$ interaction
on the invariant mass spectra \cite{fixpriv}.

Most model calculations for the production operator share the property that the
production operator was calculated within the meson exchange
picture\footnote{The only exception to the list given above is
  Ref. \cite{eta_instanton}, where the ratio of $pp$-- to $pn$--induced $\eta$
  production is explained by the instanton force. The relation between this
  result and the hadronic approach is unclear \cite{eta_instanton}.}.  This
was then convoluted with the $NN$ FSI, treated either microscopically or in
some approximate fashion. In most analyses it was found that the dominant
production mechanism is via the $S_{11}(1535)$, with the exception of
Ref.
\cite{penaeta}, where the reaction $pp\to pp\eta$ turned out to be dominated
by isoscalar meson exchanges, in analogy with the heavy meson exchanges in the
reaction $pp\to pp\pi^0$. 
The largest qualitative
differences between the various models is the relative importance of the exchanged
mesons. In 
Refs. \cite{ulfnovel,eta_moalem3,eta_santra,eta_wilkin1,eta_wilkin2} the
$\rho$--exchange turned out to be the dominant process, whereas it played a
minor role in Refs. \cite{vadim,eta_mosel,penaeta}. In Ref. \cite{eta_batinic}
the $\rho$--exchange was not even considered, but $\eta$ exchange turned out
to be significant there. 
It remains to be seen, however, if models that are dominated by isoscalar
exchanges, Refs. \cite{eta_batinic,penaeta}, are
capable of accounting for the large ratio of $pn$-- to $pp$--induced eta production,
as discussed in sec. \ref{isospinrole}.
\begin{figure}[t!]
\begin{center} 
\epsfig{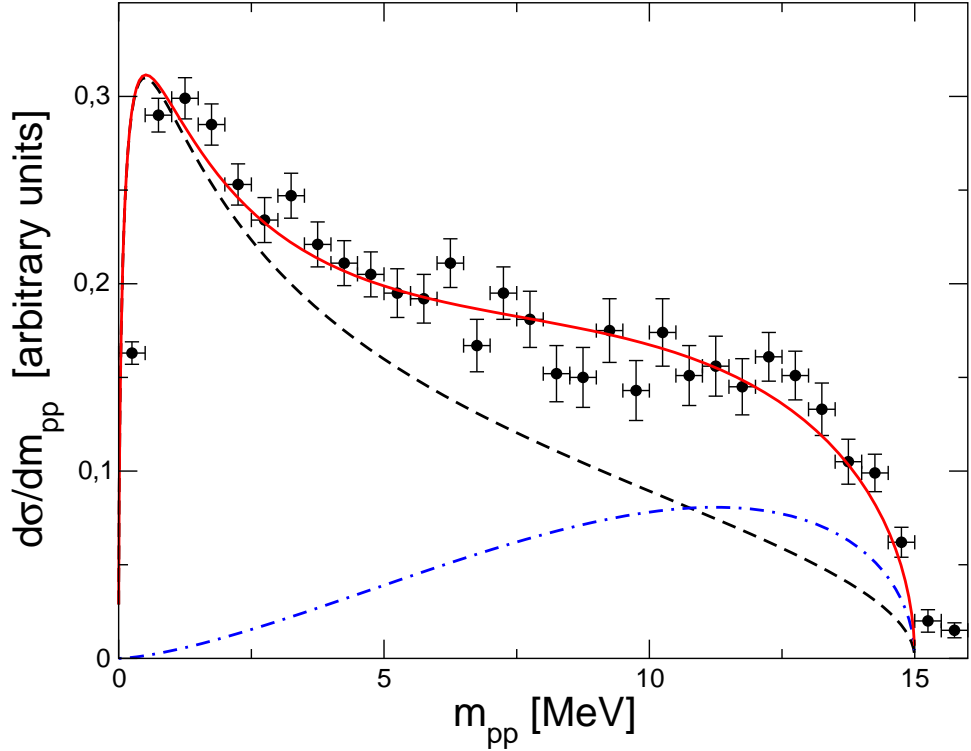}
\caption{\it Invariant mass spectrum for the two--proton system for the
  reaction $pp\to pp\eta$ at $Q$=15 MeV. The dashed line shows the $Ss$
  contribution,
the dot--dashed line the $Ps$ distribution and the solid line the incoherent
  sum of both. The data is taken from Ref. \protect\cite{eduard_eta}.}
\label{etampp}
\end{center}
\end{figure}

As stressed above, in this section we do not want to focus on the details of
the dynamics of the production operator, but instead on the physical aspect of
what we can learn about the $S_{11}$ resonance from studying eta meson
production in $NN$ collisions. Thus, in the remaining part of this section we
will restrict ourselves to a model--independent analysis of a particular set of
data based on the amplitude method introduced in sec. \ref{generalstructure}.
Here we closely follow the reasoning of 
Ref. \cite{unsereta}.

To be definite we will now concentrate on the measurement of angular
distributions and invariant mass spectra for $pp\to pp\eta$ at $Q$=15 MeV
\cite{eduard_eta}. The experiment shows that the angular distribution of both the
two--proton pair and of the $\eta$ are flat, suggesting that only
$s$--waves are present. On the other hand,
the invariant mass distribution $d\sigma /dm_{pp}$ deviates significantly from
what is predicted based on the presence of the strong $pp$ FSI only
(c.f. discussion of sec. \ref{fsisec}). This is illustrated in
Fig. \ref{etampp}, where the distribution for $pp$ $S$--wave, $\eta$ $s$--wave 
($Ss$) is shown as the dashed line. Since the $\eta N$ interaction is known to
be strong, its presence appears as the natural explanation for the deviation
of the dashed line from the data. The structure, however, is even
more pronounced at $Q=42$ MeV (c.f. Ref. \cite{unsereta}) in contrast to what
should be expected for the $\eta N$ interaction.

On the other hand, the discrepancy between the data and
the dashed curve can be well accounted for by a $Ps$ configuration (given by
phase--space times a factor of $p'\, ^2$, c.f. sec. \ref{heur}). This is also
illustrated in Fig. \ref{etampp}, where the pure $Ps$ configuration is shown
as the dashed--dotted line and the total result---the incoherent sum of $Ps$
and $Ss$ cross section---as the solid line.

How is this compatible with the angular distribution of the two nucleon system
being flat? This question can be very easily addressed in the amplitude method
described in sec. \ref{generalstructure}. As long as we resrict ourselves to
$\eta$ $s$--waves and at most $P$ waves in the $NN$ system, the only 
non--vanishing terms in Eq. \eqref{ampdef} are $\vec Q$, containing the amplitude
for the $Ss$ final state, and $\vec A$, containing the two possible amplitudes
for the $Ps$ final state. The explicit expressions for the amplitudes were
given previously in Eqs.
\eqref{a11amps}.

In this particular case we find for example (c.f. Eqs \eqref{si0def}-\eqref{aiidef})
\be
\nonumber
\sigma_0 &=& \frac14\left(|\vec Q|^2+|\vec A|^2\right) \\
\nonumber
A_{0i}\sigma_0 &=& 0 \\
\nonumber
A_{xx}\sigma_0 &=&  \frac14\left(
|\vec Q|^2-|\vec A|^2\right) \ 
\ee
where 
\be
\nonumber
|\vec Q|^2&=& |a_1|^2 \ , \\
\nonumber
|\vec A|^2&=&p' \, ^2\left(|a_2-(1/3)a_3|^2+ x^2\{|a_3|^2-(2/3)\mbox{Re}(a_3^*a_2)\}\right) \ ,
\ee
and $p'x=(\vec p \, '\cdot \hat p)$. Here the amplitudes $a_1$, $a_2$ and
$a_3$ correspond to the transitions $^3P_0\to ^1S_0s$, $^1S_0\to ^3P_0s$, and
$^1D_2\to ^3P_2s$, respectively.
From these equations we directly read that
\begin{itemize}
\item if $a_3$ is negligible, the $pp$ differential cross section $d\sigma/dx$ is flat
 \item the observable $\sigma_0(1-A_{xx})$ directly measures the $NN$ $P$--wave
  admixture
\item any non--zero value for the analysing power is an indication for higher
  partial waves for the $\eta$.
\end{itemize}
It is easy to see that for $a_3=0$ the differential cross section has to be
flat, since then the only partial wave that contributes to the $NN$ $P$--waves
is  $^1S_0\to ^3P_0s$
and a $J=0$ initial state does not contain any information about the beam direction.

Certainly, the $\eta N$ final state interaction has to be present in this
reaction channel as well and it is important to understand its role in
combination with the two--nucleon continuum state. It should be clear,
however, that
in order to understand quantitatively the role of the $\eta N$ interaction, it
is important to pin down the $NN$ $P$--wave contribution first. This is why a
measurement of $A_{xx}$ for $pp\to pp\eta$ is so important.

What does it imply if the conjecture of a presence of $NN$ $P$--waves at
rather low excess energies were true? In Ref. \cite{unsereta} it is
demonstrated that
the need to populate predominantly the $^1S_0\to ^3P_0s$ instead of the 
$^1D_2\to ^3P_2s$ very strongly constrains the $NN\to S_{11}N$ transition
potential. This indicates that a detailed study of $NN\to NN\eta$ should
reveal information about the $S_{11}$ in a baryonic environment that might
prove valuable in addressing the question of the nature of the lowest
negative--parity nucleon resonance.

It should be mentioned that in Ref. \cite{deloff} an alternative
explanation for the shape of the invariant mass spectrum was given, namely an
energy dependence was introduced to the production operator. Based
on the arguments given in the previous chapters, we do not believe that this
is a natural explanation. However, as we just outlined, with $A_{xx}$ an
observable exists that allows to unambiguously distinguish between the two
possible explanations. The experiment is possible at COSY \cite{frankpriv2}.

%
%
%
\subsection{Associated strangeness production $or$
the hyperon--nucleon interaction from production reactions}
\label{yn}

\begin{figure}[t]
\begin{center}
\epsfig{file=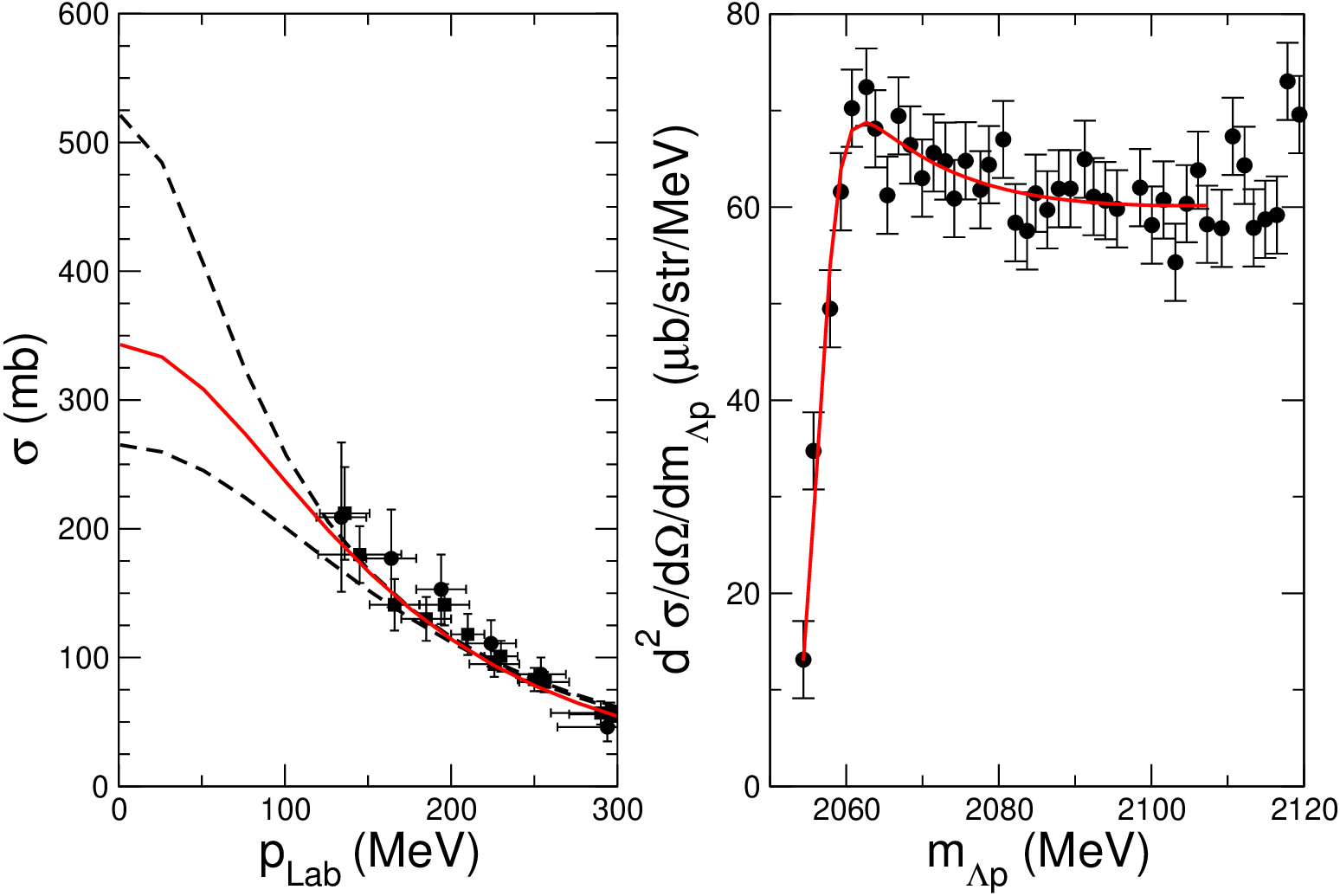, height=9cm}
\caption{\it Comparison of the quality of available data for the reactions
  $p\Lambda$ elastic scattering (data are from
  \protect\cite{sechi,eisele,alexander68}) and $pp\to K^+\Lambda p$ at
  $T_{Lab}$=2.3 GeV \protect\cite{saclay}\protect\footnote{The data was taken
    inclusively (only the kaon was measured in the final state), however, for
    the small invariant masses shown only the $\Lambda$K channel is open.}.
In both panels the red curve corresponds to a best fit to the data. In the left
  panel the dashed lines represent the spread in the energy behavior allowed
  by the data, according to the analysis of Ref. \cite{alexander68}; analogous
  curves in the right panel would lie almost on top of the solid line and are
  thus not shown explicitly.}
\label{elastprodcomp}
\end{center}
\end{figure}

In sec. \ref{syms} the small breaking of $SU(2)$ isospin symmetry present in
the strong interaction was discussed. If we include strange particles in the
analysis we can also study  the breaking of flavor $SU(3)$. It is well--known
that the light mesons and baryons can be arranged according to the irreducible
representations of the group $SU(3)$. The mass splittings within the multiplet
can be well accounted for by the number of strange quarks in some baryon
or meson. However, not much is known about the dynamics of systems that
contain strangeness. Many phenomenological models for, e.g., hyperon--nucleon
scattering \cite{nimYN,rijken,juel1,juel2} use the flavor $SU(3)$ to fix the
meson baryon--meson couplings.  The remaining parameters, like the cut--off
parameters, are then fit to the data. As we will discuss below, so far the
existing data base for hyperon--nucleon scattering is insufficient to judge,
if this procedure is appropriate.

As was stressed previously, effective field theories provide the bridge
between the hadronic world and QCD. In connection with systems that contain
strangeness there are still many open questions. Up to now it is not clear if
the kaon is more appropriately treated as heavy or as light particle. In
addition, in order to establish the counting rules it is important to know the
value of the $SU(3)$ chiral condensate. For a review this very active field of
research as well as the relevant references we refer to Ref. \cite{strangeulf}.

To further improve our understanding of the dynamics of systems that contain
strangeness, better data is needed. The insights to be gained are relevant
not only for few--body physics, but also for the formation of hypernuclei \cite{nogga},
and might even be relevant to the structure of neutron stars (for a recent
discussion on the role of hyperons in the evolution of neutron stars see Ref.
\cite{neutronstars}). Naturally, the hyperon--nucleon scattering lengths are 
the quantities of interest in this context.

\begin{figure}[t!]
\begin{center} 
\epsfig{file=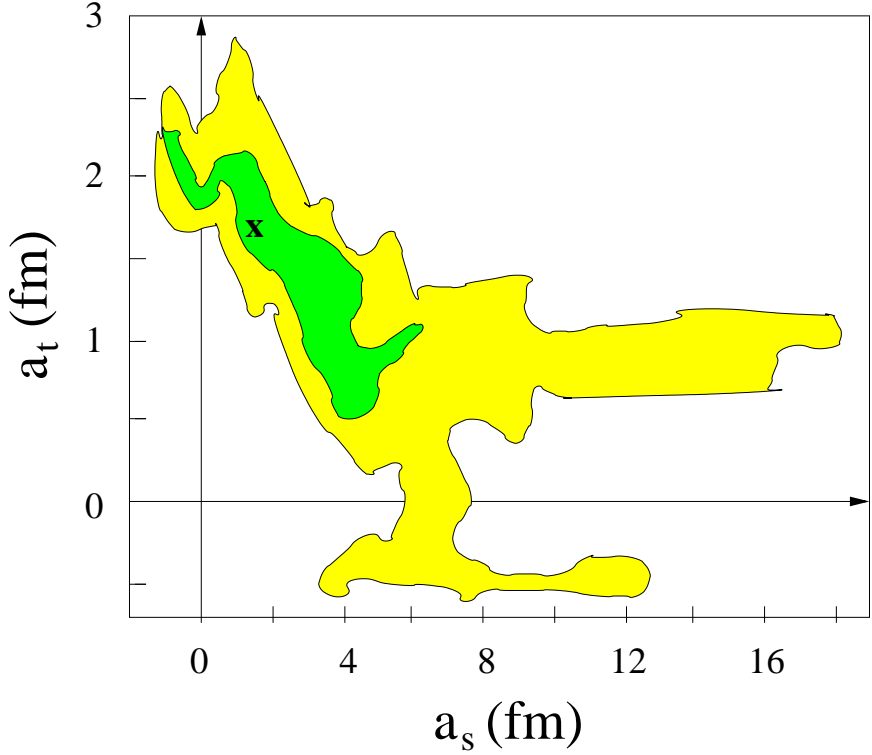, height=10cm}
\caption{\it Values allowed for the spin singlet and spin triplet scattering
  length by the $\Lambda N$ elastic scattering data according to Ref.
  \protect\cite{alexander68}. The dark shaded area denotes the 1$\sigma$ range
  for the parameters and the light shaded area the 2$\sigma$ range. The cross
  shows the best fit value ($a_s=1.8$ fm and $a_t=1.6$ fm).}
\label{lamnelast}
\end{center}
\end{figure}
In the left panel of Fig. \ref{elastprodcomp} we show the world data set for
 elastic $\Lambda N$ scattering.
 In Ref. \cite{alexander68} a
Likelihood analysis based on the elastic scattering  data was performed in order to extract the
low energy $\Lambda N$ scattering parameters. The resulting contour levels are shown in Fig.
\ref{lamnelast}, clearly demonstrating that the available elastic
hyperon--nucleon scattering data do not significantly constrain the scattering
lengths: the data allows for values of (-1, 2.3) as well as (6,1) (all in
fermi) for ($a_s,a_t$) respectively. Later models were used to extrapolate the
 data. However, also in this way the scattering lengths could not be pinned
 down accurately. For example, in 
Ref. \cite{rijken} one can find six different models that equally well
describe the available data but whose ($S$-wave) scattering lengths 
range from 0.7 to 2.6 fm in the singlet channel and from 1.7 to 2.15 fm 
in the triplet channel.
Production reactions are therefore a promising alternative. In the literature
the reactions $K^- d\to \gamma \Lambda n$ \cite{gibbsln}, $\gamma  d\to
 K^+\Lambda n$ (Ref. \cite{gammad5} and references therein) and $pp\to
 pK^+\Lambda$ \cite{jan} were suggested.
Therefore the method described in sec. \ref{secdisp}, that applies to the
 latter two reactions, is an important
step towards a model--independent extraction  of the hyperon--nucleon
scattering lengths.

A natural question that arises is the quality of data needed e.g. for the
reaction $pp\to pK^+\Lambda$ in order to significantly improve our knowledge
about the hyperon--nucleon scattering lengths.
In Ref. \cite{ynfsi} it is
demonstrated that data of the quality of the Saclay experiment
for $pp\to K^+X$ \cite{saclay}, shown in  right panel of
Fig. \ref{elastprodcomp}
\footnote{Shown is only the low $m^2$ tail
of the data taken at $T_{Lab}=2.3$ GeV and an angle of 10$^o$} that had a
mass resolution of 2 MeV, allows for an extraction of a scattering length with
an experimental uncertainty of only 0.2 fm. Note, however, that the actual value of the
scattering length extracted with Eq. \eqref{final} from those data is
not meaningful, since the analysis presented can be applied only if 
just a single partial wave contributes to the invariant mass spectrum.
The data set shown, however, represents the incoherent sum of the $^3S_1$ and the
$^1S_0$ hyperon--nucleon final state.

The two spin states can be separated using polarization measurements. In sec.
\ref{hnpol} as well as in Ref. \cite{ynfsi} it was shown, that for the spin
singlet final state the angular distributions of various polarization
observables are largely constrained. This is sufficient to extract the spin
dependence of the $\Lambda$--nucleon interaction from the reaction $pp\to
pK^+\Lambda$.
%
%
%
%
One more issue is important to stress: to make sure that the structure seen
in the invariant mass spectrum truly stems from the final state interaction
of interest, a Dalitz plot analysis is necessary, for resonances can well distort
the spectra. This is discussed in detail in sec. \ref{unpolobs} as well as at
the end of sec. \ref{secdisp}.

\subsection{Production of scalar mesons $or$
  properties of the lightest scalar}
\label{a0f0}

The lightest scalar resonances $a_0(980)$ and $f_0(980)$ are two
well--established states seen in various reactions \cite{pdb}, but their internal
structure is still under discussion.  Analyses can be found in the literature
identifying these structures with conventional $q\bar{q}$ states (see Ref.\ 
\cite{anisovich} and references therein), compact $qq$-$\bar{q}\bar{q}$ states
\cite{mitbag,achasov} or loosely bound $K\bar K$ molecules
\cite{weinstein,janssen}. In Ref. \cite{osetoller} a close connection between
a possible molecule character of the light scalar mesons and chiral symmetry
was stressed. It has even been suggested that at masses below 1.0 GeV a
complete nonet of 4-quark states might exist \cite{close}.

Resolution of the nature of the light scalar resonances
is one of the most pressing questions of current hadron physics. First of
all we need to understand the multiplet structure of the light scalars in order to
identify possible glueball candidates. In addition, it was pointed
out recently that  $a_0(980)$,  $f_0(980)$ as well as the newly discovered
$D_s$ \cite{ds} might well be close relatives \cite{eef}. Thus, resolving the nature of
the light scalar mesons allows one simultaneously to draw conclusions about the
charmed sector and
 will also shed light on both the confining
mechanism in  light--light as well as light--heavy systems.
 
\begin{figure}[t]
\begin{center}
\epsfig{file=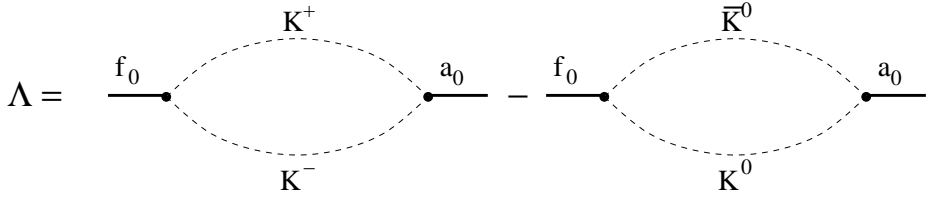, height=2.5cm}
\caption{\it Graphical illustration of the leading contribution to the $f_0-a_0$ mixing
    matrix element $\Lambda$ defined in Eq. (\protect\ref{llam}).}
\label{lcsb} 
\end{center}
\end{figure}

Although predicted long ago to be large \cite{achasovmix}, the phenomenon of
$a_0-f_0$ mixing has not yet been established experimentally. In Ref.
\cite{achasovmix} it was demonstrated that the leading piece of the $f_0-a_0$
mixing amplitude can be written as\footnote{Here we deviate from the original
  notation of Achasov et al. in order to introduce dimensionless coupling
  constants in line with the standard Flatt\'e parameterization.}
\begin{equation}
\Lambda = \langle f_0 |T| a_0\rangle = ig_{f_0K\bar K}g_{a_0K\bar
    K}\sqrt{s}\left( p_{K^0}-p_{K^+} \right) \ + {\cal
    O}\left(\frac{p_{K^0}^2-p_{K^+}^2}{s}\right) \ ,
\label{llam}
\end{equation}
where $p_K$ denotes the modulus of the relative momentum of the kaon pair and
the effective coupling constants are defined through $\Gamma_{xK\bar
  K}=g_{xK\bar K}^2p_K$.
 Obviously, this leading contribution is just that of
the unitarity cut of the diagrams shown in Fig. \ref{lcsb} and is therefore
model--independent. Note that  in Eq.
(\ref{llam}) electromagnetic effects were neglected, because they are expected
to be small \cite{achasovmix}. 

\begin{figure}[t]
\begin{center}
\epsfig{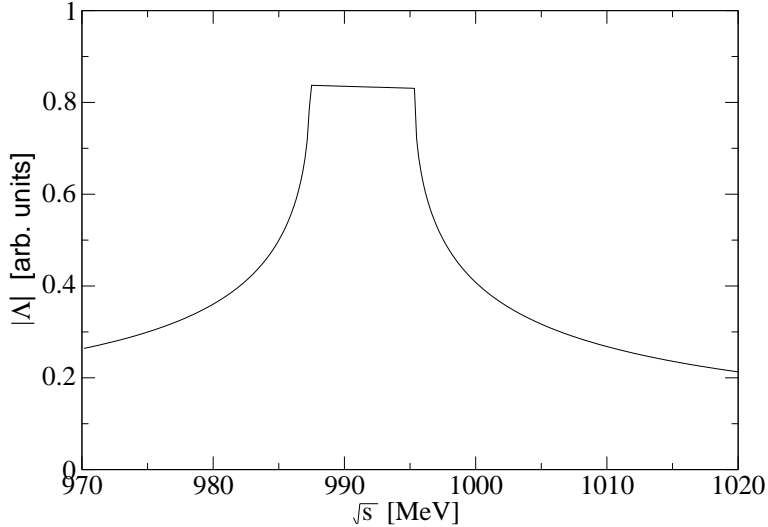}
 \caption{\it Modulus of the leading piece of the mixing amplitude
    $\Lambda$ defined in Eq. (\protect\ref{llam}). The two kinks occur at the
     $K^+ K^-$ (at 987.35 MeV) and the $\bar K^0 K^0$ (995.34 MeV) threshold respectively.}
\label{mixsdep}
\end{center}
\end{figure}

The contribution shown in Eq.  (\ref{llam}) is
unusually enhanced between the $K^+ K^-$ and the $\bar K^0 K^0$ thresholds, a
regime of only 8 MeV width. Here it scales as (this formula is for
illustration only---the Coulomb interaction contributes with similar strength
to the kaon mass difference \cite{gassermasses})
$$\sqrt{\frac{m_{K^0}^2-m_{K^+}^2}{m_{K^+}^2+m_{K^0}^2}}\sim
\sqrt{\frac{m_u-m_d}{\hat m+m_s}}\ ,$$
where $m_u$, $m_d$ and $m_s$ denote the current quark mass of the up, down
and strange quark,
respectively, and $\hat m=(m_u+m_d)/2$. This is
in contrast to common CSB
effects\footnote{Here we denote as common CSB effects those that occur at the
  Lagrangian level.} which scale as $(m_u-m_d)/(\hat m+m_s)$, since they have to
be analytic in the quark masses. It is easy to see that away from the kaon
thresholds $\Lambda$ returns to a value of natural size.  This $\sqrt{s}$
dependence of $\Lambda$ is depicted in Fig. \ref{mixsdep}.

Within a microscopic model for $\pi \pi$ and $\bar K K$ scattering the mixing
of $a_0$ and $f_0$ was studied in Ref. \cite{oli}.
Within this model both resonances are of dynamical origin.
 The only mixing mechanism
considered was the meson mass differences. Within this model the
predictions of Ref. \cite{achasovmix} were confirmed.

As was demonstrated by Weinberg for the case of the deuteron \cite{wein_deut},
the effective couplings of resonances to the continuum states contain valuable
information about the nature of the particles. In the case of the deuteron this
analysis demonstrated that the effective coupling of the deuteron to the $pn$
continuum, as  can be derived from the scattering length and the effective
range, shows that the deuteron is a purely composite system. Recently it was
demonstrated that, under certain conditions that apply in the  case of $a_0$ and
$f_0$,  the analysis can be extended to unstable scalar states as well
\cite{unserprep}. Accurate data on the effective couplings of the scalars to
kaons should therefore provide valuable information about their nature. As was
argued in sec. \ref{syms}, the occurrence of scalar mixing dominates the CSB
effects in production reactions. Therefore quantifying the mixing matrix
element might be one of the cleanest ways to measure the effective decay
constants of $a_0$ and $f_0$.
In sec. \ref{syms} it was demonstrated that from studies of $\pi \eta$
production in $NN$ and $dd$ collisions one should be able to extract the
$a_0-f_0$ mixing amplitude.
Those arguments were supported in sec. \ref{a0amp}, where it was shown, that
the reaction  $pp\to d\bar K^0 K^+$ is indeed dominated by $\bar KK$
$s$--waves.
We should therefore expect a significant signal for the mixing as well.
In the years to come we can thus expect the experimental information about the
scalar mesons to be drastically improved.

%
%


\section{Meson production on light nuclei}

In this section we wish to make a few comments concerning meson production
on light nuclei. 
Note, $dd$ induced meson production was discussed 
to some extent in section \ref{syms}.
A more extensive discussion can be found in Ref. \cite{oelertrep}.

\subsection{Generalities}

Almost all general statements made for meson production in $NN$ collisions
apply equally well for meson production involving light nuclei. Naturally, now
the 
selection rules are different (see discussion at the end of sec. \ref{sr})
and the possible breakup of the nuclei introduces additional thresholds that must
be considered in theoretical analyses.

A comparison of meson production in two--nucleon collisions and in few--nucleon
systems should improve our understanding of few--nucleon dynamics. As a result of a
systematic study of, for example, both $NN$ and $pd$ induced reactions, a
deeper insight into the importance of three--body forces should be gained.
For example, a recent microscopic calculation using purely two--body input to
calculate the reaction $\pi^+ {}^3He\to ppp$, found that the data \cite{pionabsd1,pionabsd2} call for
three--body correlations \cite{pionabs}, confirming earlier studies \cite{pionabsold}.
It should be stressed, however, that this field is still in its infancy and
a large amount theoretical effort is urgently called for.

\begin{figure}[t]
\begin{center}
\epsfig{file=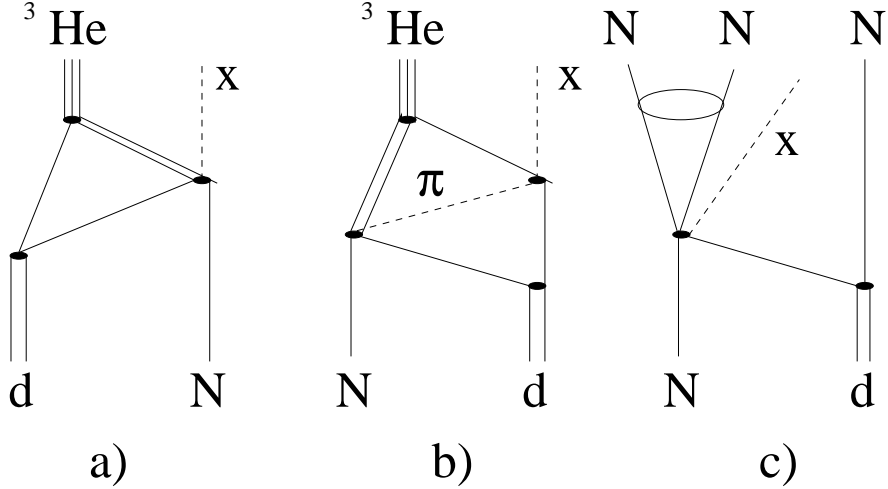, height=6.5cm}
\caption{\it Possible diagrams to contribute to the reaction $Nd\to 3Nx$.
In diagram a) and b) the three nucleons in the final state from a bound
state, whereas diagram c) shows the dominant diagram in quasi--free
kinematics.
Here it is assumed that only a particular nucleon pair (to be identified
through proper choice of kinematics) interacts in the final state.}
\label{pdind}
\end{center}
\end{figure}

\subsection{$2\to 2$ reactions}

The reaction channel studied best up to date is $pd\to ^3\mbox{He}x$. Data
exist mainly from SATURNE for $x=\pi$ \cite{pdpi_exp}, $x=\omega$
\cite{pdomega_exp}, $x=\eta$ \cite{pdeta_exp,pdeta2_exp}, as well as $x=\eta'$ and $\phi$
\cite{pdetapphi_exp}. Most of the experiments were done using a polarized deuteron beam.  In
addition the reactions $dd\to \alpha \pi^0$ \cite{dd2alphapi0}\footnote{Note that
this reaction is charge symmetry breaking and was discussed in
sec. \ref{syms}.} and $dd\to
\alpha \eta$ \cite{ddalphaeta_exp} were measured.

The reaction $pd\to ^3\mbox{He}\pi^0$ could be understood quantitatively from
diagram a) of Fig. \ref{pdind} \cite{wilkin_pdpi}.
However, it turned out that diagram a) alone largely under predicted the data
\cite{laget_eta,wilkin_pd} for the reaction $pd\to ^3\mbox{He}\eta$, whereas
diagram b) contributes sufficiently strongly to allow description of the data
\cite{laget_eta,wilkin_pd2,jain_pdeta}. In all these approaches the individual amplitudes
were taken from data directly; so far no microscopic calculation exists. 
The relevance of the pion exchange mechanism was explained in Ref. \cite{kun}
as what the authors called a kinematic miracle: in the near--threshold regime
the kinematics for eta production almost exactly matches that for $pp\to
d\pi^+$ followed by $\pi^+ n\to \eta p$, with the intermediate pion on--shell.

The striking feature of reactions with an $\eta$--nucleus final state is the
pronounced energy dependence that was already described in sec. \ref{eta} for the reaction
$pn\to d\eta$---often interpreted as a signal of an existing
(quasi)bound state in the $\eta$--nucleus system. Indeed, since the quark
structure of the $\eta \sim \bar uu+\bar dd$ is proportional to the number
operator, one should expect the $\eta$--nucleus interaction to get stronger
with increasing number of light quark flavors present in the interaction
region. For a more detailed discussion of this issue we refer to Ref. \cite{oelertrep}.

\begin{figure}[t]
\begin{center}
\epsfig{file=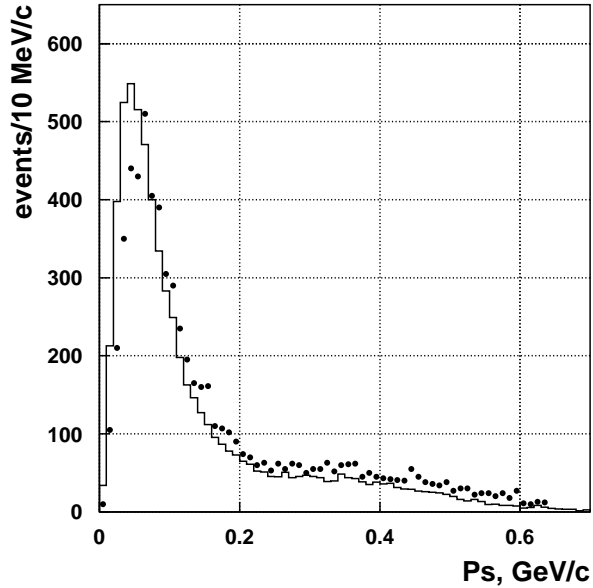,width=9.cm}
\caption{\it Momentum
distribution of proton-spectators in the reaction
$\bar{p}d{\to}3\pi^+2\pi^-p$. The
data are taken from Ref.~\cite{Ahmad} while the solid line is
is based on a calculation that includes the quasi--free production as well as
meson rescattering. The figure is taken from \protect\cite{omegaelast}, where also
details about the corresponding calculation can be found.}
\label{fig:anpi}
\end{center}
\end{figure}

\subsection{Quasi--free Production}

A deuteron is a loosely bound state of a proton and a neutron. For properly
chosen kinematics, the deuteron can therefore be viewed as an effective
neutron beam/target as alternative to neutron beams (see Ref. \cite{oelertrep}
and references therein). The corresponding diagram is shown in Fig.
\ref{pdind}c).

Obviously, the existence of a three--nucleon bound states already shows that
there must be a kinematic regime, where the interaction of all three
nucleons in the final state is significant, namely, when all
three have small relative momenta. On the other
hand, if the spectator nucleon escapes completely unaffected, its momentum
distribution should be given by just half the deuteron momentum, convoluted with
the deuteron wavefunction times the phase--space factors.
 Thus, we should expect a momentum distribution for
the spectator that shows in addition to  a pronounced peak from the
quasi--free production a
long tail stemming from rescattering in the final state. 

Experimental data exist for the proton--spectator momentum distribution
for the reaction  $\bar{p}d{\to}3\pi^+2\pi^-p$  \cite{Ahmad}. In
Fig. \ref{fig:anpi} this data is shown together with the results of a
calculation \cite{omegaelast} that considers both the quasi--free piece as well as a
rescattering piece. The data clearly shows the quasi--elastic peak. Thus, when
considering spectator momenta that are of about 100 MeV or less, to a
good approximation, the reaction should be quasi--free. 

Experimentally this conjecture can be checked by comparing data for $pp$
induced reactions with those stemming from a  $pd$ initial state with a
neutron spectator. Those comparisons were carried out at TRIUMF for pion
production \cite{triumf1,triumf2} as
well as at CELSIUS for $\eta$ production \cite{calen2,calen3},  showing that
the quasi--elastic assumption is a valid approximation.

In sec. \ref{syms} it was argued that a forward--backward asymmetry in the
reaction $pn\to d(\pi^0 \eta)_{s-wave}$ is a good system from which to extract the
$f_0-a_0$ mixing matrix element. If investigated at COSY, this reaction can
only be studied with a deuteron as effective neutron target. Since we are
after a CSB effect, the expected asymmetry is of the order of a few percent.
Thus we need to ensure a priori that the spectator does not introduce an
asymmetry of this order through its strong interaction with, say, the
deuteron.  Theoretically it would be very demanding to control such a small
effect in a four--particle system. Fortunately, we can test experimentally to
what extent the spectator introduces a forward--backward asymmetry by
studying, for example, the reaction $pd\to \pi^+ d n$. If the reaction were
purely quasi--elastic, the angular distribution of the $\pi^+ d$ system in its
rest frame would be forward--backward symmetric, for it would be stemming from
a $pp$ initial state. However, any interaction of the
spectator with either the pion or the deuteron, should immediately
introduce some forward--backward asymmetry. Experimental investigations of
this very sensitive test of the spectator approach are currently under way
\cite{ralfpriv}.


\section{Summary}

The physics of meson production in nucleon--nucleon collisions is very
rich. The various observables that are nowadays accessible experimentally are,
for example, 
influenced by baryon resonances and final state interactions as well as their interference.
It is therefore inevitable to use polarized observables to disentangle the
many different physics aspects.

From the authors personal point of view, the most important issues for the
field are:
\begin{itemize}
\item that $NN$ and $dd$ induced reactions are very well suited for
studies of charge--symmetry breaking \cite{csbrep}. Especially, 
investigation of the reactions  $pn\to d(\pi^0 \eta)_{s-wave}$ and $dd\to \alpha (\pi^0\eta)_{s-wave}$
should shed light on the nature of the scalar mesons. For the experiments
planned in this context we refer to Ref. \cite{a0f0css};
\item that sufficiently strong final state interactions can be extracted from
  production reactions with large momentum transfer, as, for example, $pp\to
  pK^+\Lambda$. The condition for an accurate extraction is a measurement with
  high resolution, as should be possible with the HIRES experiment at COSY \cite{hires};
\item that an effective field theory was developed for large momentum
  transfer reactions such as $NN\to NN\pi$. Once pushed to sufficiently high
  orders, those studies will not only provide us with a better understanding
  of the phenomenology of meson production in $NN$ collisions but also allow us
  to identify the charge symmetry breaking operators that lead to the cross
  sections reported in Refs. \cite{allenapi0,dd2alphapi0};
\item that resonances can be studied in $NN$ induced reactions. Those studies
  are complementary to photon induced reactions, since the resonances can be
  excited by additional mechanisms, which can be selected using, for example,
  spin 0 particles as spin filter \cite{morsch}. In addition, also the properties of
  resonances in the presence of another baryon can be systematically investigated.
\end{itemize}
There are very exciting times to come in the near future, for not only does
the improvement of experimental apparatus in recent years permit the
acquisition of a great deal of high--precision data but also because or
theoretical understanding has now also improved to a point that 
important physics questions can be addressed systematically.

{\bf Acknowledgments}

This article would not have been possible without the strong support,
the lively discussions as well as the fruitful collaborations of the
author with V. Baru, M. B\"uscher, J. Durso, Ch. Elster, A. Gasparyan, 
J. Haidenbauer, B. Holstein, V. Kleber,
 O. Krehl, N. Kaiser, S. Krewald, B. Kubis, 
A.E.  Kudryavtsev,  U.-G. Mei\ss ner, K. Nakayama, J. Niskanen, 
A. Sibirtsev, and J. Speth. 
Thanks to all of you!
I am especially grateful for the numerous editorial remarks by J. Durso.

\vspace{1cm}

{\bf Appendices}


\appendix
\section{Kinematical variables}
\label{kin}

In this report we only deal with reactions, in which
the energies of the final states are so low that
a non--relativistic treatment of the baryons is 
justified. This greatly simplifies the kinematics.
Thus, in the center of mass system we
may write for the reaction $NN\to B_1B_2x$,

\begin{eqnarray}
\nonumber
E_{tot} &=& 2\sqrt{\vec{p}\, ^2+M_N^2} \\
\nonumber
&=& \sqrt{M_1^2+{p'_1}^2}+
\sqrt{M_2^2+{p'_2}^2}+\sqrt{m_x^2+{q'}^2} \\
& \approx & M_1 + M_2 + \frac{p' \ ^2}{2\mu_{12}} + \frac{q' \ ^2}{2(M_1+M_2)}
 +\sqrt{m_x^2+{q'}^2} \ ,
\label{eformeln}
\end{eqnarray}
where $p$ denotes the momentum of the initial nucleons, $p'$ the
relative momentum of the final nucleons and $q'$ the center
of mass momentum of the meson of mass $m_x$.
The reduced mass of the outgoing two baryon system is denoted
by  $\mu_{12} = M_1M_2/(M_1+M_2)$.

The kinematical variable traditionally used for the total energy
of a meson production reaction is  $\eta$,
 the  maximum meson momentum in units of the meson mass.
Then one gets from Eq. (\ref{eformeln}):

\begin{equation}
E_{tot}(\eta)
 \approx M_1 + M_2  + \eta ^2 \frac{m_x^2}{2(M_1+M_2)} +m_x\sqrt{1+\eta ^2}
\end{equation}

\renewcommand{\arraystretch}{1.0}
\begin{table}[h!] 
\begin{center}
\begin{tabular}{|c||c|c||c|c||c|c|}
    \hline
 & \multicolumn{2}{c||}{$pp \to d\pi^+$} &
\multicolumn{2}{c||}{$pp \to pp\pi^0$} &
\multicolumn{2}{c|}{$pp \to pn\pi^+$} \\
    \hline
$\eta$ & \multicolumn{1}{c}{$T_{Lab}$} & \multicolumn{1}{c||}{$Q$} &
 \multicolumn{1}{c}{$T_{Lab}$} & \multicolumn{1}{c||}{$Q$} &
 \multicolumn{1}{c}{$T_{Lab}$} & \multicolumn{1}{c|}{$Q$} \\
    \hline
  0.00 &  287.6 &   0.00 &  279.6 &   0.00 &  292.3 &   0.00 \\
  0.05 &  288.0 &   0.19 &  280.0 &   0.18 &  292.7 &   0.19 \\
  0.10 &  289.2 &   0.75 &  281.2 &   0.72 &  293.9 &   0.75 \\
  0.15 &  291.2 &   1.68 &  283.1 &   1.62 &  295.9 &   1.68 \\
  0.20 &  293.9 &   2.97 &  285.8 &   2.87 &  298.7 &   2.97 \\
  0.25 &  297.5 &   4.62 &  289.2 &   4.46 &  302.3 &   4.62 \\
  0.30 &  301.8 &   6.61 &  293.3 &   6.38 &  306.6 &   6.61 \\
  0.35 &  306.8 &   8.94 &  298.1 &   8.62 &  311.6 &   8.94 \\
  0.40 &  312.5 &  11.58 &  303.6 &  11.17 &  317.3 &  11.58 \\
  0.45 &  318.9 &  14.53 &  309.8 &  14.02 &  323.7 &  14.53 \\
  0.50 &  325.9 &  17.77 &  316.5 &  17.14 &  330.7 &  17.77 \\
  0.55 &  333.5 &  21.29 &  323.9 &  20.53 &  338.3 &  21.29 \\
  0.60 &  341.7 &  25.06 &  331.8 &  24.18 &  346.6 &  25.06 \\
  0.65 &  350.5 &  29.09 &  340.2 &  28.05 &  355.3 &  29.08 \\
  0.70 &  359.8 &  33.34 &  349.1 &  32.16 &  364.6 &  33.34 \\
  0.75 &  369.5 &  37.81 &  358.5 &  36.47 &  374.4 &  37.81 \\
  0.80 &  379.8 &  42.49 &  368.4 &  40.98 &  384.6 &  42.48 \\
  0.85 &  390.5 &  47.36 &  378.6 &  45.67 &  395.3 &  47.35 \\
  0.90 &  401.6 &  52.40 &  389.3 &  50.54 &  406.5 &  52.40 \\
  0.95 &  413.1 &  57.62 &  400.4 &  55.57 &  418.0 &  57.62 \\
  1.00 &  425.0 &  63.00 &  411.8 &  60.75 &  429.9 &  62.99 \\
  1.10 &  449.8 &  74.19 &  435.7 &  71.54 &  454.8 &  74.18 \\
  1.20 &  476.0 &  85.91 &  460.9 &  82.83 &  481.0 &  85.90 \\
  1.30 &  503.4 &  98.10 &  487.1 &  94.57 &  508.4 &  98.09 \\
  1.40 &  531.9 & 110.71 &  514.5 & 106.72 &  536.9 & 110.69 \\
  1.50 &  561.4 & 123.69 &  542.8 & 119.23 &  566.4 & 123.68 \\
    \hline
  \end{tabular}
\end{center}
\caption{
{\it The values for $T_{Lab}$, $Q$ and $\eta$ for the
different channels for pion production.}}
\label{tlabqeta}
\end{table}

Another often used variable for the energy of a meson production
reaction is 
\begin{equation}
Q = \sqrt{s}-(M_1+M_2+m_x) \ .
\end{equation}
It is straightforward to express $\eta$ in terms of $Q$:

\begin{eqnarray}
\nonumber
\eta &=& \frac{1}{2m_x}\sqrt{\frac{(s-M_f ^2-m_x ^2)^2-4(m_x M_f)^2}{s}} \\
&\simeq&  \frac{1}{m_x}\sqrt{2\mu Q}\left(1+\frac{Q}{4\mu}\left(
1-3\frac{\mu}{M_f+m_x}\right)\right) \ .
\label{etarelat}
\end{eqnarray}
where $M_f=M_1+M_2$ and the reduced mass of the full system is given by 
$\mu = M_fm_x/(M_f+m_x)$. For the latter approximation we used that in the
close to threshold regime $Q \ll (M_f+m_x)$.

Traditionally $\eta$ is used for the pion production only, whereas $Q$
is used for the production of all heavier mesons. To simplify the
comparison of results for those different reaction channels, we present in table
\ref{tlabqeta} the various values for $\eta$, $Q$ as well
as $T_{Lab}$ for the different pion production channels.

For completeness we also give the relation to the Laboratory variables

\begin{eqnarray}
T_{Lab} &=& \frac{s-4M^2}{2M} \ , \\
p_{Lab}^2 &=& T_{Lab}^2 + 2MT_{Lab} \ .
\end{eqnarray}


\section{Collection of useful formulas}

\subsection{Definition of Coordinate system}
\label{vecdef}
In this appendix we give the explicit expressions for the vectors
relevant for the description of a general $2\to 3$ reaction. 
We work in the coordinate system, where the $z$--axis is given by
the beam momentum $\vec p$.
These
formulas are particularly useful for sec. \ref{generalstructure}.
Thus we have
\begin{displaymath}
\vec p = p\left(\begin{array}{c} 0 \\ 0 \\ 1 \end{array}\right) \ , \qquad
\vec p\, ' = p'
\left(\begin{array}{c} \sin(\theta_{p'})\cos(\phi_{p'}) \\  
\sin(\theta_{p'})\sin(\phi_{p'}) \\ \cos(\theta_{p'}) \end{array}\right) \ , \qquad 
\vec q\, ' = q'
\left(\begin{array}{c} 
 \sin(\theta_{q'})\cos(\phi_{q'}) \\  
\sin(\theta_{q'})\sin(\phi_{q'}) \\ \cos(\theta_{q'}) \end{array}\right)
\end{displaymath}
From these one easily derives
\begin{eqnarray}
\nonumber
i(\vec p\times \vec p \, ') &=&
 ipp'\left(\begin{array}{c} -\sin(\theta_{p'})\sin(\phi_{p'})
 \\ \sin(\theta_{p'})\cos(\phi_{p'}) \\ 0 \end{array}\right) \ , \\
i(\vec p\, '\times \vec q\, ') &=& ip'q'
\left(\begin{array}{c} \cos(\theta_{q'})\sin(\theta_{p'})\sin(\phi_{p'})
-\cos(\theta_{p'})\sin(\theta_{q'})\sin(\phi_{q'}) \\  
-\cos(\theta_{q'})\sin(\theta_{p'})\cos(\phi_{p'})
+\cos(\theta_{p'})\sin(\theta_{q'})\cos(\phi_{q'})
 \\ \nonumber -\sin(\theta_{p'})\sin(\theta_{q'})\sin(\phi_{q'}-\phi_{p'}) \end{array}\right) \ , \\
\label{crossproducts}
\end{eqnarray}
as well as the analogous expression for $i(\vec p \times \vec q \, ')$.

\subsection{Spin traces}
\label{sptr}

In this appendix some relations are given, that are useful to evaluate 
expression that arise from in the amplitude method appearing described in
sec. \ref{generalstructure}.

\begin{equation}
\sigma_y \chi_i^*\chi_i^T\sigma_y = \frac{1}{2}\sigma_y(1+\vec P_i \cdot \vec \sigma^T)
\sigma_y= \frac{1}{2}(1-\vec P_i \cdot \vec \sigma) \ .
\end{equation}

The sum over the spins of the external particles leads to traces in spin
space, such as

\begin{eqnarray}
\mbox{tr}(\sigma_i )&=&0 \, , \\
\mbox{tr}(\sigma_i \sigma_j)&=&2\delta_{ij} \, , \\
\mbox{tr}(\sigma_i \sigma_\beta \sigma_j)&=& 2i\epsilon_{i\beta j} \, , \\
\mbox{tr}(\sigma_\alpha \sigma_i \sigma_\beta \sigma_j)&=&2
(\delta_{i \alpha}\delta_{j \beta}+\delta_{j \alpha}\delta_{i \beta}
-\delta_{ij}\delta_{\alpha \beta})
\end{eqnarray}
To identify dependent structures the following reduction formula is useful
\be
\nonumber
\vec a \, ^2 \, \vec b \cdot (\vec c \times \vec d)
&=& \\
& &\!\!\!\!\!\!\!\!\!\!\!\!\!\! (\vec a  \cdot \vec b) \, \vec a \cdot (\vec c \times \vec d)
+(\vec a  \cdot \vec d) \, \vec a \cdot (\vec b \times \vec c)
+(\vec a  \cdot \vec c) \, \vec a \cdot (\vec d \times \vec b) \ .
\label{reduc}
\ee To recall the sign or, better, the order of the vectors appearing, 
observe that on the right hand side the vectors other than $\vec a$ are
rotated in cyclic order. To prove Eq. \eqref{reduc} we use \be \nonumber
\delta_{ij}\epsilon_{klm}-\delta_{im}\epsilon_{klj} &=& \epsilon_{kl\gamma}
\left(\delta_{ij}\delta_{m\gamma}-\delta_{im}\delta_{j\gamma} \right) \\
\nonumber
&=&\epsilon_{kl\gamma}\epsilon_{i\gamma \alpha}\epsilon_{jm\alpha} \\
\nonumber
&=& \epsilon_{jm\alpha}
\left(\delta_{k\alpha}\delta_{li}-\delta_{ki}\delta_{l\alpha}
\right)=\delta_{il}\epsilon_{jmk}-\delta_{ki}\epsilon_{jml} \ .
\ee

\section{Partial wave expansion}
\label{pw}

In this appendix we give the explicit relations between the partial wave
amplitudes for reactions of the type $NN\to NNx$ and the spherical tensors
defined in Eq. \eqref{deft}, where $x$ is a scalar
particle. The relations between the spherical tensors and the various
observables is given in tables \ref{tensobs} and  \ref{tensobs2}.

In terms of the
partial wave amplitudes, we can write for two
spin--$\frac{1}{2}$ particles in the initial state

\begin{eqnarray}
\nonumber 
T_{k_1q_1,k_2q_2}^{k_3q_3,k_4q_4} &=& \frac{1}{16\pi}\sum
\langle S'M_S',\vec p \, ',\vec q \, '|M|SM_S, \vec p \rangle \  
\langle \bar S \bar M_S,\vec p |
M^\dagger|\bar S'\bar M_S',\vec p \, ',\vec q \, ' \rangle \ \\
\nonumber
& &\phantom{.........}
\times \ 
\langle S M_S|\tau_{k_1q_1}^{(b)}\tau_{k_2q_2}^{(t)}|\bar S \bar M_S \rangle
\langle \bar S' \bar M_S'|\tau_{k_3q_3}^{(f_1)\,
  \dagger}\tau_{k_4q_4}^{(f_2)\, \dagger }|S'  M_S' \rangle
\\
\nonumber
&=& \frac{1}{4}\sum \sqrt{\frac{(2\bar L+1)(2L+1)}{(2\bar J+1)(2J+1)}} \\
\nonumber & & \phantom{\frac{1}{16P}\sum} \  \times \
\langle S'M_S',L'M_L'|j'M_j'\rangle \langle j'M_j',l'm_l'|JM_J \rangle
\langle S M_S,L0|JM_J \rangle \\
\nonumber & & \phantom{\frac{1}{16P}\sum} \  \times \
\langle S'M_S',\bar L'\bar M_L'|\bar j'\bar M_j'\rangle
 \langle \bar j'\bar M_j',\bar l'\bar m_l'|\bar J\bar M_J \rangle
\langle \bar S \bar M_S,\bar L0|\bar J \bar M_J \rangle \\
\nonumber & & \phantom{\frac{1}{16P}\sum} \  \times \
\langle S M_S|\tau_{k_1q_1}^{(b)}\tau_{k_2q_2}^{(t)}|\bar S \bar M_S \rangle 
\langle \bar S' \bar M_S'|\tau_{k_3q_3}^{(f_1)\,
  \dagger}\tau_{k_4q_4}^{(f_2)\, \dagger }|S'  M_S' \rangle \\
\nonumber & & \phantom{\frac{1}{16P}\sum} \  \times \
Y_{l'm_l'}(\hat q \, ')Y_{L'M_L'}(\hat p \, ')
Y_{\bar l'\bar m_l'}(\hat q \, ')^*Y_{\bar L'\bar M_L'}(\hat p \, ')^* \\
& & \phantom{\frac{1}{16P}\sum} \  \times \
M^\alpha(s,\epsilon) M^{\bar \alpha}(s,\epsilon)^\dagger \ .
\label{startingpoint}
\end{eqnarray}

In order to proceed the following identities are useful \cite{edmonds}:

\begin{eqnarray}
\nonumber
Y_{l_1m_1}(\hat p)Y_{l_2m_2}(\hat p) &=& \sum_{lm} \sqrt{\frac{(2l_1+1)(2l_2+1)}{(2l+1)4\pi}}
\\ & &  \qquad \times \
\langle l_1m_1,l_2m_2|lm \rangle\langle l_1 0,l_2 0|l 0 \rangle
Y_{lm}(\hat p) \\
\langle \sigma | \tau_{kq}| \sigma '\rangle
&=& (-)^q \sqrt{2k+1} \langle 
\frac{1}{2} \sigma , k \ (-q)|\frac{1}{2} \sigma ' \rangle \ .
\end{eqnarray}

The latter, for instance, allows evaluation of the matrix element of the
spin operators:

\begin{eqnarray}
\nonumber
\langle S M_S &|&  \tau_{k_1q_1}^{(b)} \tau_{k_2q_2}^{(t)} \ | \ \bar S \bar M_S \rangle \\
\nonumber
 \ & & \!\!\!\!\! =  \ \sum
\langle S M_S|\tau_{k_1q_1}^{(b)}|m_1 m_2 \rangle 
\langle m_1 m_2 | \tau_{k_2q_2}^{(t)}|\bar S \bar M_S \rangle 
 \ \\
\nonumber
& & \!\!\!\!\! = \sum \langle SM_S|\frac{1}{2} (M_S - m_2), \frac{1}{2} m_2 \rangle
\langle \bar S \bar M_S|\frac{1}{2} m_1, \frac{1}{2} (\bar M_S-m_1) \rangle \\
\nonumber
& & \!\!\!\!\! \ \times \langle \frac{1}{2} (M_S-m_2)|\tau_{k_1q_1}^{(b)}|m_1 \rangle 
\langle m_2 | \tau_{k_2q_2}^{(t)}|\frac{1}{2} (\bar M_S-m_1) \rangle \
\\
\nonumber
 \ & & \!\!\!\!\! = 
\sum (-)^{q_1+q_2}\sqrt{(2k_1+1)(2k_2+1)} \\
\nonumber
& & \!\!\!\!\! \ \times 
\langle SM_S|\frac{1}{2} (M_S - m_2), \frac{1}{2} m_2 \rangle
\langle \bar S \bar M_S|\frac{1}{2} m_1, \frac{1}{2} (\bar M_S-m_1) \rangle \\
& & \!\!\!\!\! \ \times 
\langle \frac{1}{2} (M_S-m_2), k_1 \ (-q_1) |\frac{1}{2} m_1 \rangle
\langle \frac{1}{2} m_2, k_2 \ (-q_2) |\frac{1}{2} (\bar M_S-m_1) \rangle \ .
\end{eqnarray}

It is convenient to couple the remaining spherical harmonics to a common
angular momentum and to define

\begin{equation}
\frac{1}{4\pi}
Y_{\tilde L \tilde M_L}(\hat p)Y_{\tilde l \tilde m_l}(\hat q)
=: \sum_\lambda \langle \tilde L \tilde M_L, \tilde l \tilde m_l|
\lambda Q\rangle B^Q_{\tilde L \tilde l, \lambda}(\hat q, \hat p)
\end{equation}

where we used the fact that the sum of the projections turns out to be equal
to $q_1+q_2 = Q$;
$B$ is then

\begin{equation}
B^Q_{\tilde L \tilde l, \lambda}(\hat q, \hat p)
=\sum _{\mu_L,\mu_l}
 \frac{1}{4\pi}\langle \tilde L \mu_L, \tilde l \mu_l|
\lambda Q\rangle 
Y_{\tilde L \mu_L}(\hat p)Y_{\tilde l \mu_l}(\hat q) \ .
\end{equation}
and normalized such that 
\begin{equation}
\int d\Omega_pd\Omega_q
B^Q_{\tilde L \tilde l, \lambda}(\hat q, \hat p) = \delta_{\lambda 0}
\delta_{\tilde L 0}\delta_{\tilde l 0} \delta_{\tilde Q 0} \ .
\label{bint}
\end{equation}

Some properties of $B$ are derived in the next section.

After putting together the individual pieces we arrive at the final result:

\begin{equation}
\nonumber
{\cal T}_\rho (\hat p, \hat q) = \frac{1}{4}\sum_{\tilde L \tilde l \lambda}
B^Q_{\tilde L \tilde l, \lambda}(\hat q, \hat p) {\cal A}
^\rho
_{\tilde L \tilde l,\lambda} \ ,
\end{equation}
where $\rho = \{k_1q_1,k_2q_2,k_3q_3,k_4q_4\}$, 
and
\begin{equation}
{\cal A}^\rho
_{\tilde L \tilde l,\lambda} = \sum_{\alpha,\bar \alpha}
C^{\alpha,\bar \alpha,\rho}_{\tilde L \tilde l,\lambda}M^\alpha 
(M^{\bar \alpha})\, ^\dagger 
\end{equation}
with
\begin{eqnarray}
\nonumber
C^{\alpha,\bar \alpha,\rho}_{\tilde L \tilde l,\lambda} &=& 
\frac{1}{4\pi}\sum (-)^{M_S+M_S'}\kappa \\
\nonumber & & \phantom{.} \  \times \
\langle S'M_S',L'M_L'|j'M_j'\rangle \langle j'M_j',l'm_l'|JM_J \rangle
\langle S M_S,L0|JM_J \rangle \\
\nonumber & & \phantom{.} \  \times \
\langle \bar S' \bar M_S',\bar L'\bar M_L'|\bar j'\bar M_j'\rangle
 \langle \bar j'\bar M_j',\bar l'\bar m_l'|\bar J\bar M_J \rangle
\langle \bar S \bar M_S,\bar L0|\bar J \bar M_J \rangle \\
\nonumber & & \phantom{.} \  \times \
\langle L' M_L',\bar L' -\bar M_L'|\tilde L \tilde M\rangle
\langle l' m_l',\bar l' -\bar m_l'|\tilde l \tilde m\rangle
\langle \tilde L \tilde M, \tilde l \tilde m|\lambda Q \rangle \\
\nonumber & & \phantom{.} \  \times \
\langle L'    0,\bar L'     0|\tilde L        0\rangle
\langle l'    0,\bar l'     0|\tilde l        0\rangle \\
\nonumber & & \phantom{.} \  \times \
\langle SM_S|\frac{1}{2} (M_S - m_2), \frac{1}{2} m_2 \rangle
\langle \bar S \bar M_S|\frac{1}{2} m_1, \frac{1}{2} (\bar M_S-m_1) \rangle \\
\nonumber
 & & \phantom{.} \  \times \
\langle \frac{1}{2} (m_1+q_1), k_1 \ (-q_1) |\frac{1}{2} m_1 \rangle
\langle \frac{1}{2} m_2, k_2 \ (-q_2) |\frac{1}{2} (\bar M_S-m_1) \rangle  \ \\
 & & \phantom{.} \  \times \
\langle \frac{1}{2} (m_1'-q_3), k_3 \ q_3 |\frac{1}{2} m_1' \rangle
\langle \frac{1}{2} m_2', k_4 \ q_4 |\frac{1}{2} ( \bar M_S'-m_1') \rangle  \ , 
\label{cdef}
\end{eqnarray}
where the sum runs over $\{M_S,M_S',M_L',\bar M_L',m_1,m_1'\}$ and
$$
\kappa = \sqrt{\frac{(2l'+1)(2\bar l'+1)
(2L+1)(2\bar L+1)(2\bar L'+1)(2L'+1)(2k_1+1)(2k_2+1)}
{(2\tilde L+1)(2\tilde l+1)(2J+1)(2\bar J +1)} } \ .
$$

\section{On the non--factorization of a strong final state interaction}
\label{fsistuff}

In this appendix we will  demonstrate the need for a consistent
treatment of both the $NN$ scattering and production amplitudes in order to obtain 
quantitative predictions of meson--production reactions.

Let us assume a separable $NN$ potential
\begin{equation}
V(p', k) = \alpha g(p') g(k) \ ,
\end{equation}
where $\alpha$ is a coupling constant and $g(p)$ an arbitrary real function of 
$p$. With this potential the $T$--matrix scattering equation can be readily solved 
to yield
\begin{equation}
T(p', k) = {V(p', k)\over 1 - R(p') + i\kappa (p')V(p', p')} \ ,
\label{sepT0}
\end{equation}
with
\begin{equation}
R(p') \equiv  m{\bf P} \int_0^\infty \ dk' \ {{k'}^2 V(k',k')
                                                 \over {p'}^2 - {k'}^2} \ 
\label{Rint}
\end{equation}
and  $\kappa(p)=p\pi\mu$ denotes the
phase space density here expressed in terms of the reduced mass of the
outgoing two nucleon pair $\mu=m_N/2$.

Note that for an arbitrary
function $g(k)$, such as $g(k) \equiv 1$ as discussed 
below, $R(p')$ may be divergent. In this case $R$ is to be understood as properly regularized.
The principal value integral $R(p')$ given above is therefore a model-dependent quantity, 
for it depends on the regularization scheme used.
 The condition that the 
on--shell $NN$ scattering amplitude should satisfy Eq. \ref{phsft} relates this 
 to the on--shell potential, 
$V(p', p')$:
\begin{equation}
R(p') = 1 + \kappa(p') \cot(\delta(p')) V(p', p') \ ,
\label{RV}
\end{equation}
where it is assumed that $\eta(p')=1$. This shows that, for a given potential, the
regularization should be such that Eq. (\ref{RV}) be satisfied in order to reproduce 
Eq. \ref{phsft}. Indeed, in conventional calculations based on meson exchange models,
where one introduces form factors to regularize the principal value integral, the 
cutoff parameters in these form factors are adjusted to reproduce the $NN$ 
scattering phase shifts through Eq. \ref{phsft}. Conversely, for a given 
regularization scheme, the $NN$ potential should be adjusted such as to obey 
Eq. (\ref{RV}). This is the procedure used in effective field theories \cite{KSW}, 
where the coupling constants in the $NN$ potential are dependent on the regularization.

We also assume that the production amplitude $M$ is given by a separable form,
\begin{equation}
M(E, k)  =  \beta g(k) h(p) \ ,
\end{equation}
where $\beta$ is a coupling constant and $h(p)$ an arbitrary function of the 
relative momentum $p$ of the two nucleons in the initial state. 
With this we can express the total transition 
amplitude as 
\begin{equation}
A(E,p') = - {1\over \kappa (p')} e^{i\delta(p')}\sin(\delta(p'))
          \left({M(E,p')\over V(p',p')}\right) \ .
\label{sepA}
\end{equation}

Equation \ref{sepA} is the desired formula for our discussion. It allows us to
study the relationship between the $NN$ potential and the production amplitude
$M(E,p')$ explicitly as different regularization schemes are used. For this
purpose let us study the simplest case of a contact $NN$ potential (setting
the function $g=1$) in the limit  $p' \rightarrow 0$. If we regularize the
integrals by means of the power divergent subtraction (PDS) scheme \cite{KSW}
we get
$$
R = - \frac{a\mu}{1-a\mu} \ ,
$$
where $\mu$ denotes the regularization scale. Substituting this result into 
Eq. (\ref{RV}), we obtain  
$$
\alpha =\left( {2a \over \pi m} \right) {1 \over 1 - a\mu}
$$
for the $NN$ coupling strength. Note that for $\mu = 0$ the PDS scheme
reduces to that of minimal subtraction \cite{KSW}. Since the total production 
amplitude $A$ should not depend on the regularization scale we immediately read off 
Eq. \ref{sepA} that
$$
\beta \propto (1-a\mu )^{-1} \ .
$$
Therefore the model clearly exhibits the point made in section \ref{fsisec}: Namely, the 
necessity of calculating
 both the production amplitude and the FSI consistently in 
order to allow for quantitative predictions. 
                                  


\section{Chiral Counting for Pedestrians}
\label{chped}

In this appendix we demonstrate how to estimate the
size of a particular loop integral. This is a necessary
step in identifying the chiral order of a diagram. 
It should be clear, however, that the same methods can be used
to estimate the size of any integral. However, the importance of the chiral
symmetry is that it ensures the existence of an ordering scheme that
suppresses higher loops.

The necessary input are the expressions for the
vertices and propagators at any given order. For
the chiral perturbation theory those can be found in Ref. \cite{ulfbible}.
In addition we need an estimate for the measure of the integral.

\begin{figure}[t!!]
\begin{center}
\epsfig{file=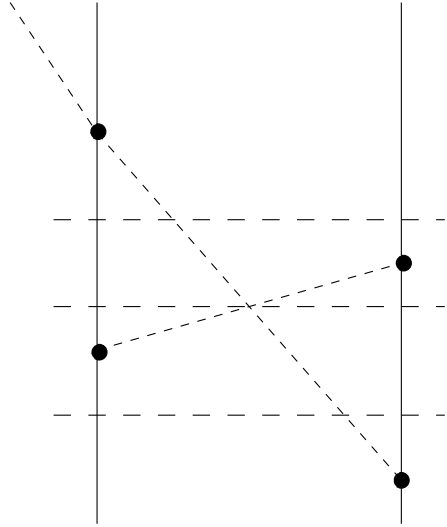, height=7cm}
\caption{\it{A typical loop that contributes to pion production
in nucleon--nucleon collisions. Solid lines denote nucleons and
dashed lines pions. The solid dots show the points of interactions.
 The horizontal dashed lines indicate
equal time slices, as needed for the evaluation of the
diagram in time--ordered perturbation theory.}}
\label{countexample}
\end{center}
\end{figure}

Once each piece of a 
diagram is  expressed in terms of the typical momenta/energies, one
gets an estimate of the value of the particular diagram. 
The procedure works within both time--ordered perturbation theory and
covariant theory.  Obviously, for each irreducible
diagram both methods have to give the same answer. If a diagram
has a pure two--nucleon intermediate state, as is the
case for the direct production, the covariant counting can only give
the leading order piece of the counting within TOPT.

In this appendix we study only diagrams that are three--particle irreducible;
i.e., the topology of the diagram does not allow an intermediate two--nucleon state to
go on--shell. The reducible diagrams require a different treatment and are discussed in 
detail in the main text.
There is another group of diagrams mentioned in the main text that is not
covered by the counting rules presented, namely radiation pions. Those occur
if a pion in an intermediate state goes on--shell. It is argued in the main
text that these are suppressed, because the pion, in order to go on--shell, is
only allowed to carry momenta of the order of the external momenta, and thus
the momentum scale within the loop is of order of the pion mass and not of the
order of the initial momentum. Therefore we do not consider radiation pions
any further.

\subsection{Counting within TOPT}

As mentioned above, if we want to assign a chiral order to a diagram,
 all we need to do is to replace each piece in the
complete expression for the evaluation of the matrix element by its value
when all momenta are of their typical size. In case of meson production
in nucleon--nucleon collisions this typical momentum is given by the initial
momentum $p_i$. Time--ordered perturbation theory contains only three--dimensional integrals
and thus we do not need to fix the energy scale in the integral.

The counting rules are
\begin{itemize}
\item the energy of virtual pions is interpreted as ${\cal
    O}(p_i)$\footnote{There is one exception to this rule: if a time
    derivative acts on a pion on a vertex, where all other particles
are on--shell, then energy conservation fixes the energy.} and thus
\begin{itemize}
\item every time slice that contains a virtual pion is interpreted as $1/p_i$
  (see Fig. \ref{countexample})
\item for each virtual pion line put an additional $1/p_i$ (from the vertex factors),
\end{itemize}
\item interpret the momenta in the vertices as $p_i$
\item every time slice that contains no virtual pion is interpreted as $1/m_\pi$
; most of these diagrams, however,
are reducible (c.f. main text)
\item the integral measure is taken as $p_i^3/(4\pi)^2$.
\end{itemize}
Here we used that $p_i^2/M_N \simeq m_\pi \ll p_i$, in accordance with
Eqs. (\ref{count}),  and thus nucleons can be
treated as static in the propagators if there is an additional pion present.
However if there is a time slice that contains two nucleons only, the
corresponding propagator needs to be identified with the inverse of the
typical nucleon energy $1/m_\pi$ and the static approximation is very bad \cite{ulfnovel}.

In the diagram of Fig. \ref{countexample} three $\pi NN$ vertices, each  $\simeq p_i/f_\pi$, appear
as well as the $\pi\pi NN$ Weinberg--Tomozawa vertex $\simeq p_i/f_\pi^2$. 
In addition the three time slices give a factor $1/p_i^3$ and we also need to
include a factor $1/p_i^2$, since there are two virtual pions.
We 
therefore find
$$
M^{TOPT} \simeq \left(\frac{p_i}{f_\pi}\right)^3\frac{p_i}{f_\pi^2}\left(\frac{1}{p_i}\right)^5
\frac{p_i^3}{(4\pi)^2} \simeq \frac{1}{f_\pi^3}\left(\frac{m_\pi}{M_N}\right) \ .
$$
Here we used $4\pi f_\pi \simeq M_N$ and $p_i \simeq \sqrt{M_Nm_\pi}$.

\subsection{Counting within the covariant scheme}

Naturally, as TOPT and the covariant scheme are equivalent, the
chiral order that is to be assigned to some diagram needs to be
the same in both schemes. The reason why we demonstrate both
is that here we are faced with a problem in which the typical
energy scale $m_\pi$ and the typical momentum scale $p_i$ are
different. In the covariant approach a four--dimensional integral
measure enters and naturally the question arises whether $m_\pi$ or $p_i$ is
appropriate
 for the
zeroth component of this measure $p_0$.
For example, in Ref. \cite{chiral2} it was argued, that one should choose $m_\pi$, although
this choice is by no means obvious from the structure of the integrals.
However, given the experience we now have in dealing with loops
in time--ordered perturbation theory, where these ambiguities do
not occur, the answer is simple:
we just have to assign that scale to $p_0$ that will reproduce 
the same order for any diagram as in the counting within TOPT \cite{withnorbert}.
Once the choice for $p_0$ is fixed, the much easier to use covariant counting
can be used to estimate the size of any loop integral.
Thus have the following  rules:
\begin{itemize}
\item the energy of virtual pions is interpreted as ${\cal O}(p_i)$\footnote{With the 
same exception as in the time ordered situation (c.f. corresponding footnote).}
\item each pion propagator is taken as ${\cal O}(1/p_i^2)$
\item each nucleon propagator that cannot occur in a two--nucleon cut
is taken as ${\cal O}(1/p_0)$ (the leading 
contribution of a nucleon propagator that can occur 
in a two--nucleon cut is ${\cal O}(1/m_\pi)$; most of these diagrams, however,
are reducible (c.f. main text))
\item interpret the momenta in the vertices as $p_i$
\item the integral measure is taken as $p_0p_i^3/(4\pi)^2$ (when
 the diagram allows for a two--nucleon cut
the measure reads $(m_\pi p_i^3)/(4\pi)^2$).
\end{itemize}
Thus we have for the diagram of Fig. \ref{countexample}
$$
M^{cov} \simeq \left(\frac{p_i}{f_\pi}\right)^3\frac{p_i}{f_\pi^2}
\left(\frac{1}{p_i^2}\right)^2\left(\frac{1}{p_0}\right)^2
\frac{p_0p_i^3}{(4\pi)^2} \simeq
\frac{1}{f_\pi^3}\left(\frac{m_\pi}{M_N}\right)\left(\frac{p_i}{p_0}\right) \ .
$$
Thus, we need to assign $p_0\sim p_i$ in order to get the same result in
both schemes. As a side result we also showed, that the nucleons are indeed
static in leading order inside loops that do not have a two--nucleon cut, as
 pointed out in Ref. \cite{ando}.


\listoffigures
\listoftables

\bibliography{refs}
\bibliographystyle{unsrt}

\end{document}